\documentclass[authorversion,screen,acmsmall]{acmart}

\AtBeginDocument{%
  \providecommand\BibTeX{{%
    \normalfont B\kern-0.5em{\scshape i\kern-0.25em b}\kern-0.8em\TeX}}}

\setcopyright{acmcopyright}
\copyrightyear{2021}
\acmYear{2021}
\acmDOI{10.1145/3433000}

\usepackage{multirow}
\usepackage{bbm}
\usepackage{colortbl}

\renewcommand{\C}{\cellcolor[gray]{0.9} Y}

\acmJournal{CSUR}
\acmVolume{54}
\acmNumber{2}
\acmArticle{27}
\acmMonth{3}

\received{May 2020}
\received[revised]{October 2020}
\received[accepted]{October 2020}

\begin{document}

\title{A Survey of Information Cascade Analysis: Models, Predictions, and Recent Advances}

\author{Fan Zhou}
\email{fan.zhou@uestc.edu.cn}
\orcid{0000-0002-8038-8150}
\affiliation{%
  \institution{University of Electronic Science and Technology of China}
  \city{Chengdu}
  \state{Sichuan}
  \country{China}
  \postcode{610054}
}

\author{Xovee Xu}
\email{xovee@live.com}
\orcid{0000-0001-6415-7558}
\affiliation{%
  \institution{University of Electronic Science and Technology of China}
  \city{Chengdu}
  \state{Sichuan}
  \country{China}
  \postcode{610054}
}

\author{Goce Trajcevski}
\email{gocet25@iastate.edu}
\orcid{0000-0002-8839-6278}
\affiliation{%
  \institution{Iowa State University}
  \city{Ames}
  \state{IA}
  \country{USA}
  \postcode{50011}
}

\author{Kunpeng Zhang}
\email{kpzhang@umd.edu}
\orcid{0000-0002-1474-3169}
\affiliation{%
	\institution{University of Maryland}
	\city{College Park}
	\state{MD}
	\country{USA}
	\postcode{20742}
}
\thanks{Work was supported by National Natural Science Foundation of China Grants No.~62072077 and No. 61602097, and NSF Grants No. 1646107 and No. 2030249.}

\authorsaddresses{%
 Authors' addresses: Fan Zhou and Xovee Xu (Corresponding author), University of Electronic Science and Technology of China, N.4, Sect. 2, North Jianshe Rd, Chengdu, Sichuan, China, 610054; emails: fan.zhou@uestc.edu.cn, xovee@live.com; Goce Trajcevski, Iowa State University, 347 Durham, 613 Morrill Rd, Ames, IA 50011-2100; email: gocet25@iastate.edu; Kunpeng Zhang, University of Maryland, VMH4316, 7699 Mowatt In., College Park, MD, 20742; email: kpzhang@umd.edu.
}

\begin{abstract}
  The deluge of digital information in our daily life -- from user-generated content such as microblogs and scientific papers, to online business such as viral marketing and advertising -- offers unprecedented opportunities to explore and exploit the trajectories and structures of the evolution of information cascades. Abundant research efforts, both academic and industrial, have aimed to reach a better understanding of the mechanisms driving the spread of information and quantifying the outcome of information diffusion. This article presents a comprehensive review and categorization of information popularity prediction methods, from \textit{feature engineering and stochastic processes}, through \textit{graph representation}, to \textit{deep learning-based approaches}. Specifically, we first formally define different types of information cascades and summarize the perspectives of existing studies. We then present a taxonomy that categorizes existing works into the aforementioned three main groups as well as the main subclasses in each group, and we systematically review cutting-edge research work. Finally, we summarize the pros and cons of existing research efforts and outline the open challenges and opportunities in this field. 
\end{abstract}



\maketitle

\section{Introduction}\label{sec:intro}The rapid development of wireless communication technologies and the Internet, along with miniaturization and availability of mobile devices, have dramatically changed the way people obtain data and information and use it when interacting with each other. Understanding how information is spread, which factors drive the success of information diffusion, and making predictions about the population size that information can affect are challenging but critical necessities in many real-world application domains, e.g., viral marketing~\cite{leskovec2007dynamics}, advertising~\cite{kempe2003maximizing}, scientific impact quantification~\cite{wang2013quantifying}, recommendation~\cite{wu2019dual}, campaign strategy \cite{bond201261}, and epidemic prevention~\cite{Zhao2019}. 

The trajectories and structures of information diffusion, as well as the adopters/participants in information spreading, form the so-called \textit{information cascades}. Many researches and enterprises have put significant efforts into modeling and learning information diffusion in cellular networks, online social networks, paper citation networks, and content sharing networks, among many others \cite{adamic2016information,asur2011trends,vosoughi2018spread,wang2018attention,zhang2016dynamics}, demonstrating that the ability to predict information items or cascades is of interest from both academic and business perspectives. The word \textit{prediction} may have different meanings in different applications. 
For example, it could refer to predicting the popularity of tweets/hashtags in microblogs~\cite{kobayashi2016tideh,mishra2016feature,zhao2015seismic}, 
the number of ``likes'' for a photo/video in Facebook \cite{cheng2014can}, 
views of videos in YouTube \cite{cheng2013understanding,gursun2011describing,wu2019estimating}, 
ratings of movies in IMDB~\cite{oghina2012predicting}, 
citations of academic papers~\cite{wang2013quantifying},
votes on stories in Digg~\cite{lerman2010using}, 
comments on news articles \cite{tatar2014popularity}, 
or social influence~\cite{qiu2018deepinf} (but few examples). 

We provide a comprehensive review of existing information cascade prediction methods, which is a challenging task both because of the huge amount of publications in this field as well as the lack of uniform standards to classify the existing works. 
To begin with, the specification of the prediction problem itself varies in different works. 
According to the \textit{problem formulation} in different applications, prediction may refer to either binary/multi-class classification or regression. For example, it could be predicting the exact size of a cascade at a future moment~\cite{ahmed2013peek,bakshy2011everyone,kupavskii2013predicting,szabo2010predicting} or just estimating whether a cascade would grow beyond a threshold~\cite{cheng2014can,cui2013cascading}. 
Moreover, \textit{strategically}, based on the range of observed information, the prediction can be made prior to the publication~\cite{martin2016exploring} or by peeking into the early cascade evolution~\cite{shulman2016predictability}. 
From the perspective of \textit{analytical levels}, popularity prediction can be made by focusing on different diffusion levels. For example, \textit{macro-level} models~\cite{li2017deepcas} learn the collective behavior of cascades, while \textit{micro-level} models~\cite{qiu2018deepinf,yang2019neural} focus more on individual user actions/responses to specific information items. 

From the perspective of \textit{methodology}, there are plentiful choices of algorithmic approaches for modeling and predicting information cascades as well as data types associated with the cascades. Traditionally, various features (e.g., temporal and structural) associated with information items can be extracted with feature engineering -- while typical machine learning models (e.g., linear/logistic regressions, na\"{i}ve Bayes, SVM and decision trees) or stochastic process models (e.g., Poisson and Hawkes point processes) are readily applicable for modeling and predicting information popularity. With the recent advances in deep neural networks, especially the techniques for learning graph representation (e.g., DeepWalk~\cite{perozzi2014deepwalk}, node2vec~\cite{grover2016node2vec}, and graph convolutional network (GCN)~\cite{Kipf2017,zhang2019graph}) and sequential data (e.g., recurrent neural network (RNN) and its many variants~\cite{Hochreiter1997,Chung2014}), various deep learning-based information diffusion models have emerged~\cite{cheng2014can,li2017deepcas,zhao2015seismic}. 

Several works have reviewed the information diffusion models in recent years~\cite{guille2013information,tatar2014survey,moniz2019review,gao2019taxonomy}. Earlier works~\cite{guille2013information,tatar2014survey} mainly focus on various feature engineering approaches and classical machine learning methods for modeling and predicting the popularity of information items. More recent surveys~\cite{moniz2019review,gao2019taxonomy} concentrate on web content and/or microblog information diffusion, while emphasizing different aspects of information cascade modeling. For example, Reference \cite{moniz2019review} focuses on web content popularity prediction and reviews \textit{a priori} and \textit{a posteriori} prediction approaches, as well as  evaluation protocols and classification/regression methods. However, the authors also give an overview of feature-based approaches and time series modeling methods. Similarly, Reference \cite{gao2019taxonomy} focuses on microblogs 
-- more specifically, Twitter and Weibo information popularity prediction -- while prioritizing generative models that rely on modeling time series events with stochastic processes. 

In comparison with previous surveys, this article makes the following distinct contributions:
\begin{enumerate}
	\item A more general perspective:  
	we extend the realm of modeling online web content, e.g., user-generated content microblogs~\cite{gao2019taxonomy}, to a general definition of information, i.e., \textit{any} measurable entities that can be propagated in \textit{any} networks, including scientific publications and citation networks, user published text/photos/videos and online sharing networks, and so on.
	
	\item A wider range of networks and cascades: as opposed to existing surveys that focus on a single type of social network (e.g., Twitter and Weibo in Reference \cite{gao2019taxonomy}), this work reviews information  spreading in various networks, including but not limited to Facebook, Flickr, YouTube, Netflix, IMDB, Instagram, Wikipedia, Reddit, and DBLP.
	
	\item A broader and balanced in-depth survey of the methodologies for predicting the outcome or popularity of cascades: complementing the previous surveys' composite of methods at different analytical levels~\cite{guille2013information,tatar2014survey,moniz2019review}, we provide finer-grained analysis of trade-offs, advantages and limitations of existing methods. We also consider a wider spectrum of features, methodologies and interpretations, compared to a recent survey that also focused on popularity prediction~\cite{gao2019taxonomy}.
	\item More comprehensive recent literature:
	We currently provide the most comprehensive literature review of methods of modeling information diffusion, which not only spans a broad range of traditional feature engineering approaches and generative models, but also contains recent advances in modeling and predicting information popularity with graph representation learning and deep learning techniques. 
	
\end{enumerate}

\begin{figure}
    \centering
    \includegraphics[width=.55\textwidth]{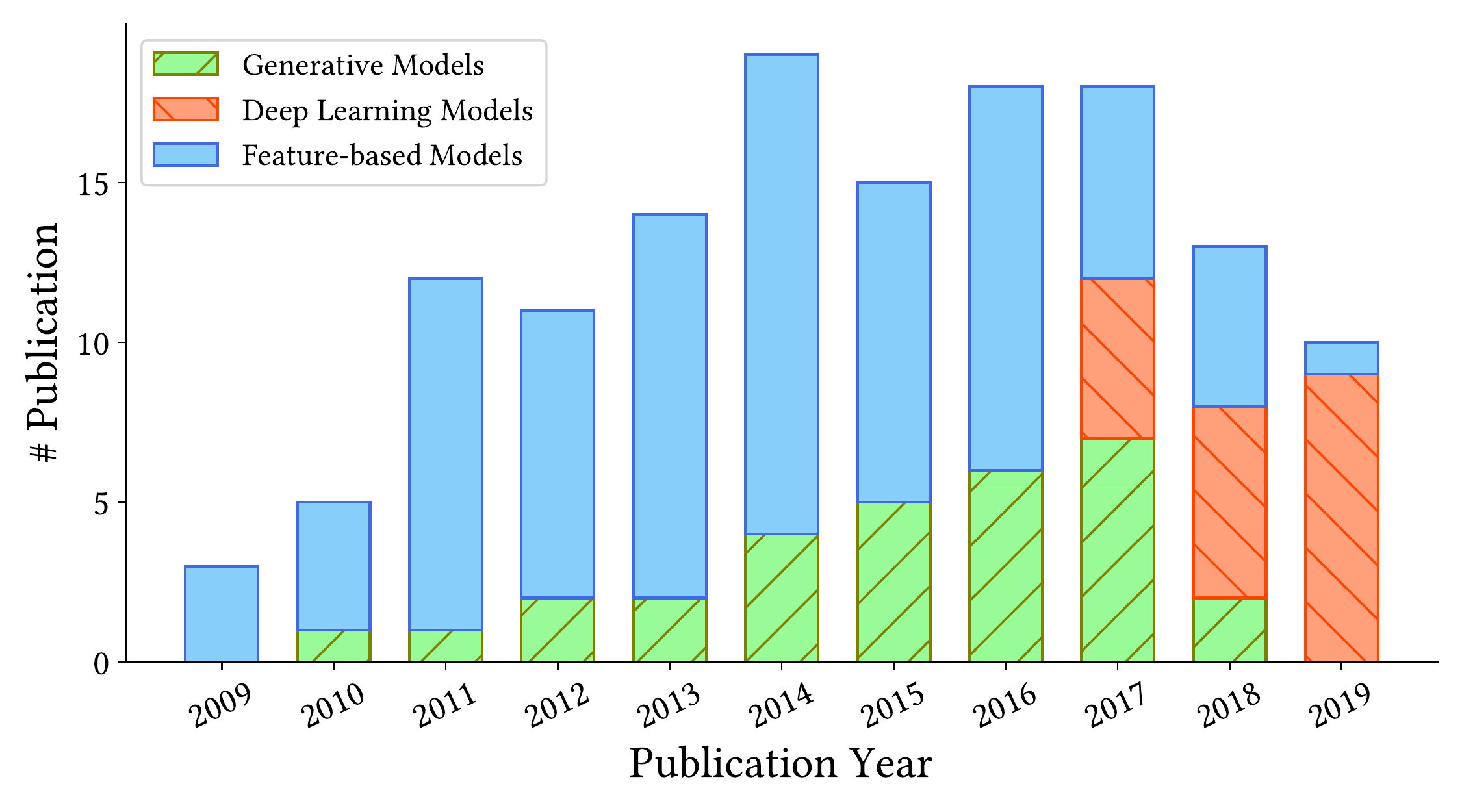}
    \includegraphics[width=.35\textwidth]{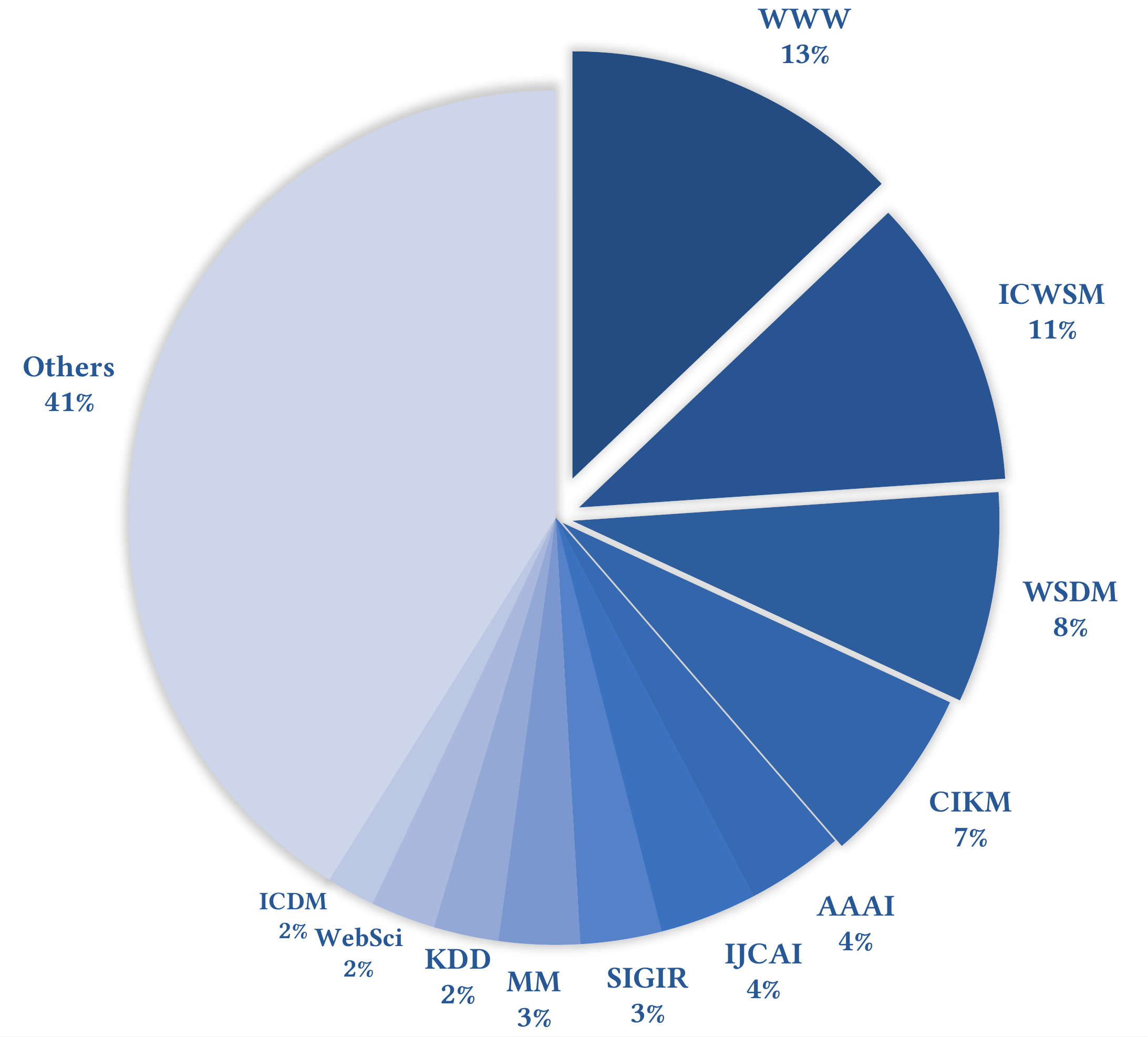}
    \Description[Publication Year]{Publication Year}
    \caption{Left: Number of publications over the last decade. Right: Distribution of venues of publications.}
    \label{fig:pub-year}
\end{figure}

\noindent Figure~\ref{fig:pub-year} provides a summary statistics of the papers included in this survey. As shown, information cascade modeling and prediction have consistently spurred great interest in recent years, although the methodologies employed varied over time -- e.g., most recently, deep learning-based methods have emerged as the more popular techniques.  
Research works that we reviewed were published in high-ranking conferences/journals related to data mining, social networks and information management, such as WWW, ICWSM, KDD, CIKM, WSDM, SIGIR, MM, TKDE, TKDD, ICDM, AAAI, and IJCAI. We note, though, that more than 40\% of papers come from a variety of conferences/journals outside of the typical venues, further attesting that this is an interdisciplinary topic spanning computer science, artificial intelligence, sociology, economics, marketing, statistics, and so on.

\begin{figure}
    \centering
    \includegraphics[width=.95\textwidth]{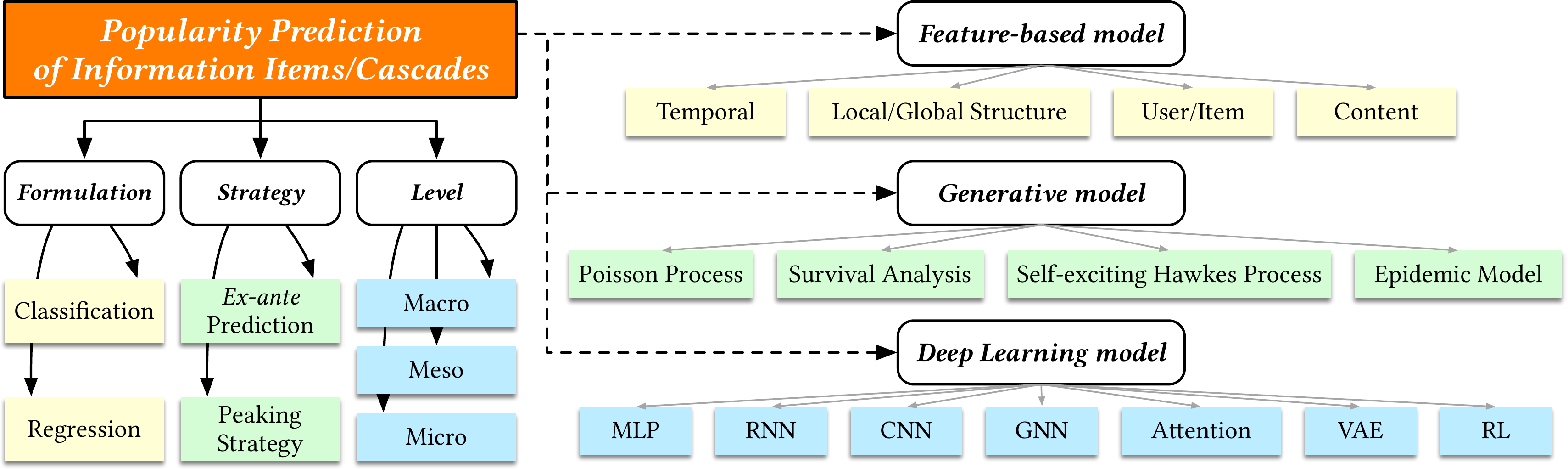}
    \Description[Survey Framework]{This figure sketches the taxonomy and organizations of this survey.}
    \caption{Taxonomy of the information diffusion models.}
    \label{fig:survey-framework}
\end{figure}

We build a taxonomy of methods from three different aspects, as illustrated in Figure~\ref{fig:survey-framework}. \textit{First}, popularity prediction can be considered as either a \textit{classification} or a \textit{regression} problem, according to the problem formulation and target application tasks. For example, information outbreak prediction and users' joint behavior prediction perform the classification tasks, while predicting the exact volume of items/cascades at a specific future time requires regression learning. \textit{Second}, prediction can be made before or after the publication of the information. According to the prediction time, existing methods can be categorized as \textit{ex-ante} prediction or prediction after \textit{peeking} into the early stage of information cascades. \textit{Third}, information prediction works can be further classified as \textit{micro-level}, \textit{macro-level} and \textit{meso-level} predictions by the granularity of the tasks. \textit{Methodologically}, existing works are grouped into three categories based on the methods modeling information cascades, including feature-based methods, generative models and deep learning approaches.

The rest of this survey is structured according to the hierarchies of the method taxonomy. In Section~\ref{sec:pre}, we introduce the basic formulations, problem definitions, and the commonly used evaluation protocols and benchmark datasets. 
In Section~\ref{sec:feat}, we illustrate basic characteristics of information cascades and review corresponding feature-based models. 
In Section~\ref{sec:gene}, we review another major group of models relying on time series of cascades and stochastic processes.  Section~\ref{sec:deep}  summarizes recent advances in modeling information cascades with deep learning-based techniques. We discuss the open challenges and opportunities and conclude this survey in Section~\ref{sec:con}.

\section{Problem Definition, Evaluation, Datasets, and Taxonomy}\label{sec:pre}Modeling information items/cascades has long been an active field in both academia and industry~\cite{rogers2010diffusion,tatar2014survey}. Due to the natural diversity and broad scope of information, existing works vary in their item/cascade definitions. 
In this section, we first review the frequently studied information items and cascades -- examples ranging from traditional news articles and academic papers to newly emerged microblogs and online photos/videos. Then we group the problems into three categories: (i) predictions  via classification or regression; (ii) predictions before or after publication, and (iii) the level/scope of the prediction. Along with introducing problem formulations, we present the necessary preliminaries associated with information items/cascades. 

\subsection{Types of Information Items/Cascades}
\label{subsec:information-cascade}

In this survey, we consider information items as \textit{any} measurable entities in terms of their \textit{popularity}, whereas an information cascade is constituted by a propagation sequence of information items. The best-known and most heavily studied information items are user-generated content (UGC), made possible by the prevailing Web 2.0 services and mobile devices. UGC is no longer produced by editors or publishers only, as traditional audiences have become the consumers and creators of content, such as posts in social networking platforms, threads in discussion forums, and photos and videos on content sharing websites. Such content has thoroughly changed the way users interact with information and other users and how information is created, presented, disseminated, and dies out. Even in conventional information fields, content such as magazines, newspapers, and academic journals are digitized and consumed in online platforms. Understanding the inner drive that determines the spread of information items is nontrivial and important for many real-world applications such as advertising, decision making, and caching strategy. 

The spread and diffusion of information items are featured by various \textit{endogenous} or \textit{exogenous} factors.
For example, in social networking platforms like Twitter and Weibo, users spontaneously post tweets, follow other users, ``like'' or comment on tweets, and most importantly, retweet other users' (re)tweets, which will show in the feed of her/his followers (potential future retweeters of this tweet). In contrast, user-driven resharing mechanisms can also be triggered by social influence~\cite{anagnostopoulos2008influence} and by external influences like burst events \cite{crane2008robust}. 

\subsection{Problem Definitions}
\label{subsec:problem-formulation}

Information items/cascades are transient, sparse, and biased data. The prediction of information item/cascade is challenging and in some cases, they are in fact unpredictable~\cite{cha2009analyzing,moniz2019review}. To better characterize the complex nature of information items/cascades and to explore under which circumstances the problem can be tackled, here we apply threefold categorization: (i) classification/regression; (ii) before/after publication; and (iii) granularity of the prediction, i.e., a macro-level prediction (collective behavior) or micro-level prediction (individual behavior).

\subsubsection{Classification versus Regression}
\label{subsubsec:clas-or-regr}

The popularity prediction of information items/cascades can be defined as either a classification problem or a regression problem. Consider a piece of information that can be disseminated by other users in a social network. We use \textit{information diffusion} to denote the process of dissemination. Therefore, the popularity prediction problem is to predict the final audience, attention, or influence of this information, given its initial state.

\begin{definition}{\textbf{Popularity Prediction}.}\label{def:popularity-prediction}
Given $M$ information item $\{I_1, I_2, \dots, I_M\}$ and a specific measurement of popularity for each item $I_i$ published by author $u_0$ at time $t_0$, popularity prediction aims at predicting the popularity $P_i(t_p)$ of $I_i$ at a future prediction time $t_p$.
\end{definition}

When it comes to the \textit{classification} tasks, the exact value of $P_i(t_p)$ is not required. Instead, one aims to predict whether an information item will be popular or not, given a predefined absolute/relative threshold. For example, some models predict whether an item gains reshares~\cite{naveed2011bad,petrovic2011rt}, or whether an item will burst in the near future~\cite{cheng2014can,cui2013cascading} -- both of which can be treated as a \textit{binary} classification problem. Another line of works predefine several popularity intervals, and then predict which interval an item is most likely to fall into~\cite{gao2014popularity,hong2011predicting,ma2012will,ma2013predicting}, converting the problem to a classical
\textit{multi-class} classification task. As for \textit{regression} models, they primarily predict the exact popularity value $P_i(t_p)$ (e.g., retweets and citations) that an item will gain in the future~\cite{ahmed2013peek,bakshy2011everyone,kupavskii2013predicting,szabo2010predicting}.

Generally speaking, classification tasks are relatively easier than regression tasks. Reference \cite{bandari2012pulse} showed that with only four types of features, classification accuracy can reach 84\% but perform poorly on regression task. Although the intuition behind formulating regression is more natural and provides a fine-grained scope to analyze which factors affect the future popularity and lead information items to success, precise regression prediction often requires far more information regarding items and users, which therefore implies a higher  complexity~\cite{gao2014effective}. In addition, it suffers from undesirable issues such as overfitting, inductive bias, and prediction error accumulations~\cite{xiao2016modeling}. 
We note many of the models we surveyed can be easily modified from regression to classification, or vice versa.

\subsubsection{Ex-ante Prediction versus Peeking Strategy}
\label{subsubsec:before-or-after}

Popularity prediction can also be made \textit{before} or at the time of its publication (a.k.a. \textit{ex-ante} prediction~\cite{martin2016exploring}), or \textit{after} its publication (a.k.a. \textit{peeking strategy}~\cite{shulman2016predictability}). Obviously, prediction in advance is a more challenging task, since only limited information about the items/users is available. Take tweets as an example: in \textit{ex-ante} prediction, the available information includes: tweet's content (e.g., purely text, emojis, embedded images/videos, URLs/hashtags), profile of the user who posts this tweet (e.g., name, bio, location, join date, number of followers/followees), the time when this tweet was published, and so on.
Prediction before publication is beneficial in many downstream tasks such as advertising and marketing, as one prefers to know how much attention and/or profit a specific advertisement would have in the near future \cite{sanjo2017recipe}. Based on predicted results, one can invest in items with highest (potential) influence to spread information to larger audiences, thus maximizing the business interests of corresponding products. 

\begin{figure}
    \centering
    \includegraphics[width=\textwidth]{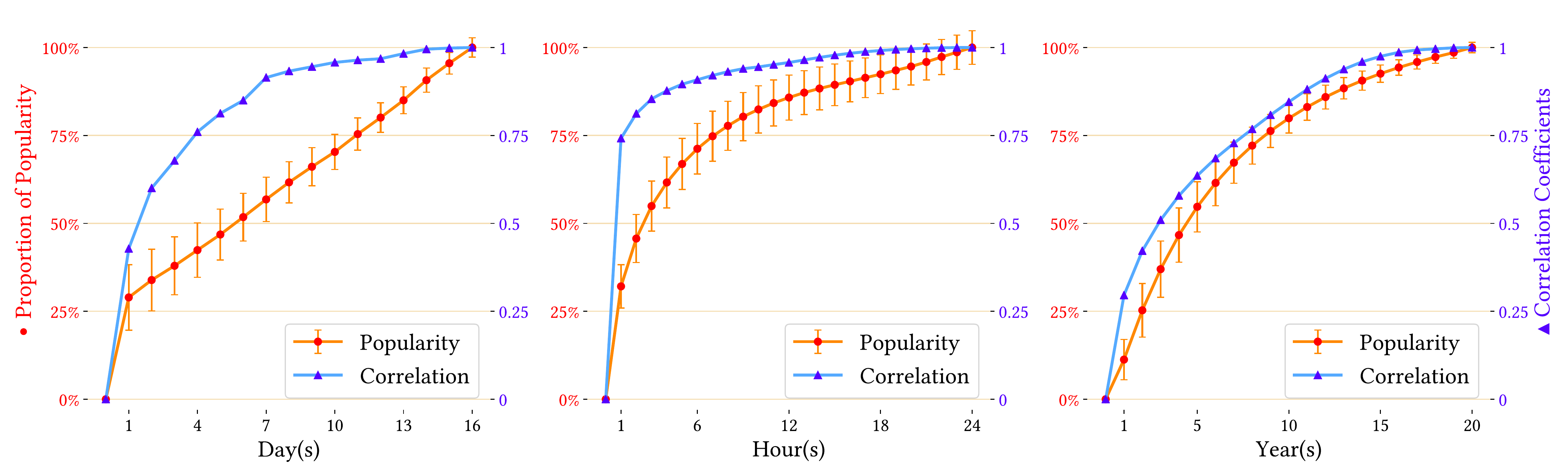}
    \Description[Line chart]{For all three datasets, the correlations between observed popularity and ground truth over time are increased. Specifically, Weibo dataset saturates the fastest, Twitter dataset at a middle speed, and APS the slowest.}
    \caption{Left axis: growth trends of cascades over time, the points are means and the bars on each point are variances. Right axis: Pearson correlation coefficients between observed popularity and ground truth popularity over time. Plots are based on three datasets: Twitter hashtags (left), Weibo tweets (middle), and APS papers (right), each have 1.3M, 119K, and 452K information cascades, respectively.}
    \label{fig:time-popularity-corr}
\end{figure}

Although \textit{a priori} popularity prediction is attractive, reliable prediction requires an exhaustive set of features, which is often unavailable, sensitive, and hard to obtain in real-world scenarios.
Instead, if one could \textit{peek} into the early stage of a cascade's evolving process, prediction would become much easier~\cite{szabo2010predicting}. Consider a time interval $[t_0, t_p]$ of the life-cycle of an information item, and let $[t_o, t_p]$ denote the interval we can observe, where $t_o < t_p$ is the observation time. We calculated the proportion of overall observed popularity $\sum_i^N P_i(t_o)$ to the overall ground truth popularity $\sum_i^N P_i(t_p)$ (values increased from $0$ to $1$ monotonically), 
and Pearson correlation coefficients between every observed $P_i(t_o)$ and prediction target $P_i(t_p)$. Figure~\ref{fig:time-popularity-corr} shows the results for three large scale datasets: Twitter hashtags~\cite{weng2013virality}, Weibo tweets~\cite{cao2017deephawkes} and APS papers~\cite{shen2014modeling}.
Clearly, for all three datasets with the increase of observation time $t_o$, the prediction task becomes simpler -- coinciding with earlier findings~\cite{szabo2010predicting} that a strong correlation exists between early and future popularity.  
Figure~\ref{fig:time-popularity-corr} also reveals different collective increasing patterns among the three different scenarios of information cascades: For Weibo tweets, the faster they gain attention from users, the faster they faded owing to the newly emerged competitors;
for Twitter hashtags, the popularity last longer among users than it does for Weibo tweets; as for APS papers, the item evolving speeds are
in the middle. 

Predicting the popularity after an item's publication indicates how it is possible for an item to be successful in spreading. The rationale is that information items that have successfully propagated in their early stages tend to become popular --  i.e., early patterns are indicative for long-term popularity~\cite{szabo2010predicting}. Previous works found that the similar, or even the same, content information may vary significantly in terms of popularity~\cite{cheng2014can}. This also raises questions about the limitations of prediction models -- e.g., is it a problem of insufficient data, or is it inherently unpredictable~\cite{cheng2014can,martin2016exploring,salganik2006experimental,shulman2016predictability}? Is this because the innate quality of content has a minor impact, while the social influence or other external factors govern the eventual popularity of items  \cite{muchnik2013social,salganik2006experimental}? Unfortunately, there is no explicit answer applicable to all situations. 
Such limitations of \textit{ex-ante} prediction drive researchers to design various peeking strategies, and/or investigate the evolution patterns of information items after the publication.

Observing who participates in forwarding (or adopts) a piece of an information item enables consideration a sequence of participants/adopters as an \textit{information cascade}, often defined as:
\begin{definition}{\textbf{Information Cascade}.}\label{def:vanilla-cascade}
Given an information item $I_i$ and its corresponding $N$ participants/adopters, an information cascade can be defined as $C_i = \{(u_j, t_j)|j\in[1, N], t_j\leq t_p\}$, where each tuple $(u_j, t_j)$ represents user $u_j$ participating in this cascade $C_i$ at time $t_j$.
\end{definition}

As an example, consider an information item $I_i$ of a tweet, and use the number of retweets as its popularity. Then, within time window $[t_0, t_o]$, we can observe a sequence of retweets which form a cascade -- framing the prediction task as: given item $I_i$'s observed cascade at observation time $t_o$, predict the future popularity $P_i(t_p) = |C_i|$ at prediction time $t_p$.

\subsubsection{Macro-, Micro-, and Meso-level}
\label{subsubsec:macro-or-micro-level}

The problem of information cascades -- especially the prediction of cascades -- is interdisciplinary, involving various expertise: text/image/video processing, social influence analysis, temporal and topological modeling, and so on. The prediction methods can be further classified according to the \textit{granularity} of tasks: \textit{macro-level} prediction aims at the collective behavior of cascades, whereas a \textit{micro-level} prediction focuses on the individual status/actions of users/items. 
Macro-level prediction~\cite{cao2017deephawkes,cheng2014can,li2017deepcas} models the cascade from a holistic and global perspective. The output of macro-level prediction is how much attention an information item will get in the future. Instead of taking a cascade as a whole, \textit{micro-level} prediction concerns the behavior of individual users/items, e.g., predicts the activation probability of a specific user, by giving the current cascade statuses, individual/group characteristics, and neighboring relationships~\cite{galuba2010outtweeting,li2017modeling,qiu2016lifecycle,qiu2018deepinf,romero2011differences,tang2017popularity,wang2017cascade,wang2017topological,wu2019dual,yang2012we,yang2019neural,zaman2010predicting}.

\begin{figure}
    \centering
    \includegraphics[width=.325\linewidth]{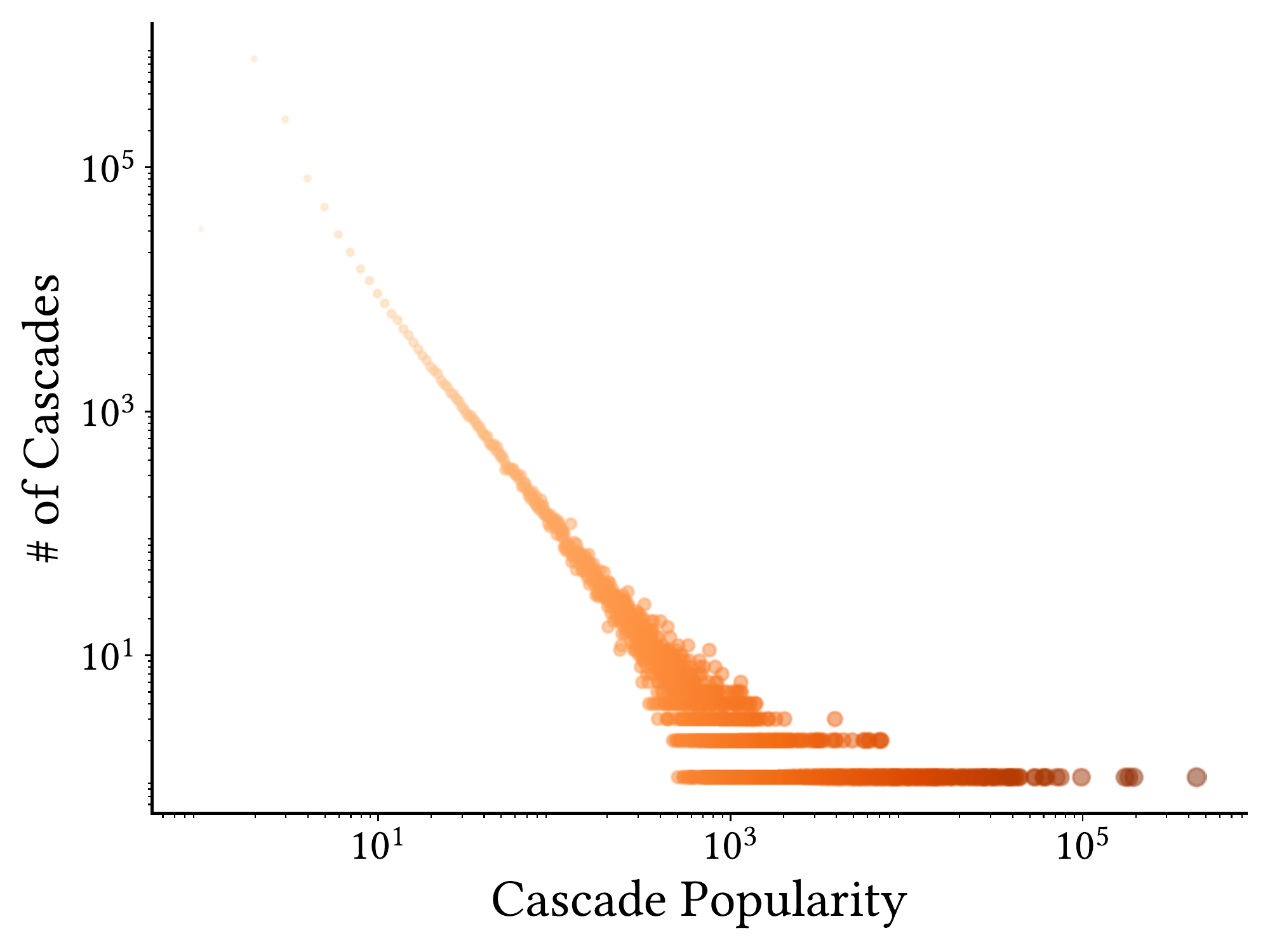}
    \includegraphics[width=.325\linewidth]{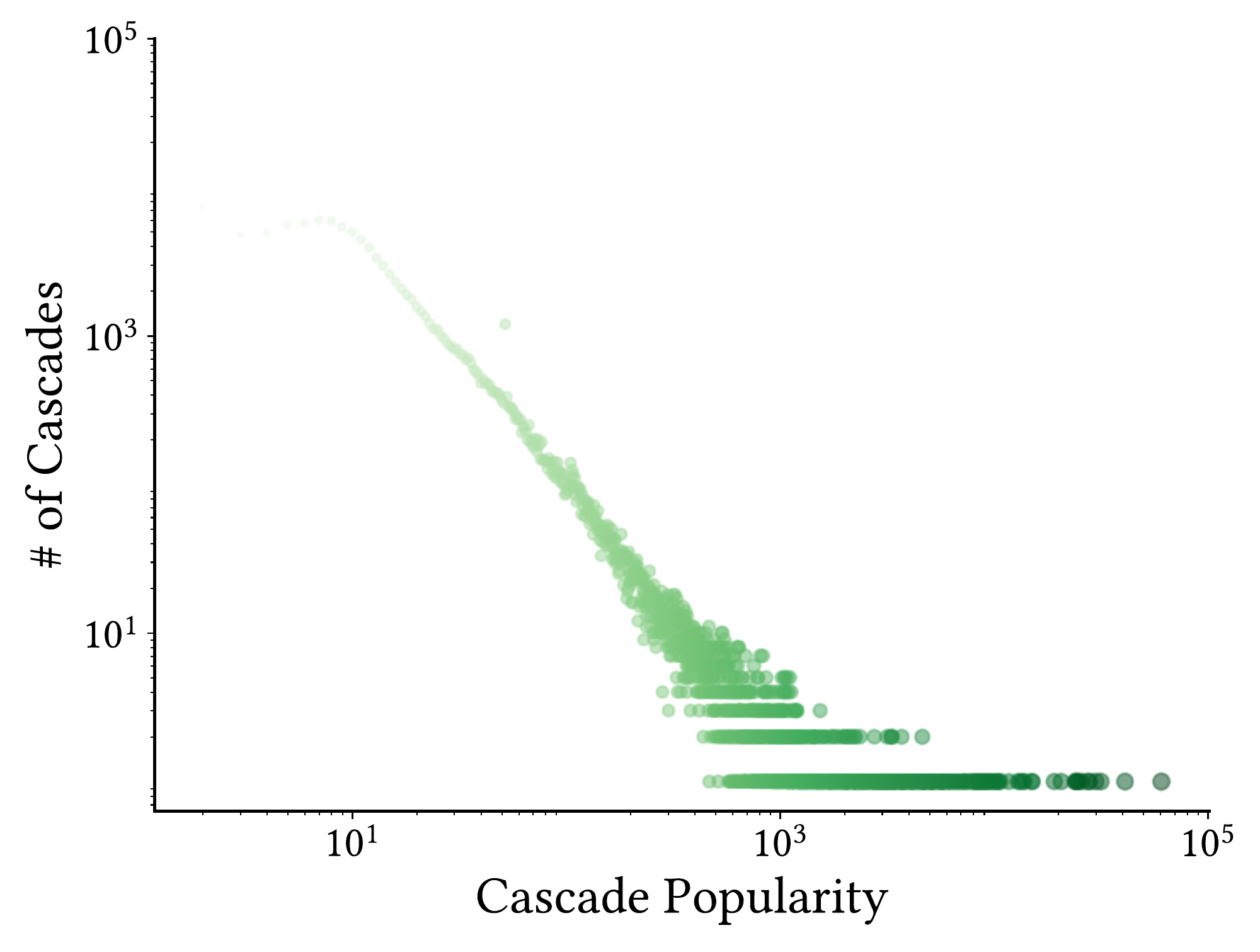}
    \includegraphics[width=.325\linewidth]{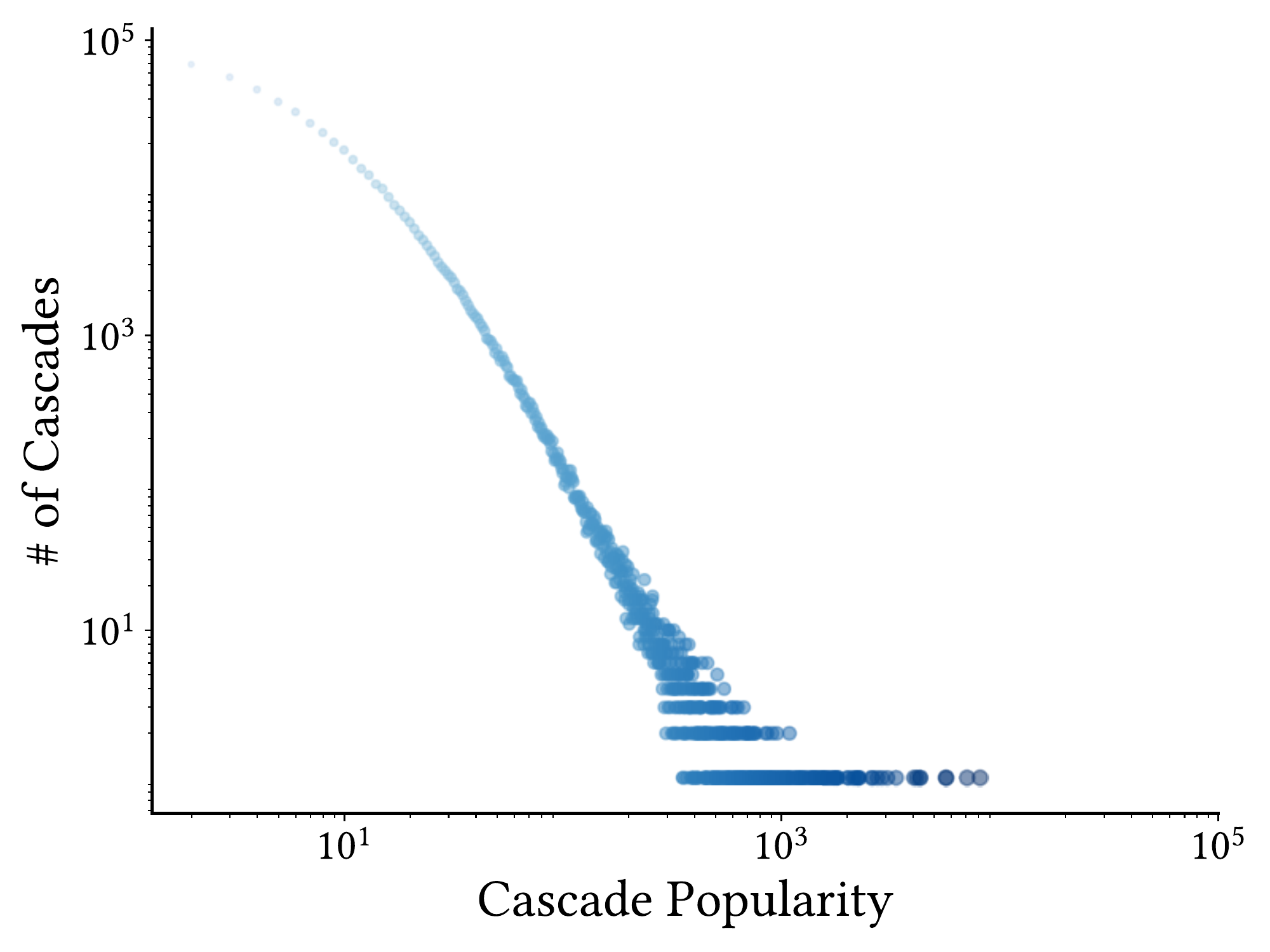}
    \Description[Popularity distribution]{Both of the datasets obey the heavy-tailed distribution.}
    \caption{Popularity distributions of Twitter (left), Weibo (middle), and APS (right) in log-log scales. By fitting the tails of power-law distribution, the exponents $\alpha$ are 1.916, 2.044 and 2.788, with minimum popularity $P_{\text{min}}$ greater than or equal to 34, 45 and 58, for Twitter, Weibo and APS datasets, respectively.}
    \label{fig:popularity-number}
\end{figure}

Another scope, referred to as \textit{meso-level} prediction \cite{gao2019taxonomy}, concerns the behaviors of communities/clusters \cite{backstrom2006group,eftekhar2013information,hoang2017gpop,hu2015community,leskovec2010empirical,romero2013interplay,tsugawa2019empirical,weng2013virality}. For example, Reference \cite{weng2013virality} studied information spreading by considering community structures; Reference \cite{hoang2017gpop} proposed a group-level prediction model. Models combining two levels of scope are also proposed, e.g., Reference \cite{yu2015micro} utilizes micro-level user behavioral dynamics (one-hop subcascades) to help in predicting the macro-level cascade popularity. In this survey, most methods are
macro-level popularity prediction of information items/cascades -- i.e., given an information item or partially observed cascade, the goal is to predict the future popularity of this item/cascade. Nonetheless, features and methods that can be used for micro-level and/or meso-level predictions are also reviewed, as they can be easily incorporated into macro-level models while providing insights into macroscopic cascade behaviors.

\subsection{Evaluation Metrics and Datasets}
\label{subsec:metrics-datasets}

\begin{table}\small
    \caption{A Review of Frequently Used Model Evaluation Metrics} 
    \label{tab:metrics}
    \begin{minipage}{\columnwidth}
    \begin{center}
    \begin{tabular}{@{} llp{7.3cm} @{}}
    \toprule
        \textbf{Metric} & \textbf{Formulation} & \textbf{Reference} \\ \midrule
        {Accuracy} & - &
        \cite{bielski2018understanding,chen2017attention,chen2019npp,cheng2014can,dong2015will,dong2016can,gao2014effective,gao2014popularity,gupta2012predicting,hong2011predicting,jamali2009digging,khabiri2009analyzing,kim2011predicting,liao2019popularity,luo2018real,mcparlane2014nobody,mishra2016feature,romero2013interplay,shafiq2017cascade,shamma2011viral,shulman2016predictability,totti2014impact,tsagkias2009predicting,wang2015burst,xie2017whats,yang2012we,zhang2016structure} \\
        Accuracy with tolerance $\tau$ & $\mathbbm{1}(|\frac{\hat{P}_i - P_i}{P_i}| \leq \tau) $ & 
        \cite{bao2017uncovering,bao2015modeling,cao2020popularity,ding2015video,gao2015modeling,gao2019taxonomy,gupta2012predicting,shen2014modeling,xiao2016modeling,yu2015micro,yu2017uncovering}, \cite{zhao2019neural}$^*$ \\
        Area under the ROC Curve & - &
        \cite{bora2015role,cheng2014can,dong2015will,dong2016can,gao2014effective,gao2014popularity,jamali2009digging,jia2018predicting,naveed2011bad,qiu2018deepinf,vallet2015characterizing,wang2015burst,yi2016mining} \\
        Coefficient of Determination & - &
        \cite{bakshy2011everyone,dong2016can,ma2013predicting,lakkaraju2011attention,lakkaraju2013s,martin2016exploring,romero2011influence,stieglitz2012political,tatar2011predicting,wu2018beyond}, \cite{abisheva2014watches,bandari2012pulse,castillo2014characterizing}$^\dag$ \\
        Coefficient of Correlation & - & 
        \cite{ding2019social,he2014predicting,jamali2009digging,oghina2012predicting,samanta2017lmpp,tatar2011predicting,trzcinski2017predicting,wang2017linking,zhao2015seismic}, \cite{bielski2018understanding,gelli2015image,lv2017multi,tsur2012s,wu2016unfolding,wu2017sequential}$^\dag$ \\
        $F_1$ or $F_\beta$ Score & - &
        \cite{alzahrani2015network,bian2014predicting,bora2015role,cheng2014can,cui2013cascading,dong2015will,dong2016can,galuba2010outtweeting,gao2014effective,gao2019taxonomy,guo2015toward,guo2016comparison,guo2018learning,hong2011predicting,jamali2009digging,jenders2013analyzing,kefato2018cas2vec,kong2014predictingbursts,kong2015towards,kong2016popularity,krishnan2016seeing,liao2019popularity,lu2017predicting,petrovic2011rt,qiu2018deepinf,romero2013interplay,shamma2011viral,subbian2017detecting,tsagkias2009predicting,tsugawa2019empirical,wang2015burst,weng2014predicting,xie2015modelling,yan2016sth,yang2010understanding,yano2010s,yi2016mining,zhang2016structure}, \cite{figueiredo2013prediction,gao2014popularity,kong2018exploring,liu2015effectively,ma2012will,ma2013predicting,yang2019neural}$^\ddag$ \\
        Mean Absolute Error & $\frac{1}{M}\sum_i^M |{\hat{P_i}-P_i}|$ & \cite{dong2016can,kobayashi2016tideh,oghina2012predicting,wang2015burst,wu2018beyond}, \cite{xie2015modelling}$^*$, \cite{bao2013popularity,lu2018collective,lv2017multi,rizoiu2017expecting,sanjo2017recipe,wang2018factorization,wang2018learning,wu2017sequential,zhang2018user}$^\dag$ \\
        Mean Abs. Percent. Error      & $\frac{1}{M} \sum_i^M |\frac{\hat{P_i}-P_i}{P_i}| $ &
        \cite{bao2017uncovering,bao2015modeling,cao2020popularity,chu2018cease,ding2015video,gao2015modeling,gao2016modeling,gao2019taxonomy,guo2016comparison,gursun2011describing,hoang2017gpop,jin2010wisdom,kong2020modeling,li2013popularity,lin2013predicting,lu2014predicting,lymperopoulos2016predicting,mishra2016feature,samanta2017lmpp,shen2014modeling,xiao2016modeling,wang2015learning,wang2017linking,yan2016sth,zaman2014bayesian,zhao2015seismic}, \cite{xie2015modelling,zhao2019neural}$^*$, \cite{kim2011predicting}$^\dag$, \cite{zhou2020variational}$^{*\dag}$ \\
        Mean Square Error & $\frac{1}{M}\sum_i^M (\hat{P_i}-P_i)^2$ & \cite{ahmed2013peek,gao2019taxonomy,guo2016comparison,rizos2016predicting}, \cite{chen2019information}$^*$, \cite{chen2016micro,ding2019social,kupavskii2012prediction,kupavskii2013predicting,lv2017multi,sanjo2017recipe,szabo2010predicting,tsur2012s,wang2018factorization,wang2018learning,xie2020multimodal,yang2019multi,yang2019neural,yu2015lifecyle,zhang2018user}$^{\dag}$, \cite{cao2017deephawkes,chen2019cascn,li2017deepcas,li2018joint,zhou2020variational}$^{*\dag}$ \\
        Precision & - &
        \cite{abisheva2014watches,alzahrani2015network,artzi2012predicting,bian2014predicting,bora2015role,cui2013cascading,dong2015will,dong2016can,galuba2010outtweeting,gao2014effective,gao2019taxonomy,guo2015toward,guo2016comparison,guo2018learning,hong2011predicting,khabiri2009analyzing,kong2015towards,krishnan2016seeing,qiu2018deepinf,romero2013interplay,shamma2011viral,tsugawa2019empirical,weng2013virality,weng2014predicting,xie2015modelling,xie2017whats,yan2016sth,yang2010understanding,yano2010s,yi2016mining,yu2014twitter,yu2015micro,zhang2016structure,zhao2018comparative}, \cite{gao2014popularity,liu2015effectively,ma2012will,ma2013predicting}$^\ddag$ \\
        Recall & - &
        \cite{abisheva2014watches,alzahrani2015network,artzi2012predicting,bian2014predicting,bora2015role,cui2013cascading,dong2015will,dong2016can,galuba2010outtweeting,gao2014effective,gao2019taxonomy,guo2015toward,guo2016comparison,guo2018learning,hong2011predicting,kefato2018cas2vec,khabiri2009analyzing,kong2015towards,krishnan2016seeing,qiu2018deepinf,romero2013interplay,shamma2011viral,tsugawa2019empirical,weng2013virality,weng2014predicting,xie2015modelling,xie2017whats,yan2016sth,yang2010understanding,yano2010s,yi2016mining,zhang2016structure,zhao2018comparative}, \cite{gao2014popularity,liu2015effectively,ma2012will,ma2013predicting}$^\ddag$ \\
        Root Mean Square Error              & ${\scriptstyle \sqrt{\frac{1}{M}\sum_i^M(\hat{P_i}-P_i)^2}}$ &
        \cite{guo2016comparison,gursun2011describing,lerman2010using,lymperopoulos2016predicting,matsubara2012rise,tatar2011predicting}, \cite{bao2013popularity,guo2016comparison,kong2014predictingbursts,kong2015towards,oghina2012predicting,yu2015micro,yu2017uncovering}$^\dag$ \\
    \bottomrule
    \end{tabular}
    \end{center}
    \scriptsize \textit{$^*$ Some models use incremental popularity, i.e., $P \to \Delta P = P(t_p) - P(t_o)$.}
    
    \textit{$^\dag$ Some models use logarithmic popularity (error), i.e., $P \to \log P$, or other similar transformations/normalizations.}
    
    \textit{$^\ddag$ Some models use macro/micro-Precision, macro/micro-Recall, or macro/micro-$F_1$.}
    \end{minipage}
\end{table}

After introducing problem definitions w.r.t. information item/cascade popularity prediction, we now turn to present evaluation metrics and datasets that frequently used in existing literature.

The most commonly used classification metrics are Accuracy, Precision, Recall, and $F$-measure. Given a predefined threshold, an item/cascade can be classified as \textit{outbreak} or \textit{viral} if its popularity (size) exceeds that threshold. One known limitation of accuracy metrics is the class imbalance problem, due to the highly skewed popularity distribution~\cite{kong2014predicting}. For example, in one Twitter dataset~\cite{weng2013virality}, more than 92.8\% tweets have retweets $\leq$10, while only 0.114\% tweets have retweets $>$1,000. They follow heavy-tailed distributions, which are consistent with the findings of many previous studies~\cite{cao2017deephawkes,cheng2014can}, as shown in Figure \ref{fig:popularity-number}. 
To address such limitations, researchers reformulate the problem definitions or adopt other types of evaluation metrics, e.g., Reference \cite{cheng2014can} adopts a class balanced binary classification, by predicting whether a cascade will exceed the median size of all cascades; References \cite{qiu2018deepinf} and~\cite{kong2014predicting} filter out a majority of cascades in order to obtain a balanced dataset (undersampling). Most of the \textit{peeking strategy} models requiring a sufficient number of early adopters, implicitly ignore those small samples, consequently yielding balanced classes. 

As for the regression task, mean square error (MSE) and its variants are the most popular metrics. Popularity is often shown in logarithmic scale, e.g., MSLE or RMSLE, to prevent the loss functions or metrics being affected by extreme values and assure numerical stability~\cite{cao2017deephawkes,chen2019cascn}. Coefficient of determination/correlation and their variants, rankings~\cite{tatar2014popularity}, $k$-top coverage \cite{nwana2013latent,zhao2015seismic}, are also frequently used metrics in some specific scenarios. 

The choice of a metric is often subjective, even under a well-defined problem formulation. Previous works found that a model might perform well in one metric but significantly drop in another~\cite{guo2016comparison}, making it difficult to do a fair comparison between various approaches. We summarize frequently used metrics as well as their adopters in Table~\ref{tab:metrics}.

In Table~\ref{tab:datasets}, we list widely used and publicly available datasets. The scope of information items ranges broadly, from news articles, academic papers, posted images, music and videos, and these diverse scenarios of popularity prediction cause difficulties in the design of prediction models. Whether for feature extractions, problem formulations, evaluation selections, devising of generative processes, or deep learning architecture designs, it is difficult and sometimes even impossible to fully generalize one model from one specific platform to another. The datasets used throughout this survey contain two microblog datasets, Twitter\footnote{\url{https://carl.cs.indiana.edu/data/\#virality2013}} hashtags~\cite{weng2013virality} and Weibo\footnote{\url{https://github.com/CaoQi92/DeepHawkes} or \url{https://bit.ly/weibodataset}} tweets~\cite{cao2017deephawkes}, and one scientific dataset APS\footnote{\url{https://journals.aps.org/datasets}} papers~\cite{shen2014modeling}.
The basic statistics of the three datasets are shown in Table~\ref{tab:three-datasets}.

\begin{table}\small
    \caption{Frequently Used Scenarios in Popularity Prediction Literature}
    \label{tab:datasets}
    \begin{tabular}{@{}llp{9.4cm}@{}}
    \toprule
        \textbf{Platform} & \textbf{Category} & \textbf{Reference} \\ \midrule
        Digg            & News Aggregator   &  \cite{ahmed2013peek,jamali2009digging,khabiri2009analyzing,lerman2010using,qiu2018deepinf,szabo2010predicting} \\
        Facebook        & Social Networking & \cite{bakshy2011everyone,castillo2014characterizing,cheng2014can,lakkaraju2011attention,subbian2017detecting,tang2017popularity,trzcinski2017predicting,ugander2012structural} \\
        Flickr          & Image Sharing     & \cite{gelli2015image,he2014predicting,khosla2014makes,lymperopoulos2016predicting,mcparlane2014nobody,shulman2016predictability,wu2016unfolding,wu2017sequential,zhang2018user,ding2019social} \\
        Twitter         & Social Networking & \cite{abisheva2014watches,alzahrani2015network,artzi2012predicting,bandari2012pulse,bakshy2011everyone,bora2015role,castillo2014characterizing,chen2017attention,chen2019npp,crane2008robust,figueiredo2013prediction,figueiredo2014does,galuba2010outtweeting,gao2019taxonomy,guo2016comparison,guo2018learning,hoang2017gpop,jenders2013analyzing,kefato2018cas2vec,kobayashi2016tideh,kong2014predictingbursts,kong2015towards,kong2020modeling,krishnan2016seeing,kwak2010twitter,kupavskii2012prediction,kupavskii2013predicting,li2017deepcas,li2018joint,luo2018real,lymperopoulos2016predicting,ma2012will,ma2013predicting,martin2016exploring,matsubara2012rise,mishra2016feature,mishra2018modeling,nwana2013latent,oghina2012predicting,petrovic2011rt,pinto2013using,qiu2018deepinf,rizoiu2017expecting,rizoiu2018sir,romero2011differences,romero2013interplay,rowe2011predicting,ruan2012prediction,samanta2017lmpp,shafiq2017cascade,shulman2016predictability,stieglitz2012political,subbian2017detecting,tsugawa2019empirical,tsur2012s,vallet2015characterizing,wang2018learning,weng2013virality,weng2014predicting,xie2015modelling,yan2016sth,yang2010modeling,yang2010predicting,yang2010understanding,yang2011patterns,yang2012we,yang2019multi,yang2019neural,yu2014twitter,zadeh2014modeling,zaman2010predicting,zaman2014bayesian,zhao2015seismic,zhao2018comparative} \\
        Weibo           & Social Networking & \cite{bao2017uncovering,bao2013popularity,bao2015modeling,bian2014predicting,cao2017deephawkes,cao2020popularity,chen2019information,chen2019cascn,cui2013cascading,gao2014effective,gao2014popularity,gao2015modeling,gao2016modeling,gao2019taxonomy,guo2015toward,guo2016comparison,guo2018learning,kefato2018cas2vec,kong2016popularity,kong2018exploring,lu2014predicting,lu2018collective,qiu2018deepinf,wang2015burst,wang2015learning,yi2016mining,yu2015micro,yu2017uncovering,zhang2016structure,zhou2020variational} \\
        YouTube         & Video Sharing     & \cite{abisheva2014watches,ahmed2013peek,cheng2014can,crane2008robust,ding2015video,figueiredo2011tube,figueiredo2013prediction,figueiredo2014does,figueiredo2014dynamics,gursun2011describing,he2014predicting,mishra2018modeling,oghina2012predicting,rizoiu2017expecting,shamma2011viral,szabo2010predicting,trzcinski2017predicting,vallet2015characterizing,wu2018beyond,yu2014twitter,yu2015lifecyle} \\
        -               & Academic          & \cite{bao2017uncovering,cao2017deephawkes,chen2019information,chen2019cascn,dong2015will,dong2016can,li2017deepcas,li2018joint,lin2013predicting,shen2014modeling,vu2011dynamic,xiao2016modeling,yan2011citation,zhou2020variational} \\
        -               & News Articles     &
        \cite{bandari2012pulse,castillo2014characterizing,kong2014predicting,lu2017predicting,tatar2011predicting,tatar2014popularity,tsagkias2009predicting,tsagkias2010news,yang2010modeling,yang2019neural,yano2010s} \\
    \bottomrule
    \end{tabular}
\end{table}

\begin{table}\small
    \caption{Basic Statistics of Three Benchmark Datasets Used in This Survey.}
    \label{tab:three-datasets}
    \begin{tabular}{lrrrrrr}
    \toprule
    \textbf{Dataset} & \textbf{\# Cascades} & \textbf{\# Nodes} & \textbf{\# Edges} & \textbf{Avg. Depth} & \textbf{Avg. $P(t_o)$} & \textbf{Avg. $P(t_p)$} \\ \midrule
    Twitter & 1,345,913 & 595,460 & 14,430,254 & 2.198 & 21.37 ($t_o=1$d) & 118.02 ($t_p=16$d) \\
    Weibo & 119,313 & 6,738,040 & 15,249,636 & 2.337 & 56.49 ($t_o=1$h) & 117.51 ($t_p=24$h) \\
    APS & 514,609 & 616,316 & 54,614,609 & 4.096 & 4.56 ($t_o=3y$) & 7.75 ($t_p=20$y) \\
    \bottomrule
    \end{tabular}
\end{table}

\section{Characteristics of Cascades and Feature Engineering Approaches}\label{sec:feat}Feature extraction plays a crucial role in predicting the popularity of items/cascades, whether formulated as 
classification or regression, prior or posterior, micro- or macro-level. Various features and machine learning techniques have been used in a variety of subjects, e.g., illustrating diffusion patterns in network, analyzing time-series evolution trends, and building prediction models. 

We now discuss the characteristics of information items/cascades, categorized into five groups: \textit{temporal}, \textit{cascade structure}, \textit{global graph}, \textit{user/item attributes}, and \textit{content features}. Table~\ref{tab:freq} and Table \ref{tab:stra-form} summarize the strategies, formulations, and feature frequency of feature-based models in the last decade.\footnote{We refer readers to the supplementary material for the details per individual work, including strategy, formulation, and incorporated features.}
Table~\ref{tab:freq} we can see that among the five groups of features, temporal, user/item, and content features have received more attention in comparison with structural features. Models based only on features in recent years are becoming less. 
Table \ref{tab:stra-form} showed that peeking is a more popular strategy than \textit{ex-ante}, and some models evaluate the prediction performance in both classification and regression. 

\begin{table}\small
    \caption{Frequency of Five Feature Groups Used by Feature-based Models} 
    \label{tab:freq}
    \begin{tabular}{lcccccccccccr}
    \toprule 
        & 2009 & 2010 & 2011 & 2012 & 2013 & 2014 & 2015 & 2016 & 2017 & 2018 & 2019 & \\ 
        Temporal       & \cellcolor[rgb]{1, .90, .90}2 & \cellcolor[rgb]{1, .85, .85}3 & \cellcolor[rgb]{1, .75, .75}5 & \cellcolor[rgb]{1, .70, .70}6 & \cellcolor[rgb]{1, .75, .75}5 & \cellcolor[rgb]{1, .50, .50}10 & \cellcolor[rgb]{1, .70, .70}6 & \cellcolor[rgb]{1, .50, .50}10 & \cellcolor[rgb]{1, .85, .85}3 & \cellcolor[rgb]{1, .95, .95}1 & \cellcolor[rgb]{1, 1, 1}0 & \footnotesize{22.8\%} \\
        Casc. struct. & \cellcolor[rgb]{1, .95, .95}1 & \cellcolor[rgb]{1, 1, 1}0 & \cellcolor[rgb]{1, .95, .95}1 & \cellcolor[rgb]{1, .90, .90}2 & \cellcolor[rgb]{1, .70, .70}6 & \cellcolor[rgb]{1, .70, .70}6 & \cellcolor[rgb]{1, .75, .75}5 & \cellcolor[rgb]{1, .65, .65}7 & \cellcolor[rgb]{1, .90, .90}2 & \cellcolor[rgb]{1, .90, .90}2 & \cellcolor[rgb]{1, 1, 1}0 & \footnotesize{14.3\%} \\
        Glob. graph   & \cellcolor[rgb]{1, .95, .95}1 & \cellcolor[rgb]{1, .95, .95}1 & \cellcolor[rgb]{1, .90, .90}2 & \cellcolor[rgb]{1, .85, .85}3 & \cellcolor[rgb]{1, .65, .65}7 & \cellcolor[rgb]{1, .60, .60}8 & \cellcolor[rgb]{1, .75, .75}5 & \cellcolor[rgb]{1, .65, .65}7 & \cellcolor[rgb]{1, .90, .90}2 & \cellcolor[rgb]{1, .90, .90}2 & \cellcolor[rgb]{1, .95, .95}1 & \footnotesize{17.4\%} \\
        User/item    & \cellcolor[rgb]{1, .85, .85}3 & \cellcolor[rgb]{1, .95, .95}1 & \cellcolor[rgb]{1, .65, .65}7 & \cellcolor[rgb]{1, .65, .65}7 & \cellcolor[rgb]{1, .85, .85}3 & \cellcolor[rgb]{1, .50, .50}10 & \cellcolor[rgb]{1, .70, .70}6 & \cellcolor[rgb]{1, .60, .60}8 & \cellcolor[rgb]{1, .85, .85}3 & \cellcolor[rgb]{1, .80, .80}4 & \cellcolor[rgb]{1, .95, .95}1 & \footnotesize{23.7\%} \\
        Content        & \cellcolor[rgb]{1, .85, .85}3 & \cellcolor[rgb]{1, .95, .95}1 & \cellcolor[rgb]{1, .65, .65}7 & \cellcolor[rgb]{1, .55, .55}9 & \cellcolor[rgb]{1, .75, .75}5 & \cellcolor[rgb]{1, .65, .65}7 & \cellcolor[rgb]{1, .85, .85}3 & \cellcolor[rgb]{1, .70, .70}6 & \cellcolor[rgb]{1, .80, .80}4 & \cellcolor[rgb]{1, .85, .85}3 & \cellcolor[rgb]{1, .95, .95}1 & \footnotesize{21.9\%} \\ 
        & \footnotesize{4.46\%} & \footnotesize{2.68\%} & \footnotesize{9.82\%} & \footnotesize{12.1\%} & \footnotesize{11.6\%} & \footnotesize{18.3\%} & \footnotesize{11.2\%} & \footnotesize{17.0\%} & \footnotesize{6.25\%} & \footnotesize{5.26\%} & \footnotesize{1.34\%} &  \\ 
    \bottomrule
    \end{tabular}
\end{table}

\begin{table}\small
    \caption{Strategies and Formulations of Feature-based Models}
    \label{tab:stra-form}
    \begin{tabular}{llp{9.7cm}}
    \toprule 
        \textbf{Strategy} & \textbf{Formulation} & \textbf{Reference} \\ \midrule
        \textit{Ex-ante}    & Classification& \cite{artzi2012predicting,dong2015will,jenders2013analyzing,jia2018predicting,khabiri2009analyzing,luo2018real,mcparlane2014nobody,naveed2011bad,petrovic2011rt,shamma2011viral,totti2014impact,tsagkias2009predicting,yano2010s,zhao2018comparative} \\
                            & Regression    & \cite{bakshy2011everyone,chen2016micro,gelli2015image,khosla2014makes,kupavskii2013predicting,lakkaraju2013s,lv2017multi,martin2016exploring,rowe2011predicting,stieglitz2012political,tsur2012s,wu2016unfolding,wu2018beyond,yan2011citation} \\
                            & Both          & \cite{bandari2012pulse,dong2016can,kupavskii2012prediction,lakkaraju2011attention} \\ 
        Peeking             & Classification& \cite{alzahrani2015network,bian2014predicting,bora2015role,cheng2014can,cui2013cascading,figueiredo2013prediction,guo2015toward,hong2011predicting,jia2018predicting,kim2011predicting,kong2016popularity,kong2018exploring,krishnan2016seeing,lu2017predicting,ma2012will,ma2013predicting,romero2013interplay,shafiq2017cascade,shulman2016predictability,tsugawa2019empirical,weng2013virality,weng2014predicting,xie2017whats,zhang2016structure} \\
                            & Regression    & \cite{ahmed2013peek,bao2013popularity,castillo2014characterizing,gao2014popularity,gursun2011describing,he2014predicting,hoang2017gpop,kong2014predicting,kupavskii2013predicting,li2013popularity,liu2015effectively,oghina2012predicting,pinto2013using,rizos2016predicting,ruan2012prediction,szabo2010predicting,tatar2011predicting,tatar2014popularity,trzcinski2017predicting,tsagkias2010news,vallet2015characterizing,wang2015burst,yi2016mining,yu2014twitter,yu2015lifecyle} \\
                            & Both          & \cite{abisheva2014watches,dong2016can,gao2014effective,guo2016comparison,gupta2012predicting,jamali2009digging,kong2014predictingbursts,kong2015towards,kupavskii2012prediction,lerman2010using,mishra2016feature} \\
    \bottomrule
    \end{tabular}
\end{table}

\subsection{Temporal Features}
\label{subsec:temporal-features}

Temporal features in information items/cascades have been among the most crucial factors in popularity prediction~\cite{cheng2014can,gao2014effective}.  
We discuss the temporal features in the context of feature engineering.

\subsubsection{Observation Time}
Temporal features are usually extracted based on peeking strategy, i.e., observing a small number of early participants and their action time, to retrieve a sequence of timestamps that can be utilized for feature selection. 
Since the length of time series are highly irregular, e.g., in a fixed time interval, some cascades may have tens of thousands of participants, while most participants receive little attention, directly utilizing timestamps as a feature is ineffectual in practice. Calculation of time series often needs transformations in advance \cite{yang2011patterns}, e.g., dividing the time span in evenly distributed intervals and then calculating the cumulative/incremental popularity, or only observing a fixed number of early participants~\cite{cheng2014can}.

To predict the popularity $P_i(t_p)$ at prediction time $t_p$ based on the information observed at time $t_o$, Reference \cite{szabo2010predicting} analyzed the relationships between (log-transformed) popularity $P_i(t_p)$ and $P_i(t_o)$. They found a high correlation between early stage and future popularity and used a simple linear prediction model taking the early observed popularity as input to speculate the future popularity.

\subsubsection{Publication Time}

Another important temporal feature is the publication time $t_0$. As reported in previous works~\cite{tsagkias2010news,lakkaraju2013s,matsubara2012rise,szabo2010predicting,tatar2011predicting}, 
the popularity of a certain item is strongly related to its publication time, e.g., items published at midnight are less likely to be viewed while items published during daytime are generally more popular (though they have more competition) \cite{bao2013cumulative,gao2015modeling,wu2016unfolding,yan2016sth}. 
To mitigate the effect of user activity periodicity, several solutions have been proposed. Reference \cite{petrovic2011rt} designed 24 local models, each of which is trained with the samples published in a specific hour during the day. The \textit{tweet time} was used in Reference~\cite{gao2014popularity}  to eliminate the imbalanced diurnal effect of user activities, while other temporal factors such as \textit{digg time}~\cite{szabo2010predicting}, \textit{source time}~\cite{tsagkias2010news}, and \textit{user activeness variability}~\cite{wu2016unfolding} have been employed to improve model robustness. More simply, some works only explore the items published in daytime to train their models~\cite{cao2017deephawkes,chen2019cascn}.

\begin{figure}
    \centering
    \includegraphics[width=.48\textwidth]{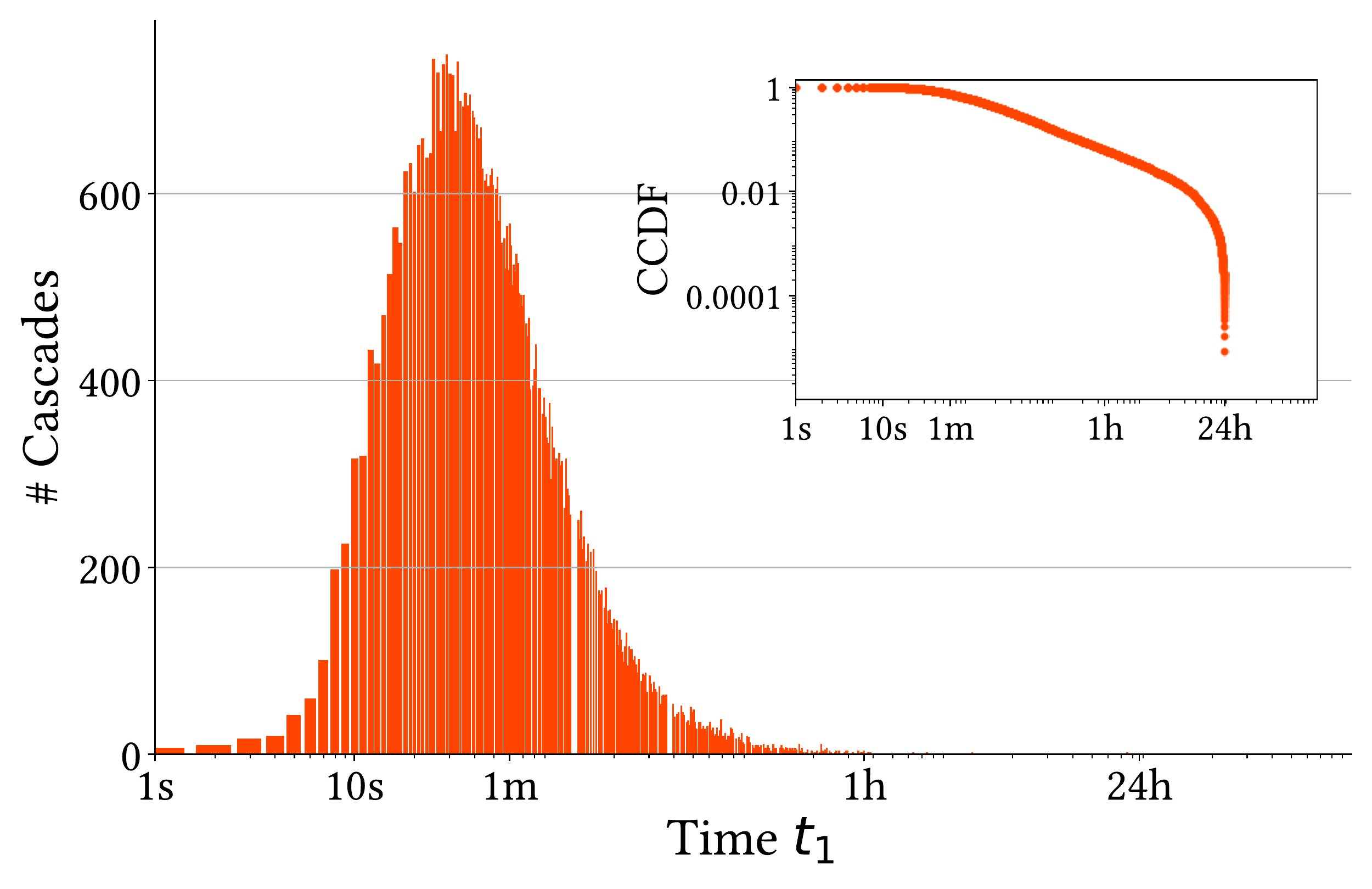}
    \includegraphics[width=.48\textwidth]{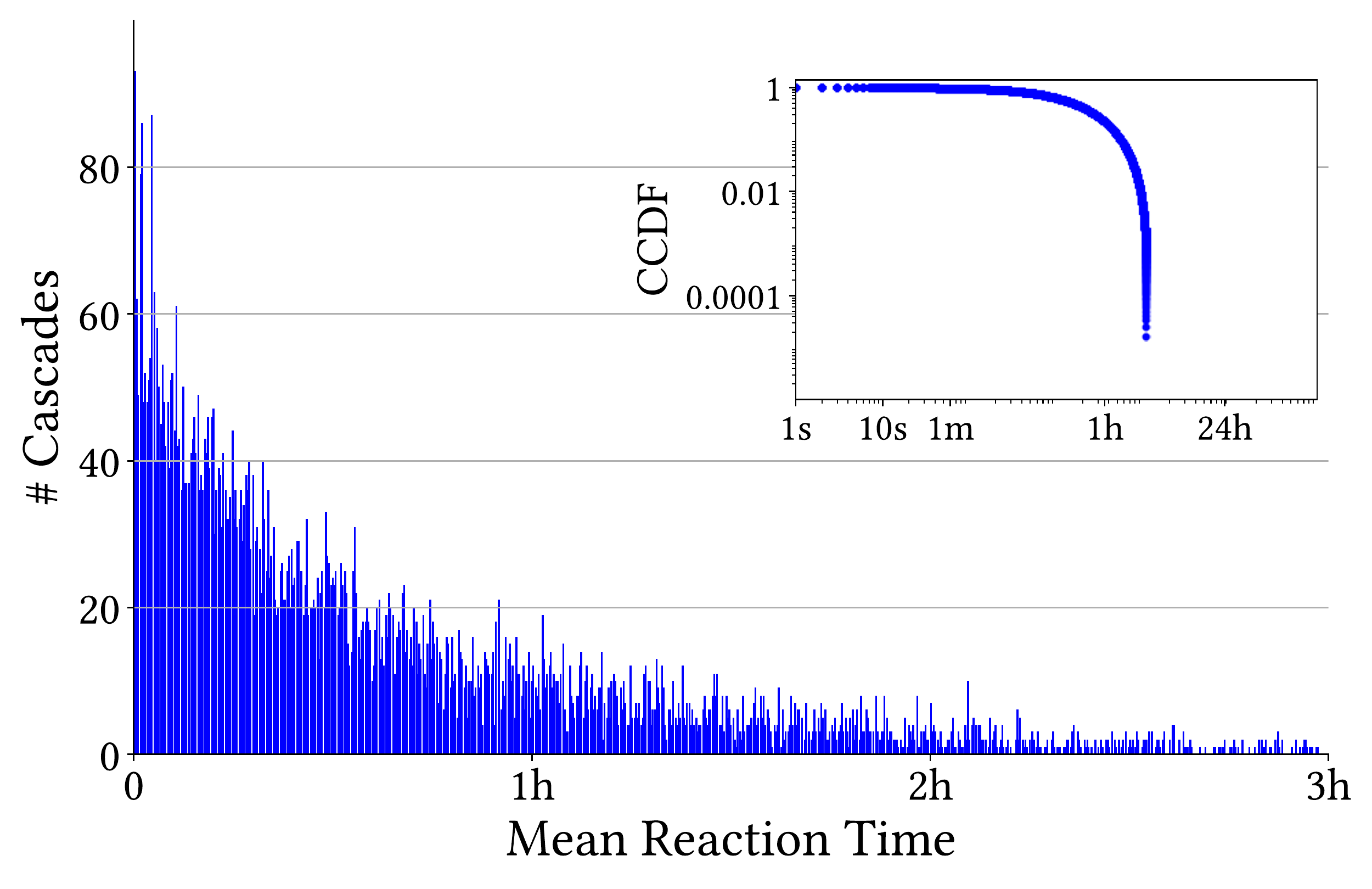}
    \Description[Statistics of two temporal features.]{Most of the arriving time $t_1$ in Weibo dataset are located between a few seconds and an hour. As for mean reaction time, its distribution is exponentially decayed.}
    \caption{Left: Histogram of arrival time $t_1$ of the first retweet in Weibo cascades. Inset: CCDF of time $t_1$. Right: Mean reaction time $\frac{1}{M-1}\sum_{j=2}^M (t_j-t_{j-1})$. Inset: CCDF of mean reaction time.}
    \label{fig:t1}
\end{figure}

\subsubsection{First Participation Time}

The time $t_1$ when the first participant arrives is also considered as an important temporal feature. As shown in the left graph of Figure~\ref{fig:t1}, most tweets with at least 10 retweets in 24 hours received the first retweet no later than 1 hour after publication. 
Similarly, more sophisticated temporal features have been considered in the literature, including \textit{mean arrival time} $\frac{1}{M}\sum_{j=1}^M t_j$, \textit{mean reaction time} $\frac{1}{M}\sum_{j=1}^M (t_j-t_{j-1})$ (right of Fig.~\ref{fig:t1}), \textit{change rate}, \textit{dormant period}, \textit{peek fraction}, etc. For instance, it has been demonstrated that the human reaction time often follows log-normal distribution, e.g., people's reactions on calls, mails, and social networks \cite{barabasi2005origin,zaman2014bayesian}. 

\subsubsection{Evolving Trends}

\begin{figure*}
    \includegraphics[width=\textwidth]{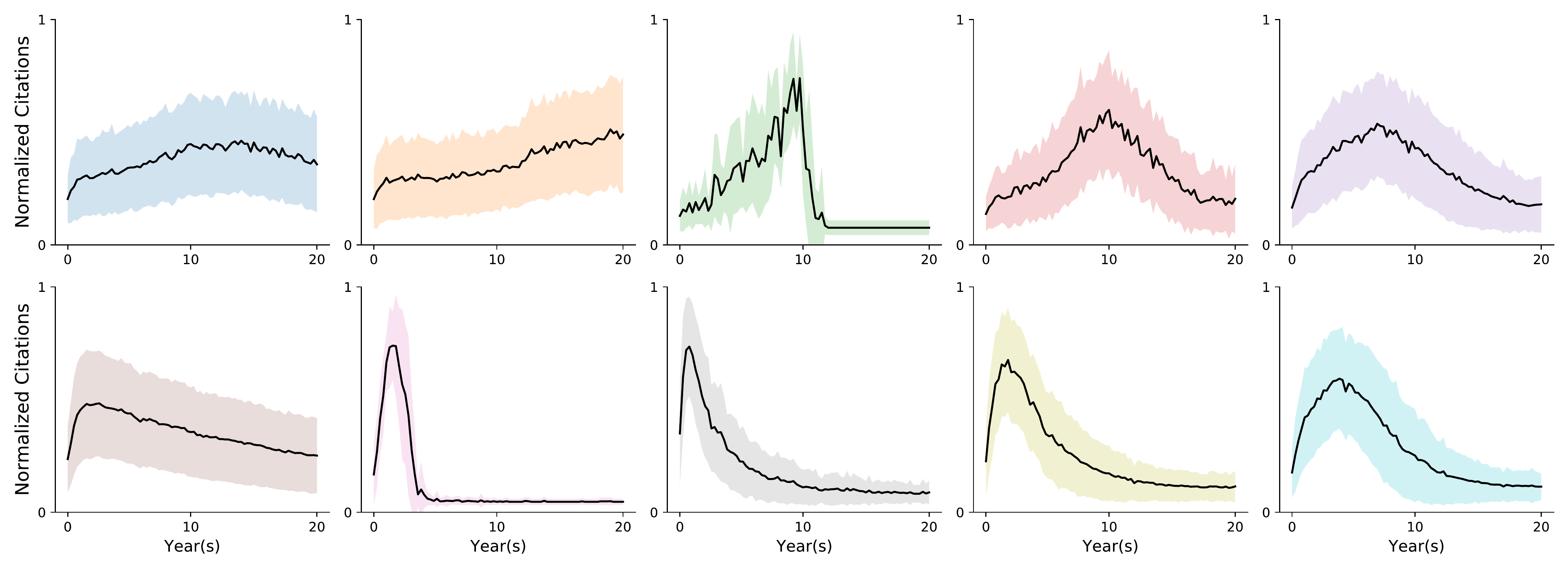}
    \Description[Time series clusters]{We clustered ten types of evolution of cascades, which revel the rise and fall patterns of cascades evolution.}
    \caption{Ten evolving patterns characterizing the citation cascades in the APS dataset. We use hierarchical clustering implemented by \textit{sicpy} package in Python to cluster the APS papers. For each cluster, we show 20 years of evolving trends of citations, by use of mean values and $\pm$ standard deviations.}
    \label{fig:dendrogram}
\end{figure*}

Characterizing the evolving trends of information cascades has been shown to yield informative signals for popularity prediction
\cite{asur2011trends,gursun2011describing,leskovec2009meme,yang2011patterns}. Such temporal patterns of time series can be categorized into several types, e.g., \textit{smoothly} increasing or \textit{suddenly} bursting/decaying, dependent on different clustering algorithms. 
In Figure~\ref{fig:dendrogram}, we show 10 evolving trends of APS paper citations spanning 20 years by applying an agglomerative hierarchical clustering algorithm \cite{hastie2009elements}. 
We can see that most of the clusters follow the trend that paper receives most of its citations in the first few years and then quickly diminishes in the following years.

\subsubsection{Discussion}

Despite their importance of temporal features, recent studies also suggested that they may not work in certain scenarios -- i.e., their advantages diminish over time~\cite{xie2017whats}, and their effect is not always comparable to other features \cite{weng2014predicting}. Note that many works try to model and analyze the temporal evolution trends of cascades, e.g., based on time series patterns~\cite{yang2010modeling,yang2011patterns}, survival analysis~\cite{lee2010approach,lee2012modeling}, and point processes~\cite{shen2014modeling,zhao2015seismic}.
A special class of works, \textit{generative} models, will be systematically reviewed in Section~\ref{sec:gene}.

\subsection{Structural Features}
\label{subsec:structural-features}

Cascade structure, sometimes also referred to as \textit{information diffusion}, has been studied extensively~\cite{bao2013popularity,cheng2014can,galuba2010outtweeting,gao2014effective,zaman2014bayesian,zhang2016structure}, and works can be categorized according to their way of modeling cascades: (i) \textit{participants only}, i.e., only \textit{cascade graphs} are involved; (ii) \textit{global graph}, i.e., both participants and non-participants are considered; and (iii) $r$-reachable graph, i.e., a compromise, extending the cascade graph within the scope of global graph.

\subsubsection{Cascade Graph}\label{subsubsec:cascade-graph}

A cascade graph is constructed based on its participants and their interactions:
\begin{definition}{\textbf{Cascade Graph}.}\label{def:cascade-graph}
Given an information item $I_i$ and the corresponding cascade $C_i$, a cascade graph is defined as $\mathcal{G}_c = \{\mathcal{V}_c, \mathcal{E}_c\}$, where nodes $\mathcal{V}_c = \{u_0, u_1, \dots, u_N\}$ are all participants of cascade $C_i$, and matrix $\mathcal{E}_c \subseteq \mathcal{V}_c\times \mathcal{V}_c$ contains a set of edges representing all the immediate relationships between $\mathcal{V}_c$ in a cascade (e.g., retweeting or citing relationship).
\end{definition}

\begin{figure}
    \centering
    \includegraphics[width=.38\textwidth]{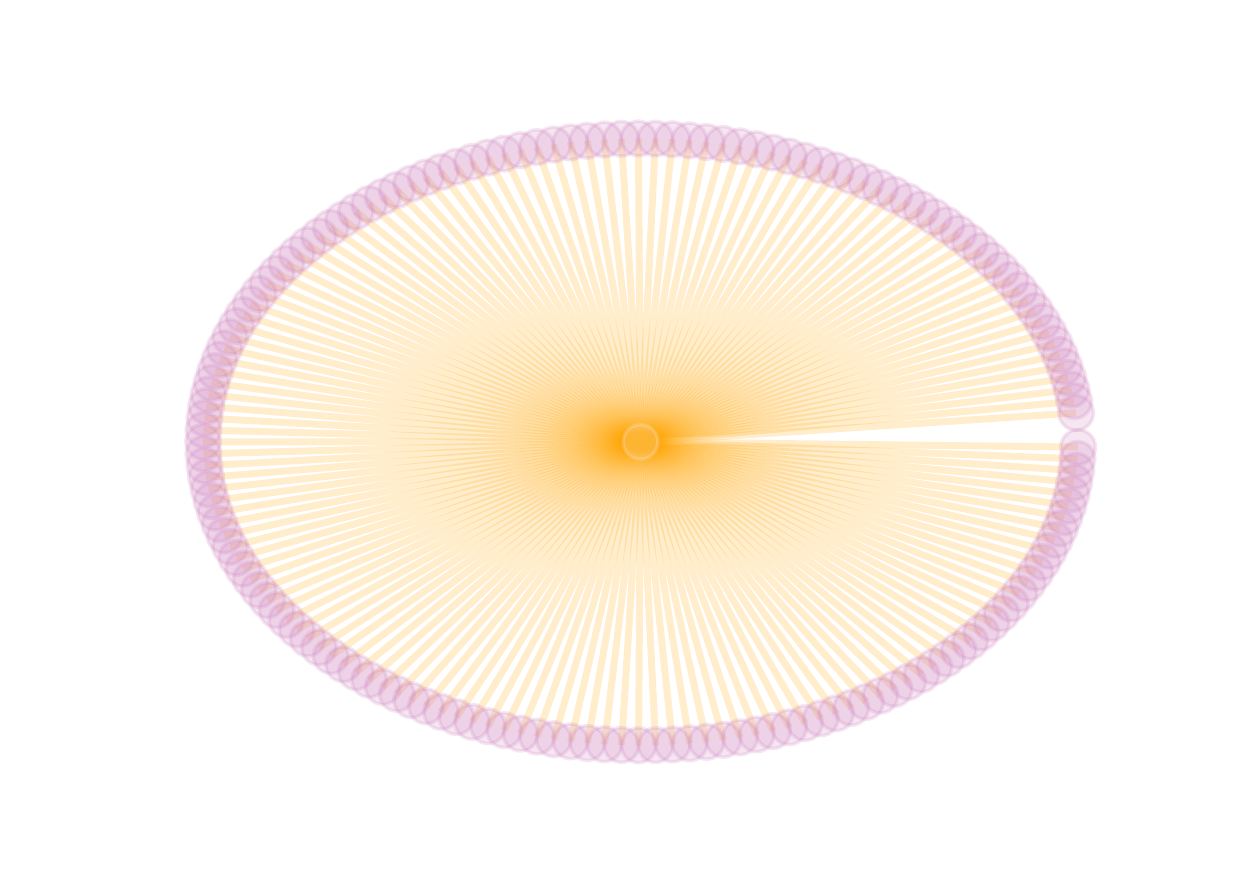}
    \includegraphics[width=.38\textwidth]{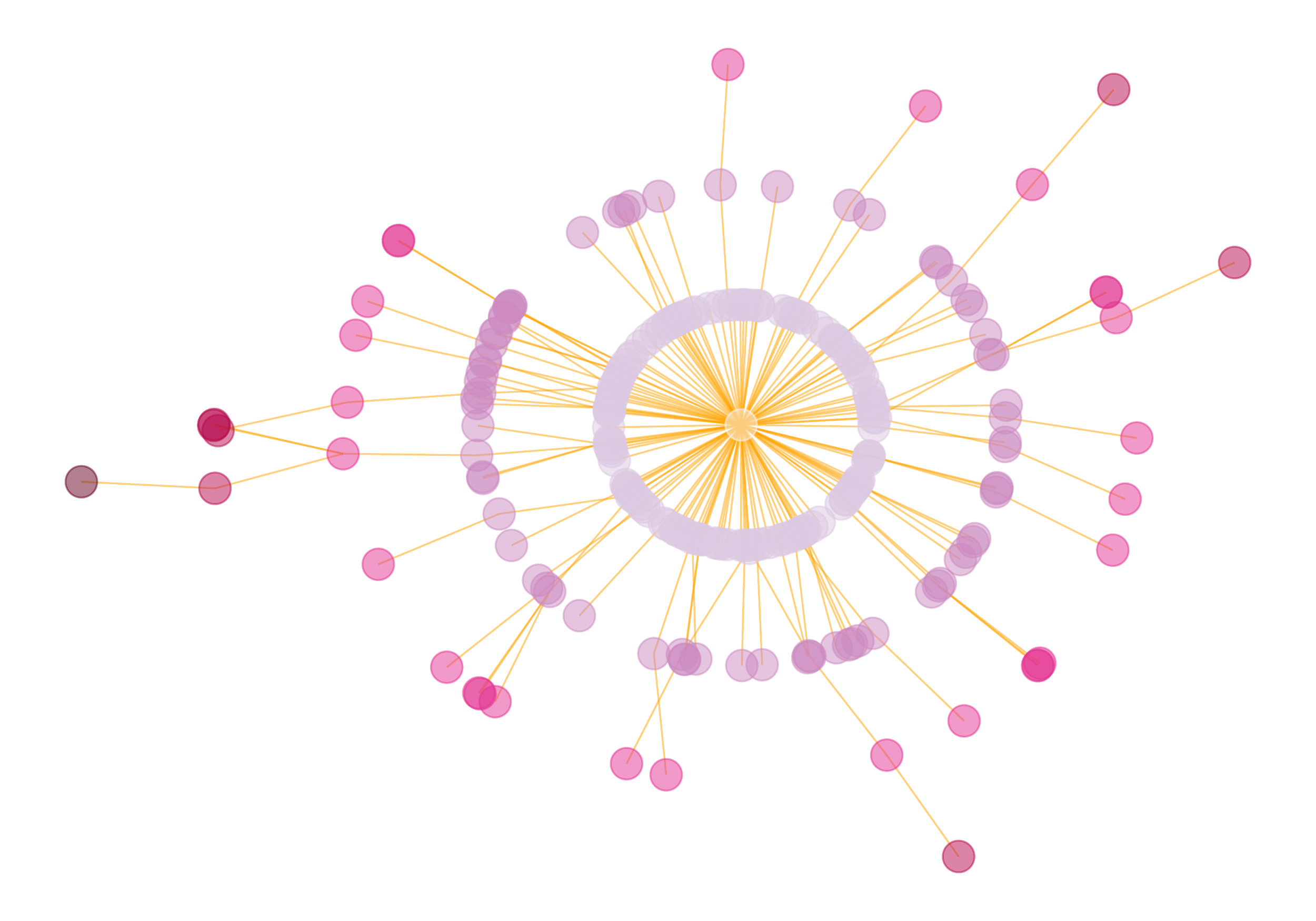}
    \Description[Two example of cascade graphs.]{Two example of cascade graphs, the first graph is broadcast, in which all nodes except the root are directly connected to the root node. The second graph is much viral than the first, a large number of nodes are indirectly connected to the root.}
    \caption{Two cascade graphs retrieved from APS dataset with popularity $P_i \approx 200$. All the nodes of left cascade graph are directly connected to the root node (Wiener index is $1.994$). In contrast, the structure of the right cascade is more viral, i.e., a large amount of nodes are connected to the root node indirectly (index is $2.605$).}
    \label{fig:cascade-graph}
\end{figure}

Cascade graph $\mathcal{G}_c$ characterizes the process of information diffusion of an item $I_i$ such as the  \textit{spreading directions} and  \textit{graph topology}.  One pioneering study employed structural features into popularity prediction to predict which Twitter users are likely to mention a specific URL~\cite{galuba2010outtweeting}. The correlation between cascade popularity and two structural features (edge density and depth) of microblogging network among early participants was analyzed in Reference \cite{bao2013popularity}. It found that with lower value of edge density and higher value of depth -- implying a diverse group of early participants -- the popularity of cascades is more likely to be large. 

The topological structure (shape) of different cascade graphs may vary significantly from each other, even if the number of nodes is the same~\cite{yi2016mining}. As shown in Figure~\ref{fig:cascade-graph}, two cascade graphs from APS have distinct shapes. Although both cascades have $\approx$200 nodes, eventually, the structure, depth, and structural virality (quantified by Wiener index~\cite{goel2015structural}) of the two cascades are totally different. 
This means that simple structure measurements, such as node degree, depth, PageRank, etc., are limited in explaining whether an information item would be popular or not. For example,  non-viral cascades (or broadcast cascades, e.g., left graph in Figure~\ref{fig:cascade-graph}) could have considerably greater popularity in the end, although the virality value is very small at the beginning. In contrast, a viral  cascade graph (e.g., right graph in Figure~\ref{fig:cascade-graph}) not always results in a larger cascade -- in fact, the Pearson correlation coefficient between cascade popularity and their structural virality is relatively low. This phenomenon implies that the initial structural features become less important as cascades grow over time \cite{cheng2014can,zhang2016structure}. 

\begin{figure}
    \centering
    \includegraphics[width=.38\textwidth]{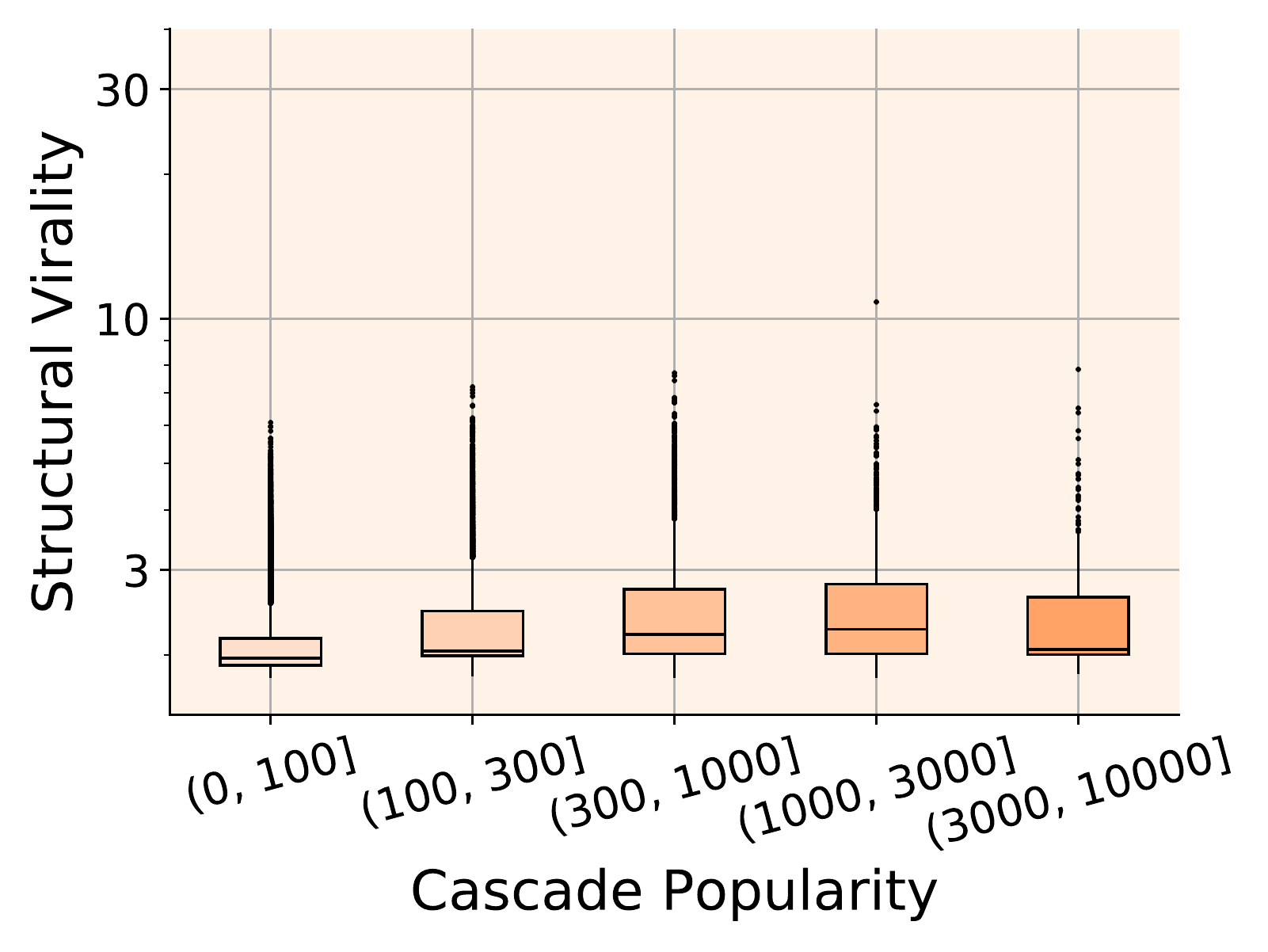}
    \includegraphics[width=.38\textwidth]{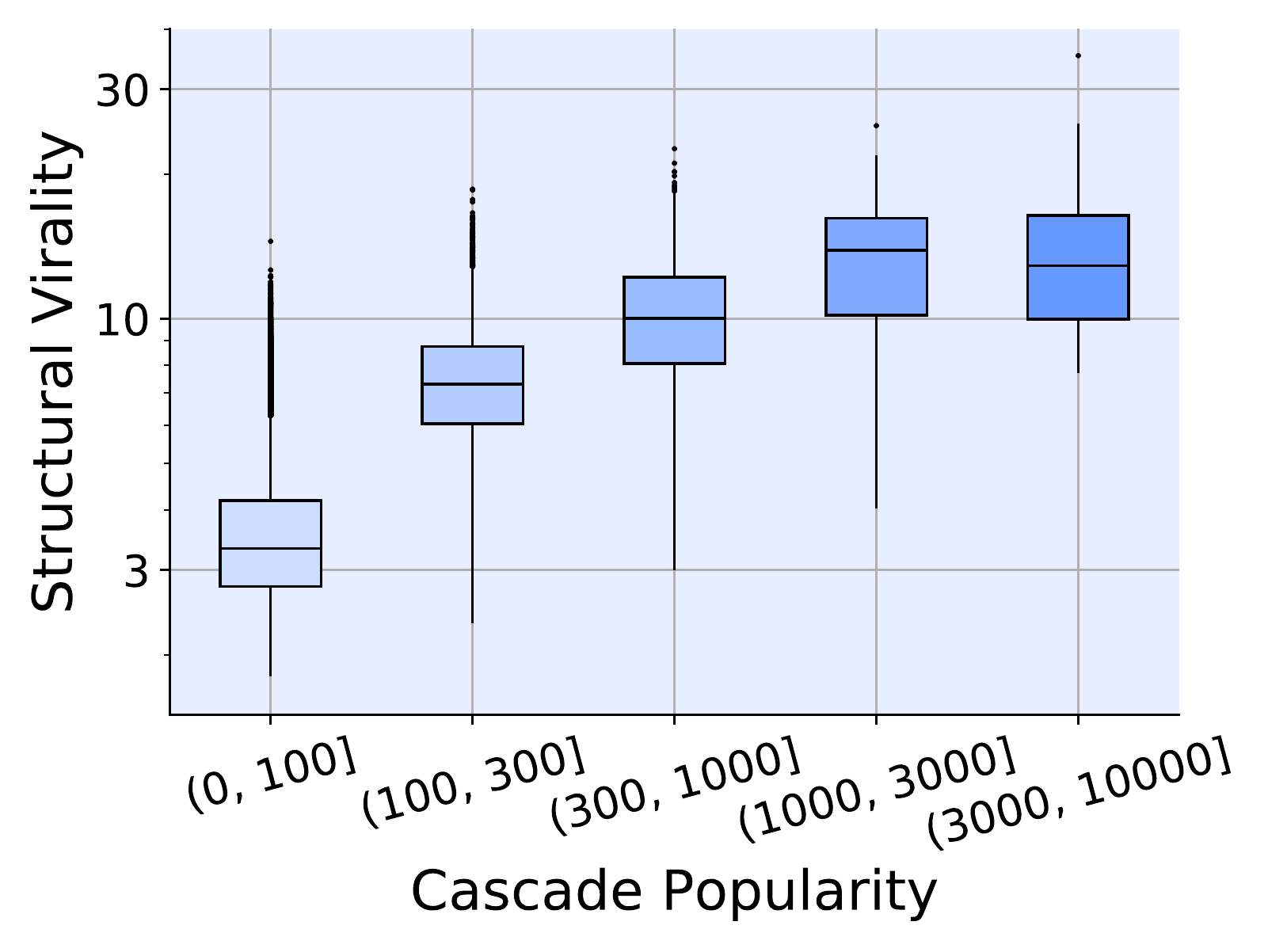}
    \Description[Box plot of structural virality grouped by cascade popularity]{Box plot of structural virality grouped by cascade popularity: as for Weibo dataset, the distributions of structural virality in each popularity group are not changed greatly. In contrast, as for APS cascades, larger cascades are often more viral.}
    \caption{Box plot of structural virality (Wiener index) by cascade popularity for Weibo (left) and APS (right). The lines in boxes are median value, whereas boxes are quartile values of structural virality from lower to upper. The relationship between cascade popularity and structural virality for Weibo tweets is neutral; on the contrary, for APS papers, the correlation is quite positive.}
    \label{fig:size-sv}
\end{figure}

To further investigate the relationship between cascade popularity and structural virality, we draw box plots over the distribution of structural virality on different ranges of cascade popularity for both Weibo and APS (cf.~Figure~\ref{fig:size-sv}), from which we make the important observation that the popularity of microblogs is dominated by two different propagation mechanisms, i.e., \textit{broadcast diffusion} and \textit{word-of-mouth}, dependent on the cascades' eventual size. For cascades smaller than 3,000, the median value of structural virality grows slightly with cascade popularity (i.e., from 1.967 to 2.258). However, for tweets with popularity $>$ 3,000, their median structural virality decreased to 2.053, which means  one of the most important drivers for cascades to succeed comes from the promotion of organizations or celebrities who have a large number of followers~\cite{bakshy2011everyone,goel2015structural}. In other words, the dominant factor for very large cascades is the \textit{broadcast diffusion} mechanism, in which case users are more likely to directly interact with information sources (or important intermediates) and thus the growth of cascade does not rely on viral spreading \cite{dow2013anatomy}. Interestingly, the same trend can also be observed in scientific paper cascades, though their structural virality is generally larger than Weibo tweets (3.779 vs. 2.164).

Other features of the cascade graph include node degree/eigenvector/closeness/betweenness centrality, authority/hub score, graph depth/density/diameter, structural diversity/virality, and various variants of the same such as mean, median, percentile, entropy, standard deviation, etc. More examples can be found in the supplementary materials.

\subsubsection{Global Graph}
\label{subsubsec:global-graph}

Apart from the cascade graph which presents the \textit{local} spread of information, many works studied global graphs~\cite{cheng2014can,gao2014effective,gill2007youtube,myers2014bursty,romero2011differences}, the definition of which depends on the application scenarios. For example, in social networks, the most common global graph is the follower/followee graph or friendship graph~\cite{huberman2008social}. There also exist many other types of global graphs -- e.g.,~\cite{huberman2008social} retrieved a sparse hidden friend graph from the dense follower/followee graph and~\cite{jamali2009digging} defined a co-participation graph from Digg comments. 
Based on the interactions between users (retweet, comment, like/dislike, etc.), from Twitter an interaction global graph can be constructed, e.g., a global graph based on historical mentioning relationships between users~\cite{gao2014effective,gao2014popularity,li2015roles,romero2013interplay,yang2010predicting}.
From APS, based on the interactions between authors (cite, collaborate, etc.), a citing global graph or collaborating global graph was constructed in Reference \cite{dong2015will}. 
These interaction global graphs can be useful in various prediction tasks, especially when the explicit social graph is unavailable, and a suitable choice to represent the actual diffusion of information from historical behaviors \cite{wang2015learning}. 

\begin{definition}{\textbf{Global Graph}}\label{def:global-graph}
A global graph $\mathcal{G}_g = (\mathcal{V}_g, \mathcal{E}_g)$ is a collection of $\mathcal{V}_g$ nodes and a set of $\mathcal{E}_g \subseteq \mathcal{V}_g\times \mathcal{V}_g$ edges representing the relationships between nodes, e.g., friends or collaborators.
\end{definition}

Based on this simple definition, other types of graph can be defined, e.g.: 
(i) a directed graph with edges indicating the direction from one to another, e.g., unilateral relationships such as follower/followee in social networks; 
(ii) a weighted graph whose nodes or edges are associated with assigned weights, e.g., multiple times of interactions between users \cite{yang2012we}; 
(iii) a heterogeneous graph whose nodes or edges have more than one type of attribute, e.g., a graph with \textit{authors}, \textit{papers}, \textit{venues} as nodes, and \textit{publish}, \textit{cite}, \textit{co-authorship} as edges; 
(iv) an attributed graph whose nodes or edges have associated features, e.g., the metadata of information items, the texts of tweets, the titles/abstracts of papers, the embeddings of images, and so on.

A global graph provides us a \textit{macro} perspective to analyze how information disseminates to individuals and to communities, and how information cascades grow their popularity in the context of their positions in the global graph. Compared to information diffusion in the cascade graph only -- which shows the local spread patterns for one specific cascade -- the global graph describes all the relationships between users and potential routes for diffusion. For example, in social networking platforms, besides various external stimuli such as recommendations from the system, key words search from users, etc., information items are primarily discovered and disseminated through the users' social networks \cite{szabo2010predicting}. 

\begin{figure}
    \centering
    \includegraphics[width=.53\textwidth]{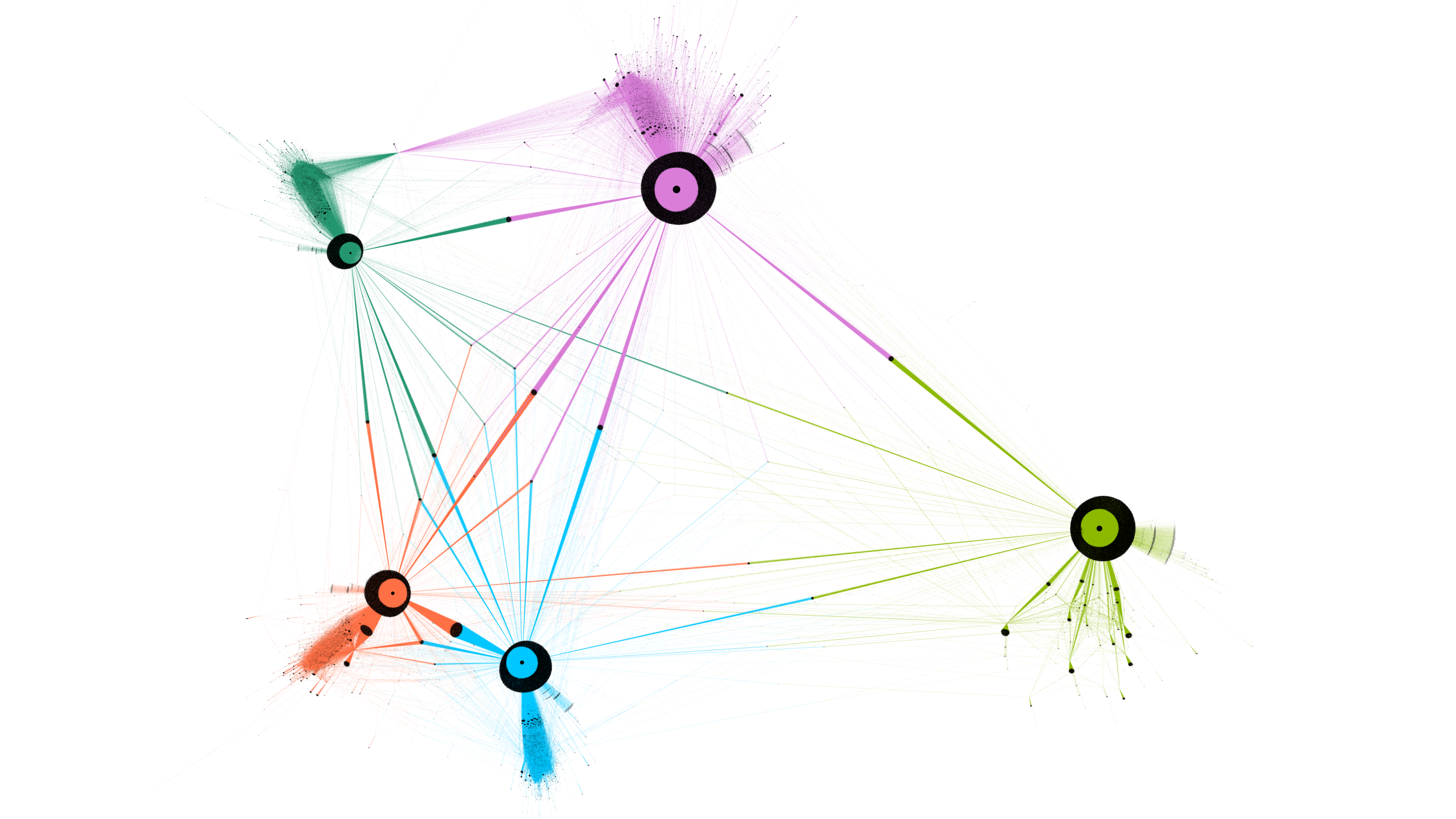}
    \includegraphics[width=.46\textwidth]{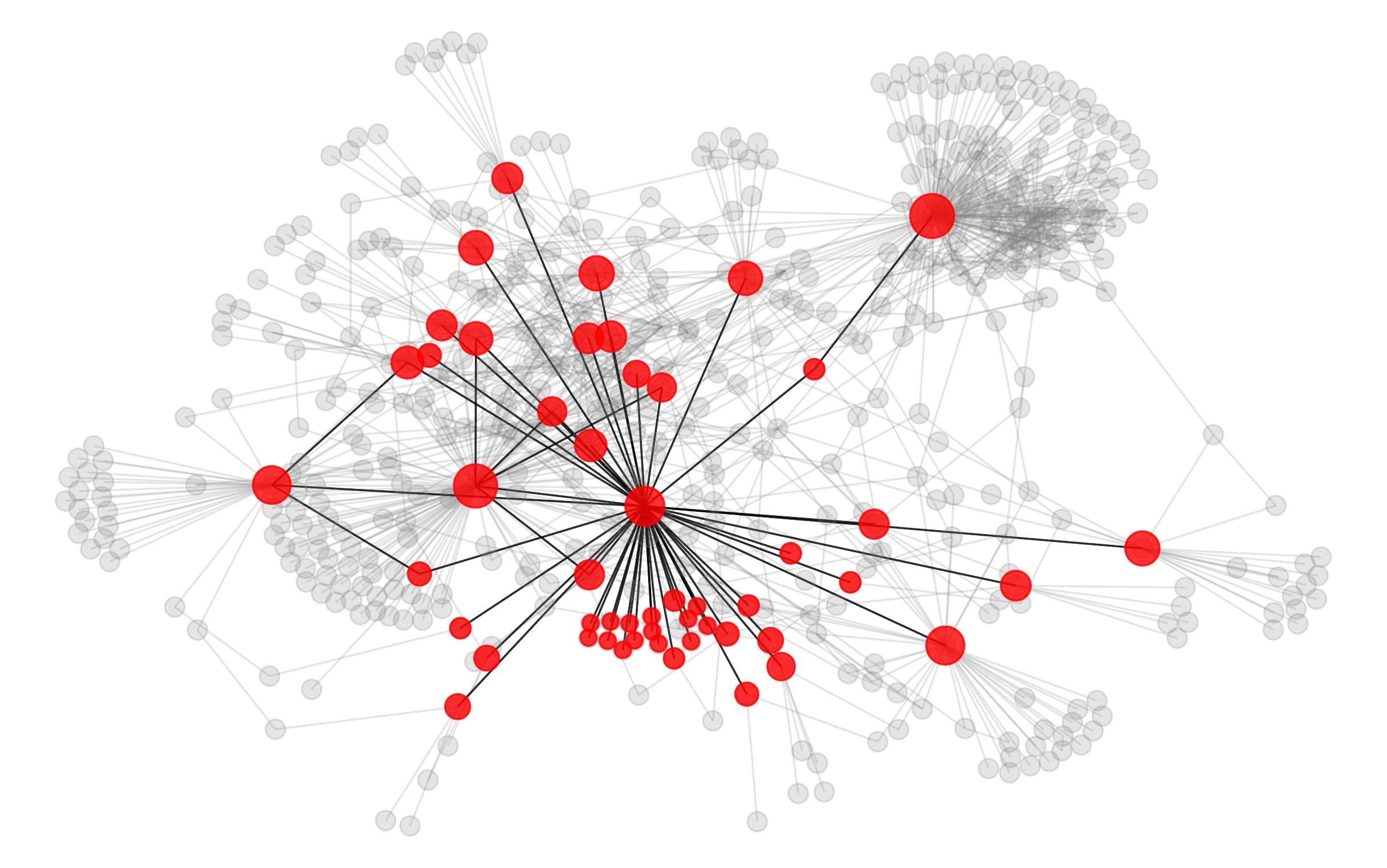}
    \Description[Global graph and $r$-reachable graph]{Global graph and $r$-reachable graph}
    \caption{Left: a global graph retrieved from the five largest cascades in Weibo dataset. It contains 262,458 nodes and 324,540 edges. Edges from different cascades are in different colors. We can clearly see that these cascades are not isolated. Right: an example of $1$-reachable graph retrieved from Weibo dataset. Dark nodes (red) come from a cascade graph $\mathcal{G}_c$, where pale nodes are neighbors of the red nodes.}
    \label{fig:graph-examples}
\end{figure}

The left portion of Figure~\ref{fig:graph-examples} shows an example of global graph, which is composed of the five largest cascades in the Weibo dataset. Nodes in the graph are individual users while edges represent the retweeting relationships between users. As it shows, a large number of users not only participate in one cascade but also act as bridges between separate cascades. The historical behaviors, personal preferences, and communities of users can further help us to identify the roles of users in the cascade graphs \cite{tsugawa2019empirical,weng2014predicting}. 
A complete global graph can be very large, e.g., a retweeting global graph retrieved from Weibo has more than 6M nodes (users) and 15M edges (retweet actions), while the citing global graph retrieved from APS has 422K nodes (authors) and 54M edges (cite actions). 
It is noteworthy that nodes in cascade graph may not necessarily appear in the global graph. For example, in Twitter, a user can retweet posts from other users who are not her/his followers or followees. This limits the usability of models which rely only on the global graph \cite{he2014predicting,yan2011citation}. 

Previous studies have also highlighted the influence of social community, e.g., social reinforcement and homophily of social contagions \cite{anderson2015global,centola2007complex,mcpherson2001birds,romero2011differences}. As shown in References \cite{weng2013virality,weng2014predicting}, the viral cascades often diffuse across more communities than non-viral cascades, i.e., they are less likely to be trapped into a low number of communities. However, the analysis of the inter-community diffusion patterns of tweets in Reference \cite{tsugawa2019empirical} demonstrated that the effects of community features are relatively small compared to other features such as past success and user degrees. 

\subsubsection{$r$-reachable Graph}\label{subsubsec:r-reachable-graph}

Based on the definition of cascade graph and global graph, we can define a sub-graph retrieved from the global graph and named $r$-reachable graph:

\begin{definition}{\textbf{$r$-reachable Graph}}
Given a global graph $\mathcal{G}_g$, and its cascade sub-graph $\mathcal{G}_c$, an $r$-reachable graph of $\mathcal{G}_c$ is defined as $\mathcal{G}_c^r = \{\mathcal{V}_c^r, \mathcal{E}_c^r\}$, where $\mathcal{V}_c^r$ contains (1) all nodes in $\mathcal{V}_c$, and (2) all nodes in $\mathcal{V}_g$ are within $r$-hops of nodes in $\mathcal{V}_c$.
For example, when $r = 1$, $\mathcal{V}_c^r$ contains all $\mathcal{V}_c$ and all immediate neighbors of nodes in $\mathcal{V}_c$.
\end{definition}

The right portion of Figure~\ref{fig:graph-examples} shows an example of $r$-reachable graph ($r = 1)$, where dark nodes and their interactions form the cascade graph. $r$-reachable graph tells us how many nodes are exposed to the active nodes and their topology. The rationale behind modeling an $r$-reachable graph is that the highly exposed nodes would potentially bring more nodes into this cascade in the future~\cite{gao2014popularity}. 

Previous works have utilized $r$-reachable graph to facilitate the prediction of cascades' popularity~\cite{cheng2014can,ma2013predicting}. 
Global graph $\mathcal{G}_g$ and several $1$-reachable graphs $\mathcal{G}_c^1$ (a.k.a. border graph) from Weibo retweet cascades were constructed in Reference \cite{gao2014effective,gao2014popularity}, extracting structural features from $\mathcal{G}_g$ and $\mathcal{G}_c^1$ to predict the popularity of tweets.
A comparison of several structural and content features in Reference \cite{ma2013predicting} revealed that the number of nodes in the cascade graph and $1$-reachable graph are the two most predictive of 53 features in predicting the popularity of Twitter hashtags. However, the construction and calculation of an $r$-reachable graph can be costly, e.g., for a cascade graph with dozens of nodes (on average), its $2$-reachable graph may contains tens of thousands of nodes.

\subsubsection{Discussion}

The influence of the cascade graph's early structure on the final item popularity remains largely unclear, with neither of the standard structural feature selection approaches making an accurate prediction. 
Analyzing structural patterns and identifying influential users in evolving graphs become paramount \cite{yang2019influence}.
In addition, different platforms have unique diffusion mechanisms that may cause dynamics to differ from the well studied social network scenarios \cite{cheng2014can,krishnan2016seeing,leskovec2007dynamics,leskovec2007cascading,li2015roles,qiu2018deepinf,zhao2015seismic}. For example, the findings in Reference \cite{anderson2015global} show that the correlation between the popularity of LindedIn signup cascades and structural virality is remarkably high, which shows a very similar structural behavior to APS citation cascades. 
Even for a particular platform, the types of items (events) and the item contents may affect the information diffusion behavior \cite{goel2015structural,zhang2016structure}.

\subsection{User/Item Features}
\label{subsec:user-item-features}

Temporal and structural features require probing into early observations, which is sometimes impractical. Therefore, a number of works have focused on the features associated with users and information items, which have unique properties and innate attractiveness that make them especially useful in predicting popularity before publication.

\subsubsection{User Features}

User behaviors play crucial roles in information dissemination and consumption (e.g., viewing, commenting, sharing, and preferring).
One of the most common user features is the number of followers as a proxy of user influence, which implies the speed and timing of future popularity \cite{zaman2014bayesian}. Those who have a large number of followers (audiences), e.g., celebrities and news organizations, are more likely to generate large cascades than normal users, since their messages are more visible in the network~\cite{bakshy2011everyone,jenders2013analyzing,suh2010want}. However, large cascades are not only produced by influential users and it is of interest to study large cascades originated by normal users instead of celebrities~\cite{dow2013anatomy}. Many other user features have been extensively studied and explored for analyzing and predicting the popularity of information items/cascades, e.g., profiles (name, age, region, education, employment, account creation date, etc.)~\cite{yuan2016will}, historical behaviors (frequency of publish items and interact with other users, active time, etc.)~\cite{bakshy2011everyone,hong2011predicting,suh2010want}, user interests \cite{yang2010understanding}, collectivity~\cite{lu2018collective}, similarity~\cite{shulman2016predictability}, past success~\cite{bakshy2011everyone}, activity/passivity~\cite{yang2012we,yu2014twitter}, discoveries~\cite{medo2016identification}, affinities and responsiveness~\cite{yuan2016will}, and so on.

\subsubsection{Item Features}

In addition to the effects of user features, the effects of item characteristics
on diffusion have also been studied in the literature. For example, Reference \cite{lerman2010using} analyzed how Digg users' interfaces affect the visibility of items; Reference
\cite{tsagkias2009predicting} showed that metadata of news articles affects the volume of comments (e.g., the publication date of the article, number of articles published at the same time, number of articles with similar content, etc.); Reference \cite{jamali2009digging} proposed to use entropy computed among information categories and topics; and Reference \cite{wu2016unfolding} used prevalence variability to analyze how popularity of different types of items changes over time. 

\subsubsection{Discussion}

Most of the user/item features are self-explanatory, and it is not too hard to understand their correlations to popularity. For instance, users who have a larger audience, who live in a densely-populated area, who speak a widely-used language, who often discuss trending topics, etc., would have a greater chance to make their information popular. However, some features require more complicated algorithms and calculations to obtain, e.g., user influences, preferences, and similarities. 
Many prior works studied how individual users affect the diffusion of information. In Reference \cite{kwak2010twitter}, number of followers, PageRank, and number of retweets were combined to rank influential users. Reference \cite{bian2014predicting} studied three types of influence during information diffusion, i.e., interest-oriented influence, social-oriented influence, and epidemic-oriented influence, and showed that each of these three surrogates of influence can contribute to the popularity prediction. However, some previous works found social influence that, merely measured by user topology in graph, reveals little about the actual influence~\cite{cha2009analyzing,cui2013cascading}. Leveraging historical evidence of user behaviors, an orthogonal sparse logistic regression (OSLOR), which predicts cascade outbreak via selecting users who are more powerful and less redundant, was proposed in Reference \cite{cui2013cascading}.

\subsection{Content Features}
\label{subsec:content-features}

Content is recognized as the inner drive and one of the key factors that leads items to success \cite{yano2010s}, e.g., breaking news, rumors/fake news, hot spots, controversial/peculiar topics, disinformation and misinformation, etc., attract significantly more attention than normal content. 

\subsubsection{Text Content}
\label{subsubsec:content-features-of-text}

Text features are widely adopted by existing popularity prediction models, since text is ubiquitous information appearing in articles, microblogs, image/audio/video captions/descriptions, and can even be retrieved from images/audio/videos. 

Existing works analyze the user-generated textual content with various language models, such as Term Frequency-Inverse Document Frequency (TF-IDF) and Latent Dirichlet Allocation (LDA)~\cite{blei2003latent}, combined with typical machine learning models such as naïve Bayes, SVM and linear regression to predict item popularity. For example, TF-IDF and LDA are used to learn the topic distributions of tweets in Reference \cite{hong2011predicting}. TF-IDF was also used in Reference \cite{yang2010understanding} to estimate the importance of keywords in user tweets, and further calculated the mutual correlation between all the historical content of her/his tweets and the content of a specific item to measure how much a user likes a piece of content, i.e., if the correlation is high, then this user may have a higher probability of adopting this item. In Reference \cite{khabiri2009analyzing}, authors
analyzed several semantic and statistical content features of Digg comments, including comment length, number of verbs/nouns, content entropy, readability, subjectivity/objectivity, polarity, etc., and they found that people prefer to retweet short, simple, and readable content. The impact of textual and semantic features on the volume of news article comments was discussed in Reference \cite{tsagkias2009predicting}. Predicting whether a tweet would be retweeted purely based on the content of tweets (including marks, terms, emoticons, sentiments and topics) from Reference \cite{naveed2011bad} was later adopted in Reference \cite{kupavskii2012prediction} to predict the size of retweet cascades, and for sentiment impact on retweetability of political tweets in Reference \cite{stieglitz2012political}. The design of content features for predicting scientific impact of academic papers, including LDA topic models which extract probability distribution from paper titles/abstracts, topic diversity calculated by entropy, and paper novelty, as well authority of authors~\cite{kleinberg1999authoritative}, etc., in References \cite{yan2011citation,dong2015will,dong2016can} showed that the content features of papers play critical roles in prediction. The sentiment and self-disclosure of tweets analysis, along with calculations of topic affinity of tweets using a Twitter-based LDA for better topic discovery and extraction, was presented in Reference \cite{yuan2016will}; another study embedded sentences in vectors to combine sentiments of the descriptions of short videos~\cite{chen2016micro}. Combining domain, spam score, tweet topic, user topic, topic interactions between tweet and user, and tweet category as the content features were studied in Reference \cite{martin2016exploring}. The comparative study of transactional (TF-IDF or LDA) and semantic (sentiment, controversy, content richness, hotness and trend momentum) models and their inter-relationships in Reference \cite{zhao2018comparative} considered three kinds of diffusion mechanism: cascades of URL, hashtag, and retweet. It was found that semantic features have better performance on hashtag cascades (with higher content complexity), while transactional features are more effective on URL and retweet cascades (with lower content complexity). 

\subsubsection{Content Features of Image}
\label{subsubsec:content-features-of-image}

The characteristics of images are significantly different from texts, in terms of modeling and retrieving image features, which often require techniques from computer vision learning. Some basic features regarding an image can be retrieved easily if available, e.g., in Reference \cite{mcparlane2014nobody} the authors analyzed the basic attributes associated with images, including size, time, date, season, orientation (landscape/portrait), device, dominant color, and even whether taken on flash, as well as image resolution, location (coordinates), caption, tag, etc. 

One seminal work~\cite{khosla2014makes} studied the correlations between features of Flickr images and their normalized popularity (i.e., view count). In particular, the content features of images -- categorized as simple human-interpretable features, low- and high-level image features -- help in  improving the prediction performance significantly, e.g., images with striking colors tend to be popular. The low-level image features, such as gist, texture, color patches, gradients, along with representative features are extracted by local binary pattern~\cite{ojala2002multiresolution} and convolutional neural networks (CNNs)~\cite{lecun1995convolutional}, while the high-level image features are retrieved from an ImageNet classifier~\cite{krizhevsky2012imagenet} among 1,000 object categories, each indicating presence or absence of a certain category in the image. The results were later extended in Reference \cite{gelli2015image} by taking sentiment features into consideration.

In Reference \cite{totti2014impact}, authors analyzed a wide spectrum of visual features of Pinterest images and evaluated their influence on popularity. Except for frequently used individual features, visual features were grouped into two sets: aesthetic and semantic features, the former accounting for the ``beautifulness'' of an image (e.g., dominant colors, saturation, brightness, contrast, texture, background area, region focus and focus centrality/density, etc.), while the latter are extracted from various computer vision techniques, e.g., SIFT local descriptor~\cite{lowe2004distinctive}. Other works include: Reference \cite{cheng2014can} which estimates the likelihood scores of categories as the content features of Facebook photos; Reference \cite{bian2014predicting}, which preprocessed images of tweets into a bag of visual words as content features; and Reference \cite{lv2017multi} which, in addition to extracting features like texture and color, also adopted VGG19~\cite{simonyan2015very} to extract deep features. 

\subsubsection{Other Content Features}
\label{subsubsec:other-content}

For online videos, basic features include the video length, resolution, and number of frames. Using several key frames of a short video to represent the video, Reference \cite{chen2016micro} added color histogram and aesthetic features, quality features, object features retrieved from CNNs, and sentiment features trained by SentiBank~\cite{borth2013large}, and used them as visual features of videos (acoustic features were also used -- including Mel-Frequency Cepstral Coefficients and Audio-Six). Detection of faces and texts in video frames in Reference \cite{trzcinski2017predicting} combined scene dynamics, clutter, rigidity, thumbnail, and features extracted from ResNet~\cite{he2016deep}. 
In Reference \cite{xie2020multimodal}, variational inference and multimodal features are used against noisy and uncertain factors, when predicting the popularity of micro-videos.

There have been some interesting experiments to collect features using crowd-sourcing. The controlled experiments on songs in~\cite{salganik2006experimental} found that social influence increases the inequality and unpredictability of a song's popularity. The experiments on Amazon Mechanical Turk~\cite{figueiredo2014does} required the respondents to watch several YouTube videos and choose which video they were willing to share and predict which video would be the most popular in the future. Although the perceptions of users are very subjective and consensus among them is rare,  the popularity of the target video became significantly higher once the respondents reached an agreement.

\subsubsection{Discussion}

Content of items has proven to be a qualified predictor, and the prediction can be made on and before the items' publication when no posterior information is observed (cold start). Many existing studies have analyzed the relationships between content of items and their popularity, including linguistic characteristics of texts \cite{tan2014effect}, color and objects in images \cite{khosla2014makes}, and so on. 
However, in analyzing the root causes of an item's final popularity, it is still hard to disentangle the effects of such descriptive factors from the effects of the item's intrinsic content~\cite{stoddard2015popularity}. 

Opinions on the effectiveness of content features differ. Some previous works consider content features as \textit{weak} predictors compared to other features such as temporal, structural, and/or individual features. For example, Reference \cite{cheng2014can} found that content features get less important when observing more participants, and Reference \cite{bakshy2011everyone} found that the model was not improved by addition of content features, which is also confirmed in Reference \cite{kupavskii2013predicting}. Authors of Reference \cite{martin2016exploring} argued that content features explain the variance of popularity poorly. Content based methods suffer from issues that hinder their performance -- e.g., despite the recent success of deep learning, natural language processing and computer vision, it is still challenging to effectively and efficiently identify, retrieve, and model the content of items, and the results are far from satisfactory. In addition, previous works found that even for items with identical content, their popularity varies greatly~\cite{borghol2012untold,cheng2014can,gilbert2013widespread,lakkaraju2013s}, raising questions about whether, if one relies only on content features, the popularity of items/cascades is inherently unpredictable or cannot be predicted \textit{a~priori}. 

\subsection{Prediction Methods}
\label{subsec:machine-learning-methods}

\begin{table}\small
    \caption{Machine Learning Methods}
    \label{tab:ml-methods}
    \begin{tabular}{@{}llp{7.5cm}@{}}
    \toprule
        \textbf{Method}                 & \textbf{Abbr.}& \textbf{Reference} \\ \midrule
        Autoregressive (–moving-average) & AR(MA)        & \cite{ding2015video,gupta2012predicting,gursun2011describing,matsubara2012rise,li2013popularity,yang2010modeling,zadeh2014modeling} \\ 
        Decision Tree                   & DT            & \cite{bandari2012pulse,cheng2014can,dong2015will,dong2016can,gao2014effective,gao2014popularity,gao2019taxonomy,gupta2012predicting,jamali2009digging,kong2014predictingbursts,kong2015towards,kong2016popularity,krishnan2016seeing,kupavskii2012prediction,kupavskii2013predicting,ma2012will,ma2013predicting,martin2016exploring,mcparlane2014nobody,vallet2015characterizing,wang2015burst,yan2011citation,yi2016mining} \\
        $k$-nearest Neighbors Algorithm & $k$-NN        &  \cite{bandari2012pulse,ding2015video,gao2014popularity,gao2019taxonomy,gupta2012predicting,jamali2009digging,kong2014predicting,kong2016popularity,ma2012will,ma2013predicting,li2013popularity,yan2011citation} \\ 
        Linear Regression               & LR            & \cite{abisheva2014watches,ahmed2013peek,bandari2012pulse,bao2013popularity,borghol2012untold,castillo2014characterizing,cheng2014can,ding2015video,dong2016can,gao2019taxonomy,guo2016comparison,gupta2012predicting,he2014predicting,khabiri2009analyzing,kong2014predicting,kong2014predictingbursts,kong2015towards,li2013popularity,lv2017multi,martin2016exploring,pinto2013using,rizoiu2017expecting,ruan2012prediction,szabo2010predicting,tatar2011predicting,tatar2014popularity,tsur2012s,weng2014predicting,wu2018beyond,yan2011citation,yano2010s,yu2015lifecyle,yu2015micro,zaman2014bayesian,zhao2015seismic} \\ 
        Logistic Regression Classifier  & LRC           & \cite{cheng2014can,cui2013cascading,dong2015will,dong2016can,gao2014popularity,hong2011predicting,jenders2013analyzing,jia2018predicting,kong2016popularity,krishnan2016seeing,ma2013predicting,mcparlane2014nobody,naveed2011bad,oghina2012predicting,qiu2018deepinf,romero2013interplay,shulman2016predictability,wang2015burst,xie2017whats,yang2010understanding,zhang2016structure} \\
        Multilayer Perceptron           & MLP           & \cite{dong2015will,gao2019taxonomy,kong2014predictingbursts,kong2015towards,wang2015burst} \\
        naïve Bayes classifier          & Bayes         & \cite{cheng2014can,dong2015will,dong2016can,gao2014popularity,gao2019taxonomy,gupta2012predicting,jenders2013analyzing,ma2012will,ma2013predicting,shafiq2017cascade,shamma2011viral,yano2010s,yi2016mining} \\
        Random Forests                  & RF            & \cite{alzahrani2015network,bora2015role,cheng2014can,dong2015will,dong2016can,gao2019taxonomy,guo2015toward,guo2016comparison,jia2018predicting,kong2016popularity,krishnan2016seeing,martin2016exploring,mishra2016feature,rizos2016predicting,shulman2016predictability,totti2014impact,tsagkias2009predicting,tsugawa2019empirical,wang2015burst,weng2013virality,weng2014predicting,xie2017whats,yi2016mining,zhang2016structure} \\ 
        Support Vector Machine          & SVM           & \cite{bandari2012pulse,chen2016micro,cheng2014can,dong2015will,dong2016can,gao2014popularity,gelli2015image,guo2016comparison,gupta2012predicting,jamali2009digging,jia2018predicting,khosla2014makes,kong2014predictingbursts,kong2015towards,kong2016popularity,kong2018exploring,krishnan2016seeing,lakkaraju2011attention,lu2017predicting,luo2018real,lv2017multi,ma2012will,ma2013predicting,mcparlane2014nobody,qiu2018deepinf,rowe2011predicting,shulman2016predictability,trzcinski2017predicting,tsur2012s,wang2015burst,yan2011citation,yang2010understanding,yang2012we,yu2014twitter,zhang2016structure} \\ 
    \bottomrule
    \end{tabular}
\end{table}

Since the main challenge of feature-based models lies in the feature engineering, improving the capability of prediction models is not the focus in related literature. For example, Reference \cite{cheng2014can} demonstrated that most of the machine learning methods have similar performance, despite time/space complexity. For completeness, we summarize common machine learning methods, or what are adopted as their main building blocks, as the prediction methods in Table~\ref{tab:ml-methods}. In addition to the methods listed in Table~\ref{tab:ml-methods}, a few learning paradigms such as inductive/transductive learning, early feature fusion, and multi-view learning approaches have been investigated to predict the popularity of information items/cascades~\cite{chen2016micro,hoang2017gpop,lv2017multi,wu2016unfolding}. 
We suggest researchers to experiment different prediction methods on their specific datasets, and techniques such as automatically selecting machine learning models and hyper-parameters \cite{kotthoff2017auto}, would greatly 
boost the training \& optimizing processes. 

\subsection{Global Overview of Pros and Cons}
\label{subsec:feature-pros-and-cons}

Feature-based models are often reported as competitive and explainable compared to others \cite{guo2016comparison}. However, the main bottleneck of feature-based models preventing their implementation in real application is the hand-crafted feature engineering. 
Some features are hard to obtain due to privacy concerns, such as preferences and viewing histories, and some (e.g., user classification and clustering) are computationally intensive, which limits the models' scalability. 
Generally speaking, most of the temporal features -- as well as user/item features -- are easily to extract and compute. On the other hand, structural features, especially for those large-scale graphs, e.g., global graph and $r$-reachable graphs, which often contain thousands or even millions of nodes and edges, require extensive computational resources. 
Content features of texts, images, audio and videos, depending on specific problem formulations, data scales, and modeling algorithms, have different time/space complexities.

In this spirit, given an exhaustive set of features, how to select a relatively small portion of representative features to maximize the marginal benefits of prediction between effectiveness and efficiency, becomes a critical consideration for designing practical prediction models~\cite{gao2014effective,yi2016mining,zhao2015seismic}. 
Also, existing models require more features like history view counts and diffusion paths, which in most cases are unavailable and, most importantly, are not generalizable to different scenarios. 

We reviewed four categories of features and corresponding prediction models. However, it is impossible to mention all features and models, or to evaluate their performance in all feature combinations. 
A comprehensive study on evaluating different features in different conditions would be beneficial in standardizing the feature engineering and feature selection/combination.

\section{Generative Models}\label{sec:gene}Many real-world phenomena -- e.g., information retweeting,  hospital admissions, and citations of scientific papers, -- can be formulated as event sequences in the continuous temporal domain. Modeling the arrival/occurrence of event sequences or the participation time series is a fundamental step towards understanding the underlying dynamics of the information diffusion.
The spreading of information items is therefore widely characterized by probabilistic statistical generative approaches such as epidemic models, survival analysis, and various stochastic point processes. 
In this section, we review these generative models for popularity prediction. 

\subsection{Poisson Processes}\label{subsec:poisson}

Models based on point processes distinguish themselves from feature-based models because of their statistical, probabilistic, and generative forms. Point processes are often used when modeling time series, e.g., the arrival rate of customers, phone calls, and mechanical failures, in queuing theory and operational research. 

A generative probabilistic model using reinforced Poisson process (RPP) was proposed in Reference \cite{wang2013quantifying} (followed by Reference \cite{shen2014modeling}) to predict items' popularity  (in their case the scientific impact of papers) using three key ingredients: (i) fitness or attractiveness of an item; (ii) temporal decay function; and (iii) reinforcement mechanism (e.g., \textit{rich-get-richer}). Specifically, the rate function of \textit{inhomogeneous} Poisson process for information item $I_i$ is defined as
\begin{equation}\label{RPP}
    \lambda_i(t) = \alpha_i \phi_i(t) P_i(t),
\end{equation}
where $\alpha_i$ is the attractiveness of item $I_i$
represented as a single value evaluated by maximum likelihood estimation~\cite{wang2013quantifying}, or follows posterior distribution from a conjugate prior~\cite{shen2014modeling};  $\phi_i(t)$ is the relaxation function or decay function to characterize the aging effect; $P_i(t)$ is the total number of popularity $I_i$ received at time $t$. With the above defined rate function, given $(j-1)$th participant arrives at $t_{j-1}$, the probability of $j$th participant arrives at $t_{j}$ is given by
\begin{equation}
    \Pr(t_{j}|t_{j-1}) = \alpha_i \phi_i(t_j) P_i(t_j) e^{-\int_{t_{j-1}}^{t_j} \alpha_i \phi_i(t_j) P_i(t_j) \text{d} t}.
\end{equation}

Based on RPP model, Reference \cite{gao2015modeling} proposed an extended model PETM. Unlike RPP, which is designed to quantify the long-term scientific impact, PETM is more applicable to the scenario of microblogging platforms. Specifically, PETM uses a power-law time decay function $\phi_i(t) = t^{-\gamma_i} (\gamma_i > 0)$ to substitute the log-normal distribution $\phi_i(t) = \frac{1}{\sqrt{2\pi}\sigma_it}\exp{\left(-\frac{(\ln t - \mu_i)^2}{2\sigma_i^2}\right)}$ in Reference \cite{shen2014modeling}, since they observe a power-law distribution in retweeting dynamics. The linear reinforcement mechanism in RPP was replaced by an exponential decay reinforcement function $p_i(t_j) = \sum_{j=0}^{P_i(t_j)} \exp{(-\delta_i j)} (\delta_i > 0)$ to regularize the large popularity, and a time mapping function 
$f: t_j \rightarrow \hat{t}_j$ to mitigate the influence of diurnal rhythm of user activities following their earlier works~\cite{gao2014effective,gao2014popularity}. 

A dynamic activeness model based on Poisson process was introduced in Reference \cite{lin2013predicting}, aiming to predict the intensity (cascade size), coverage (number of users involved), and duration of a trend 
in DBLP co-authorship network. The RepostTree model~\cite{lu2014predicting} decomposes a sequence of Weibo retweets into tree-structures based on users' follower relationships and subsequently computes a composite Poisson distribution based on early observation of retweets with maximum likelihood estimation. Another extension of RPP model~\cite{gao2016modeling} decomposed the Weibo retweet cascade graph as $k$ sub-processes (in their case $k=3$), and kept the attractiveness $\alpha_{i,j}$ and log-normal distribution relaxation function $\phi_i(\tau)$ (cf.~Equation~\eqref{RPP}) unchanged. Then the rate function was defined as
\begin{equation}
    \lambda_i(t) = \sum_{j=1}^k \alpha_{i,j} \phi_i(t, t_j) P_i(t),
\end{equation}
where $t_j$ is the time when the $j$-th sub-process starts. Recently, learning the collective user behaviors in cascades by utilizing Poisson process with a latent user interest layer was proposed in Reference \cite{lu2018collective}. 

\subsection{Survival Analysis}\label{subsec:survival}

Survival analysis is a branch of statistics that is widely used in engineering, economics, and sociology~\cite{miller2011survival}. 
Borrowing the idea from survival analysis to predict the popularity of online contents in References \cite{lee2010approach,lee2012modeling}, a model based on Cox proportional hazard regression~\cite{cox1972regression} was implemented, consisting of two components: (i) a set of explanatory risk factors $\{x_j\}_j$; and (ii) a 
baseline 
Welbull distribution function $h_0(t) = \frac{\gamma}{\lambda}(\frac{t}{\lambda})^{\gamma-1}$. After fitting the parameters $\{\beta_j\}_j$ of Cox regression and parameters $\gamma$ and $\lambda$ of Weibull distribution, the hazard function is approximated by
\begin{equation}
    h_i(t) = h_{i0}(t)^{\sum_j \beta_{ij}x_{ij}}.
\end{equation}

In similar spirit, a dynamic egocentric model of citation networks based on a counting process was proposed in Reference \cite{vu2011dynamic}, with intensity function
$
    \lambda_i(t) = h_{i0}(t)\exp\left(\sum_j\beta_{ij}s_{ij}(t)\right)
$,
where $h_0(t)$ is the baseline hazard function, $\{s_{ij}(t)\}_{j}$ is a set of risk factors dependent on item $I_i$, and $\beta_{ij}$ are parameters need to estimate. Risk factors are calculated by citation graph and LDA. 

The Weibull distribution in survival analysis was also adopted in References \cite{yu2015micro,yu2017uncovering}, introducing a networked Weibull regression model (NEWER) to characterize the information diffusion in a Weibo global graph. By modeling the retweeting event that happened on a node as the survival process, the density function is defined as $\frac{\gamma_i}{\lambda_i}(\frac{t}{\lambda_i})^{\gamma_i-1}\exp \left( -(\frac{t}{\lambda_i})^{\gamma_i} \right)$, where $\lambda_i$ and $\gamma_i$ are parameter vectors linearly associated with the user's representation vector containing user and structural features extracted from user historical behaviors and global graph, respectively. 

There are a few other survival analysis based models, e.g., References \cite{subbian2017detecting,xie2015modelling}, 
with the fundamental difference between these models lying in the design of hazard functions and survival probabilities, which should be appropriate for the target application with empirical data. 

\subsection{Self-exciting Hawkes Point Processes and Epidemic Models}\label{subsec:hawkes-point-process}

Self-exciting point process based models predict the rate of events (e.g., retweets or citations) as a function of time and the previous history of events. One of the seminal works was reported in Reference \cite{crane2008robust}, describing the dynamics of viewing behavior of YouTube videos by two factors: (i) a response function representing a power-law distribution of human activity waiting time; and (ii) an epidemic process which can be modeled by self-exciting Hawkes conditional Poisson process. The conditional intensity was defined as
\begin{equation}
    \lambda_i(t) = V(t) + \sum_{t_j < t}\mu_j \phi(t-t_j),
\end{equation}
where $V(t)$ is the exogenous source, $\mu_j$ represents the number of potential participants that will be influenced by $u_j$ to join in this cascade $C_i$ at time $t$, and $\phi(t)\sim1/t^{1+\theta} (0<\theta<1)$ is a memory kernel. 
The idea was extended in Reference \cite{zadeh2014modeling},  utilizing self-exciting Hawkes process to characterize the tweet popularity in Twitter. 
The number of followers and a Pareto distribution of the kernel $\phi(\tau)$ were used to model the magnitude $\mu_j$ of each event. 

The SpikeM model~\cite{matsubara2012rise} fits the exponential rising and power-law falling patterns of information diffusion by incorporating (addressing) both the advantages (disadvantages) of epidemic model and self-exciting Hawkes processes. The base model of SpikeM can be represented as
\begin{equation}\label{eqn:spikem}
    P(t_{j+1}) = |\mathcal{V}_u| \cdot \sum_{t=t_b}^{t_{j+1}} \mu(t) \phi(t_{j+1}-t) + \epsilon,
\end{equation}
where $|\mathcal{V}_u|$ is the number of unaffected users, $\mu(t) = P(t_j)+S \cdot \mathbbm{1}(t=t_b)$ is the available stimuli at time $t$ ($\mathbbm{1}(\cdot)$ is the indicator function), and $S$ is the influence of external shock. 
SpikeM takes periodicity of human behavior into consideration by multiplying Equation \eqref{eqn:spikem} with an additional time-dependent periodicity factor $p(t_{j+1})$. 
Extensions can be found in recent works~\cite{kong2020modeling,rizoiu2018sir}, which share the same idea of linking the epidemic model and Hawkes process. 

Dual sentimental Hawkes process (DSHP)~\cite{ding2015video} is another generative prediction model built on Hawkes process. DSHP considers sentimental impact of information items, and two self-exciting processes, i.e., self-excitation and cross-excitation. The intensity function of DSHP is
\begin{equation}
    \lambda_i(t) = V(t) + \underset{\text{self-excitation}}{\sum_{t_{j, k}<t} \mu_{ik}\phi_k(t-t_j)} + \underset{\text{cross-excitation}}{\sum_{t_{j, l}<t} \mu_{il}\phi_l(t-t_j)},
\end{equation}
where the exogenous source is formulated as a Pareto distribution $V(t) = \alpha / t^{\alpha +1 }$, and kernels ($\phi_k$ and $\phi_l$) are formulated as two Rayleigh distributions, i.e., $\phi(\tau) = \frac{\tau}{\sigma^2}\exp\left(-\frac{\tau^2}{2\sigma^2}\right)$. 

The SEISMIC (self-exciting model of information cascades) in Reference \cite{zhao2015seismic} aims to predict the retweet counts. It first fits a memory kernel $\phi(\tau)$ of human reaction time -- a constant early on, then following a power-law decay later -- and then measures the retweetability of tweet $I_i$ at time $t$ with a time dependent tweet infectiousness $p_i(t)$. The intensity is described as
\begin{equation}
    \lambda_i(t) = p_i(t) \cdot \sum_{t_j<t} \mu_j\phi(t-t_j),
\end{equation}
where $\mu_j$ is the number of followers of user $u_j$ who retweet the original tweet $I_i$. 
Self-excited Hawkes process (SEHP) proposed in Reference \cite{bao2015modeling} is another example, which defined the intensity function as: 
$
    \lambda_i(t) = V_i\phi(t) + \mu\sum_{t_j<t} \phi(t-t_j)
$
, where $\phi(\tau)$ is an exponential decay memory kernel. 

Time-dependent Hawkes process (TiDeH)~\cite{kobayashi2016tideh} is yet another extension of SEISMIC taking into account the circadian oscillations and aging of information.
Later, leveraging a combination of advantages of feature-based models and generative models, Reference \cite{mishra2016feature} proposed a hybrid model consisting of Hawkes process with a predictive layer trained from random forest and a feature-based predictor. A self-exciting Hawkes process model to predict the individual paper citation count was introduced in Reference \cite{xiao2016modeling}, with intensity  
defined as
\begin{equation}
    \lambda_i(t) = V_i\phi_{ik}(t) + \mu_i\sum_{t_j<t} \phi_{il}(t-t_j) 
\end{equation}
where $V_i$ captures the intrinsic popularity (or quality) of each paper, which is defined by paper/author-specific covariates, and where $\phi_{ik}(\tau)$ and $\phi_{il}(\tau)$ are exponential aging kernels. 

A combination of Hawkes intensity processes (HIP) with \textit{exogenous} stimuli and \textit{endogenous} triggering effect from Twitter and YouTube was proposed in Reference \cite{rizoiu2017expecting}, to predict the popularity of videos. In particular, the self-exciting Hawkes process was extended by defining the expectation of intensity of the observed events
\begin{equation}
    \xi_i(t) = \mathbb{E}[\lambda_i(t)] = V(t) + \mu_i \int_0^t \xi(t-\tau)\phi_i(\tau) \text{d}\tau,
\end{equation}
where $V(t)$ accounts for the unobserved external influence from Twitter and the number of video shares, and $\phi_i(\tau)$ is a power-law memory kernel. HIP models the video view volumes directly rather than taking individual events one after another. 

\subsection{Other Generative Models and Discussion (Pros and Cons)}\label{subsec:other-generative}

The spatial and temporal heterogeneous Bass model (STH-Bass) proposed by Reference \cite{yan2016sth} borrowed the idea of the Bass model~\cite{Bass1969} that describes the process of how new products get adopted in a population to predict popularity of tweets. Compared to the original Bass model, there is an additional consideration of spatial-temporal heterogeneity, which is more applicable for tweet popularity prediction rather than predicting the sales of product.  
Subsequently, Reference \cite{chu2018cease} used a topic-oriented feature combined Bass model with association analysis to make the popularity prediction. 

The Bayesian approach to predict the popularity of tweets in Reference \cite{zaman2014bayesian} takes features into consideration, including time-series and retweet cascade graphs. In Reference \cite{lymperopoulos2016predicting}, the popularity growth was approached as a sequence of linear and non-linear phases and the proposed model does not rely on microscopic information, while the prediction can be made efficiently without training. 
Other related works include: Reference \cite{samanta2017lmpp}, proposing LMPP model to predict the hashtag popularity by modeling the hashtag-tweet reinforcement and inter-hashtag competitions; CHESS~\cite{tang2017popularity}, predicting the popularity (watch time) of Facebook videos through an efficient and scalable Hawkes process; Reference \cite{bao2017uncovering}, using Hawkes process with survival theory to predict the popularity of tweets and papers; and Reference \cite{wang2017linking}, which devised MIC2MAC to link microscopic event data to macroscopic popularity inference, where Hawkes process and jump stochastic differential equation are used for prediction. 

Generative models generally do not need heavy feature engineering and are inherently interpretable. They mainly rely on time-series data, and their predictions can be made in real-time, once the model has been prepared and the parameters have been estimated~\cite{zhao2015seismic}. However, their performance has been questioned~\cite{cao2017deephawkes,guo2016comparison}, and they are often easily influenced by outliers \cite{mishra2016feature}. In addition, generative models usually make strong assumptions on fixed parameters, which limits their generality and model expressiveness~\cite{du2016recurrent}. Moreover, the complex underlying mechanisms governing the success of cascades are underestimated and simplified, in order to simulate/reproduce the cascade diffusion processes. Finally, most of generative models are network-agnostic -- i.e., they fail to model the important structural information that could help understand the process/paths of information diffusion. Thus, despite their efficiency and interpretability, generative models are less powerful in making precise predictions~\cite{cao2017deephawkes}. 

\section{Deep Learning Models}\label{sec:deep}The renaissance of neural networks in recent years has spurred a number of deep learning based prediction models. Deep neural networks are often shown to be more effective than linear models~\cite{tang2017popularity}. For example, models based on RNNs do not rely on explicit assumptions of the diffusion of cascades and are more flexible to capture the temporal dependencies in information cascades \cite{wang2017cascade}. Models based on graph representation learning do not require laborious hand-crafted features from the underlying graphs of cascades, e.g., the specific designs of node influence and community detection.

Existing deep learning models can be categorized as three types: (i) models based on content of the users/items, e.g., texts, images, and videos, and these models usually adopt techniques from computer vision and NLP to learn expressive representations of users/items; (ii) models based on temporal sequences, e.g., the cascading behavior in social and academic networks, and these models rely on RNN or pooling mechanisms; (iii) models based on graphs, e.g., the cascade graphs and global graph, and these models often deal with graph neural networks or graph representation learning, aiming to learn effective embeddings of nodes/edges/graphs for graph-structured data.
Other techniques, such as attention mechanism, variational inference, reinforcement learning, etc., are often employed in deep learning models. In many cases, multi-modalities \cite{sanjo2017recipe}, multi-scale \cite{yang2019multi}, and multi-task learning \cite{chen2019information} are considered to improve the prediction ability. 

One of the early approaches using representation learning and deep learning techniques is~\cite{khosla2014makes}, focusing on predicting the number of views of Flickr images. On one hand, several hand-crafted low- and high-level social and visual features were designed (e.g., past success, number of contacts, colors, and textures). On the other hand, convolutional neural networks were utilized to learn the representations of images with the last hidden layer of the model. 

DeepCas~\cite{li2017deepcas} is the first graph representation learning based method for modeling and predicting the popularity of information cascades. It borrows the idea of DeepWalk~\cite{perozzi2014deepwalk} to sample the cascade graphs with random walks. The sampled node sequences are then fed into a bidirectional gated recurrent units (Bi-GRU)~\cite{Chung2014}, along with attention mechanism~\cite{Bahdanau2015}, to obtain the node embeddings. The prediction was made in an end-to-end manner. Subsequently, DeepCas was extended to DCGT~\cite{li2018joint} by incorporating the contents associated with the nodes in the cascade. 

\begin{table}\small
    \caption{Deep Learning Models} 
    \label{tab:deep}
    \begin{tabular}{@{} lrll @{}}
    \toprule
        \textbf{Reference} & \textbf{Venue} & \textbf{Model}  & \textbf{Building blocks} \\ \midrule 
        Li et al.~\cite{li2017deepcas}                      & WWW '17    & DeepCas    & Random Walks, GRU, Attention                                     \\ 
        Sanjo et al.~\cite{sanjo2017recipe}                 & CIKM '17   & -          & AlexNet, word2vec, Deep Averaging Network \\
        Chen et al.~\cite{chen2017attention}                & ISI '17    & ANPP       & node2vec, GloVe, GRU, Attention                                     \\ 
        Wu et al.~\cite{wu2017sequential}                   & IJCAI '17  & DTCN       & ResNet, LSTM, Attention                                                 \\ 
        Cao et al.~\cite{cao2017deephawkes}                 & CIKM '17   & DeepHawkes & GRU, Pooling, Non-Parametric Time Kernel                                                    \\ 
        Zhang et al.~\cite{zhang2018user}                   & WWW '18    & UHAN       & VGGNet, LSTM, Hierarchical Attention \\ 
        Wang et al.~\cite{wang2018factorization}            & DASFAA '18 & MOOD       & Memory Network, Attention, Tensor Factorization \\ 
        Wang et al.~\cite{wang2018learning}                 & IJCAI '18  & UMAN       & Convolution, LSTM, Attention, User Memory \\ 
        Guo et al.~\cite{guo2018learning}                   & KAIS '18   & LSTMIC     & LSTM, Pooling Mechanisms \\
        Kefato et al.~\cite{kefato2018cas2vec} & SNAMS '18 & CAS2VEC & CNN, Max Pooling, Network-Agnostic \\
        Bielksi et al.~\cite{bielski2018understanding}      & ACCESS '18  & -          & ResNet, Self-Attention, Pooling, GloVe, LSTM \\ 
        Liao et al.~\cite{liao2019popularity}               & AAAI '19   & DFTC       & LSTM, 1-D CNN, Attention, HAN \\ 
        Chen et al.~\cite{chen2019cascn}                    & ICDE '19   & CasCN      & GCN, LSTM, Time Decay, Sum Pooling \\ 
        Zhao et al.~\cite{zhao2019neural}                   & PAKDD '19  & KB-PPN     & Knowledge Base, LSTM, Gate/Attention Mechanisms \\ 
        Yang et al.~\cite{yang2019neural} & TKDE '19 & NDM & User Embedding, Attention, Convolution \\
        Ding et al.~\cite{ding2019social} & MM '19 & - & ResNet, NIMA, I$^2$PA, BERT, User/Item Features \\
        Chen et al.~\cite{chen2019information}              & SIGIR '19  & DMT-LIC    & Multi-Task Learning, GCN, LSTM, Shared Gate \\ 
        Yang et al.~\cite{yang2019multi}                    & IJCAI '19  & FOREST     & Reinforcement Learning, GRU, DeepWalk \\ 
        Chen et al.~\cite{chen2019npp} & Neurocom. '19 & NPP & GRU, Text/User/Time-Series Encoders, Attention \\
        Cao et al.~\cite{cao2020popularity}                 & WSDM '20   & Coupled-GNN & GNNs, Influence/State Gate Mechanisms, DeepWalk \\ 
        Xie et al.~\cite{xie2020multimodal} & WWW '20 & MMVED & Multimodal, Variational Inference, LSTM \\
        Zhou et al.~\cite{zhou2020variational}              & INFOCOM '20& VaCas      & Spectral Graph Wavelets, Hierarchical VAEs, GRU \\
    \bottomrule
    \end{tabular}
\end{table}

DeepHawkes~\cite{cao2017deephawkes} attempts to unify the advantages of generative models and deep learning techniques.
In particular, three crucial concepts of Hawkes process -- user influence, self-exciting mechanism, and time decay effect -- are analogously transformed into the DeepHawkes model. 
Similar to DeepCas, user representations are learned in an end-to-end way. However, rather than directly modeling the structural patterns of cascade graphs, given a piece of information and its diffusion trajectories, DeepHawkes uses GRU, sum pooling, and non-parametric time kernel to aggregate the contributions of early adopters.  

ANPP~\cite{chen2017attention} utilizes GloVe~\cite{pennington2014glove} to embed the words of tweets, and leverages node2vec~\cite{grover2016node2vec} to encode user graph. GRU with attention mechanism is used to aggregate the learned embeddings, as well as the time series feature vectors.
Deep temporal context networks (DTCN)~\cite{wu2017sequential}  predict the popularity of Flickr images by jointly learning the user/photo embeddings, temporal context of resharing sequences, and multiple time-scale attention mechanism. Specifically, ResNet~\cite{he2016deep} and LSTM~\cite{Hochreiter1997} are utilized to model the visual dependencies and temporal dependencies, respectively. 

User-guided hierarchical attention network (UHAN)~\cite{zhang2018user} considered both visual and textual modalities of Flickr images by characterizing three different representations, i.e., visual representation pre-trained from VGGNet~\cite{simonyan2015very}, textual representation encoded by LSTM, and user representation learned under supervision. Then user-guided intra- and inter-attention mechanisms were used to jointly learn the importance over the aforementioned two modalities. 

The deep fusion of temporal processes and content features (DFTC) model~\cite{liao2019popularity} predicts the popularity of WeChat online articles. It first uses an LSTM to model the long-term growth trends of a cascade, and then utilizes 1-D CNN to capture the short-term fluctuations. Article content is modeled by a hierarchical attention network~\cite{yang2016hierarchical}. The final prediction is made by an attentive fusion of previously learned representations. 

Recurrent cascades convolutional network (CasCN)~\cite{chen2019cascn} samples a cascade graph as a series of sequential subcascades and adopts a dynamic multi-directional GCN~\cite{Kipf2017} to learn structural information of cascades.
The subsequent work~\cite{chen2019information} addresses the  activation prediction and popularity prediction problems jointly within a deep multi-task learning framework. To that end, attention and gated mechanisms were utilized, combined with a shared-representation layer, to capture the spatial-temporal dynamics of cascades.
Coupled-GNNs~\cite{cao2020popularity} leveraged two specifically designed GNNs -- one for node states and the other one for influence spread -- to capture the cascading effect. 

More recently, variational cascade graph learning neural networks (VaCas)~\cite{zhou2020variational} integrated graph wavelets, hierarchical variational autoencoders, and Bi-GRUs to learn the structures of cascade graphs. 
Both node- and cascade-level diffusion uncertainties, as well as the contextualized user behaviors, were modeled in an end-to-end deep learning framework. 

There are some works focusing on different aspects of performance improvement. For example,
KB-PPN~\cite{zhao2019neural} leverages the embeddings of knowledge base entities and their neighbors to enhance the popularity prediction based on a LSTM network. 
FOREST~\cite{yang2019multi} first learns an RNN-based microscopic cascade prediction model, and then performs a series of simulations guided by reinforcement learning to predict the macroscopic popularity of observed cascades.

\subsection{Discussion: Strengths and Limitations of Deep Learning Models}

Table~\ref{tab:deep} summarizes the deep learning based cascade models and their main building blocks. 
Generally, these works learn different aspects of information item/cascade with various deep learning techniques, e.g., capturing the long-term dependencies of time series and temporal characteristics of participation/citations with an RNN or its variants; learning representations of texts and images via deep language and visual models; dealing with graph-structured data by unsupervised network embedding \cite{Cai2018} or (semi-supervised) graph neural networks \cite{Zhou2018}. 
Compared to feature-based models, which depend on hand-crafted features (platform-specific and rely on prior knowledge), or generative models, which assume hard-coded diffusion protocols (lack of flexibility and also rely on human designs), deep learning models do not require heavy feature engineering and can capture non-linear representations of both user/item content and popularity accumulation trend. 
The success of deep learning in other domains seems to be continued in information cascade modeling, more and more approaches adopted techniques from deep learning and achieved state-of-the-art results. 
The main strength of deep learning models lies in their relatively simple architecture (deep stacked layers with little human designs) and powerful learning ability via backpropagation under supervision (with massive data and computational resources) \cite{lecun2015deep}. 

However, despite their improvements on the prediction performance, deep learning models still face many limitations. 
One main drawback of deep learning models is the lack of model interpretability owing to the ``black-box'' nature of neural networks. 
The computational cost of deep learning models is significantly larger than feature-based and generative models. The model tuning process, hyper-parameter selection, risk of overfitting, etc., sometimes cause a lot of efforts for engineers to obtain the satisfactory performance.

\section{Open Challenges and Opportunities}\label{sec:con}We now discuss some challenges and aspects that have not been sufficiently studied in cascade modeling and information diffusion prediction. 

\subsection{Predictability and Interpretability}
\label{subsec:pred-inte}

The prediction of popularity from information cascades -- no matter how the problem is reformulated, which datasets are used for evaluation, and even what metrics are selected as evaluation protocols -- is largely subjective to researchers' design (a.k.a. researcher degrees of freedom \cite{hofman2017prediction}). Despite the many publications and significant progress in this field, some fundamental questions have yet to be answered. For example, can we really predict the popularity of cascades? If the answer is yes, then to what extent can a particular cascade be predicted? Or, what is the glass-ceiling performance a model can achieve? To effortlessly adapt to new emerging information (e.g., online fake news and rumors) and new information propagation platforms, we may want to exploit existing knowledge and methods. Then, what conclusions drawn from previous works are applicable to another dataset/scenario? Can we build general cascade datasets that can be used for learning basic knowledge/features regarding various types of information items, such as the ImageNet and Wikipedia in computer vision and natural language processing? Answering these questions is a non-trivial task and requires more research as well as more standardized open datasets -- the latter, however, need to obey the privacy issues of many online social networks.

Reviewing different groups of prediction models helped identify the advantages and limitations of existing methods for predicting popularity -- e.g., design and selection of features vary greatly for different cascade formulations and information items, while their effectivenesses and universalities are not generalizable from one to another. Furthermore, the assumptions of diffusion mechanisms made by generative models are subject to specific scenarios, while the prediction of ``black box'' end-to-end deep learning models cannot be properly explained. To better understand the predictions of existing models, a standardized comparison environment should be emphasized, e.g., selecting appropriate evaluation protocols for classification/regression, using publicly available datasets, etc., in order to avoid overestimating (or underestimating) the performance of a particular model.

Other than purely pursuing prediction performance, or seeing at it in a new light, interpreting the behavior of predictions may enable better understanding of the disciplines governing the success of information items, maximizing social and business benefits through mastering and even managing the information diffusion process. Thus, it is desirable to unveil the inner-context driven mechanisms that guide information to spread further -- e.g.,: 
(i) investigating which topology of the diffusion network expands faster~\cite{cheng2014can}; (ii) the temporal and textual characteristics of burst public events and fake news and why they are traveled fast~\cite{naveed2011bad,vosoughi2018spread}; (iii) the visual explanations of why some images/videos become popular quickly and why others do not~\cite{khosla2014makes}; (iv) interpreting 
deep neural networks \cite{camburu2020explaining}; etc.
Successfully explaining these phenomena will improve the prediction, and help design more powerful decision making systems to detect and bound viral marketing/advertising, rumor/fake news spreading, public opinion control, epidemic prevention, and so on. 

\subsection{External Stimuli}
\label{subsec:external-stimuli}

External stimuli are one of the primary causes that lead the prediction of information cascades to be unexplainable, unreliable, and finally maybe impossible. Compared to endogenous stimuli like in-site search and share~\cite{figueiredo2011tube}, external (or endogenous) stimuli are uncertain and unforeseeable events. For example, a normal cascade with little fluctuations in its early stage may suddenly become popular due to some external stimuli which significantly increases the exposure/adoption rate of that information. As a previous work~\cite{myers2012information} 
shows, about 71\% of the information in Twitter is diffused in the network without external influence, while the remaining 29\% of tweets have been significantly affected and even manipulated by exogenous force. 

Retrieval of knowledge and propagation patterns from one platform to help the prediction of items in another platform with cross-domain real-time transfer learning (e.g., learn tweet spreading to enhance the performance of predicting YouTube video views) was studied in Reference \cite{roy2013towards}.
Another example is SoVP~\cite{li2013popularity}, which learned the shared diffusion behavior in social networks to strengthen the ability to predict video popularity. The rationale behind these works is that the trends appearing on Twitter would later drive users to search and watch related videos in YouTube. Moreover, burst events are more likely to first appear on microblogging platforms like Twitter/Weibo and then spread to other kinds of platforms such as newspapers and video sharing sites. This inspired later works to predict popularity in one domain via resorting to other information sources and dissemination platforms to model external stimuli responsible for the popularity~\cite{abisheva2014watches,rizoiu2017expecting,yu2014twitter}. 
More such examples can be found in Reference \cite{oghina2012predicting,vallet2015characterizing}, which retrieve information from Twitter and YouTube to predict the ``ratings'' and ``views'' of movies in IMDB. In Reference \cite{castillo2014characterizing}, information from Facebook and Twitter were collected to predict the views of articles. Another recent work~\cite{mishra2018modeling} uses RNNs to model asynchronous media streams of YouTube videos. 

Modeling external stimuli can significantly enrich data diversity and improve model robustness. Actually, many web sites now provide extra information that attracts external stimuli, e.g., the information of referrers/URLs that may bring new views/adoptions. For YouTube videos, a referrer can be a hyperlink embedded in a tweet pointing to a video. This provides clues to trace the sources of external stimuli. The top-$10$ referrers and the number of views each referrer brings as features to predict the evolving trends of videos were used in References \cite{figueiredo2011tube,figueiredo2013prediction}. This anatomy of sources of popularity gives more insights into the future evolution of cascades~\cite{broxton2013catching}, e.g., how many external sources are contributing to the popularity of cascades, which continuously existed and have already been diminished or even vanished.

\subsection{Cooperation and Competition}
\label{subsec:cooperation-and-competition}

Information cascades are not isolated but affect each other and are often interwoven via interactions, cooperation and competition, when spreading among users. Cooperation means that the diffusion of an information item is accelerated by other item(s), e.g., if one thing was reported by many sources many times, then this information is more credible and has a higher probability of adoption~\cite{myers2012clash}. In contrast, due to the limited attention span of users~\cite{weng2012competition}, information items (especially those with very similar content) could restrict each other's diffusion~\cite{coscia2014average}.
Thus, if cooperation happens, the expectation of an information item's future popularity should be larger than normal cases. Also in the case of competition, items with similar contents would inevitably face the problem of low popularity or at least reduced popularity as their competitors increase. 

Existing works on this topic mainly focus on discovering cooperation/competition during diffusion, but rarely on analyzing reciprocity and its effect on popularity prediction. For example, interaction matrix between items to facilitate the prediction of tweets/hashtags was used in Reference \cite{kong2016popularity}. Hawkes processes were adopted to model cooperation and competition between cascades in social networks in References \cite{zarezade2017correlated,yoo2019competition}. Modeling the interactions between information items/cascades, especially the cooperation and competition among them, can benefit popularity prediction and deepen our understanding of information diffusion in various social scenarios: e.g., figuring out when and how cooperation/competition happens and identifying the best timing and content of items to enable distributors to avoid competitions and meet more cooperation.

\section{Conclusion}\label{subsec:con}

We provided a broad overview of the field of information diffusion modeling and popularity prediction and proposed a taxonomy to categorize related literature, highlighting a number of influential papers/approaches in each category. We also discussed the advantages and disadvantages of different types of methods over time and outlined some challenges and open research problems. 
There exists a vast literature and growing research on this topic, and we systematically covered, in certain detail, selected representative works, with a hope that this knowledge can be used as a stepping-stone for future work and/or evaluations in this field. 

\begin{acks}
We thank all doctors, nurses, and medical personnel who are on the front line of the war against COVID-19. 
\end{acks}

\appendix

\section{Supplementary Material}\label{sec:app}

Here we list several tables corresponding to: Feature-based models (Table \ref{tab:feature-based-models} and \ref{tab:feature-based-models-continued}), temporal, structural, and user/item feature (Table \ref{tab:temporal-features}, \ref{tab:structural-features}, and \ref{tab:user-features}), for the ease of reference.

\bibliographystyle{ACM-Reference-Format}
\bibliography{xovee}

\begin{table}\small
    \caption{Feature-based Models. Abbreviations: \textbf{Te}. -- Temporal, \textbf{CS} -- Cascade structure, \textbf{GG} -- Global graph, \textbf{Us.} -- User, \textbf{Co.} -- Content.}
    \label{tab:feature-based-models}
    \begin{tabular}{lrccccccc}
    \toprule 
        \multirow{2}{*}[-3pt]{\textbf{Reference}}  & \multicolumn{1}{c}{\multirow{2}{*}[-3pt]{\textbf{Venue}}} & \multirow{2}{*}[-3pt]{\textbf{Strategy}}  & \multirow{2}{*}[-3pt]{\textbf{Formulation}}   & \multicolumn{5}{c}{\textbf{Feature}} \\ \cmidrule(lr){5-9} 
        &       &                    &                   & \textbf{Te.} & \textbf{CS} & \textbf{GG} & \textbf{Us.} & \textbf{Co.}       \\ \midrule
        Khabiri et al.~\cite{khabiri2009analyzing}                  & ICWSM '09  & \textit{Ex-ante}  & Classification    &     &    &    & \C  & \C         \\
        Tsagkias et al.~\cite{tsagkias2009predicting}               & CIKM '09   & \textit{Ex-ante}  & Classification    & \C  &    &    & \C  & \C         \\
        Jamali et al.~\cite{jamali2009digging}                      & WISM '09   & Peeking           & Both              & \C  & \C & \C & \C  & \C         \\
        Lerman \& Hogg~\cite{lerman2010using}                       & WWW '10    & Peeking           & Both              & \C  &    & \C & \C  &            \\
        Tsagkias et al.~\cite{tsagkias2010news}                     & ECIR '10   & Peeking           & Regression        & \C  &    &    &     &            \\
        Yano \& Smith~\cite{yano2010s}                              & ICWSM '10  & \textit{Ex-ante}  & Classification    &     &    &    &     & \C         \\
        Szabo \& Huberman~\cite{szabo2010predicting}                & Comm. ACM '10 & Peeking       & Regression        & \C  &    &    &     &            \\ 
        Hong et al.~\cite{hong2011predicting}                       & WWW '11    & Peeking           & Classification    & \C  & \C & \C & \C  & \C         \\
        Bakshy et al.~\cite{bakshy2011everyone}                     & WSDM '11   & \textit{Ex-ante}  & Regression        &     &    &    & \C  & \C         \\
        Kim et al.~\cite{kim2011predicting}                         & CIT '11    & Peeking           & Classification    & \C  &    &    &     &            \\
        Gürsun et al.~\cite{gursun2011describing}                   & INFOCOM '11& Peeking           & Regression        & \C  &    &    &     &            \\
        Tatar et al.~\cite{tatar2011predicting}                     & WIMS '11   & Peeking           & Regression        & \C  &    &    &     &            \\
        Rowe et al.~\cite{rowe2011predicting}                       & ESWC '11   & \textit{Ex-ante}  & Regression        &     &    &    & \C  & \C         \\
        Naveed et al.~\cite{naveed2011bad}                          & WebSci '11 & \textit{Ex-ante}  & Classification    &     &    &    &     & \C         \\
        Petrovic et al.~\cite{petrovic2011rt}                       & ICWSM '11  & \textit{Ex-ante}  & Classification    & \C  &    &    & \C  & \C         \\
        Shamma et al.~\cite{shamma2011viral}                        & ICWSM '11  & \textit{Ex-ante}  & Classification    &     &    &    & \C  &            \\
        Lakkaraju \& Ajmera~\cite{lakkaraju2011attention}           & CIKM '11   & \textit{Ex-ante}  & Both              &     &    &    & \C  & \C         \\
        Yan et al.~\cite{yan2011citation}                           & CIKM '11   & \textit{Ex-ante}  & Regression        &     &    & \C & \C  & \C         \\
        Stieglitz et al.~\cite{stieglitz2012political}              & HICSS '12  & \textit{Ex-ante}  & Regression        &     &    &    & \C  & \C         \\
        Tsur \& Rappoport~\cite{tsur2012s}                          & WSDM '12   & Both              & Regression        & \C  &    &    & \C  & \C         \\
        Oghina et al.~\cite{oghina2012predicting}                   & ECIR '12   & Peeking           & Regression        & \C  &    &    &     & \C         \\
        Artzi et al.~\cite{artzi2012predicting}                     & HLT-NAACL '12& \textit{Ex-ante}& Classification    & \C  &    &    & \C  & \C         \\
        Bandari et al.~\cite{bandari2012pulse}                      & ICWSM '12  & \textit{Ex-ante}  & Both              &     &    &    & \C  & \C         \\
        Ruan et al.~\cite{ruan2012prediction}                       & WebSci '12 & Peeking           & Regression        & \C  &    & \C & \C  & \C         \\
        Ma et al.~\cite{ma2012will}                                 & SIGIR '12  & Peeking           & Classification    &     & \C & \C &     & \C         \\
        Gupta et al.~\cite{gupta2012predicting}                     & ASIST '12  & Peeking           & Both              & \C  &    &    & \C  & \C         \\
        Kupavskii et al.~\cite{kupavskii2012prediction}             & CIKM '12   & Both              & Both              & \C  & \C & \C & \C  & \C         \\
        Figueiredo~\cite{figueiredo2013prediction}                  & WSDM '13   & Peeking           & Classification    & \C  &    &    &     & \C         \\
        Ahmed et al.~\cite{ahmed2013peek}                           & WSDM '13   & Peeking           & Regression        & \C  &    &    &     &            \\
        Pinto et al.~\cite{pinto2013using}                          & WSDM '13   & Peeking           & Regression        & \C  &    &    &     &            \\
        Ma et al.~\cite{ma2013predicting}                           & JASIST '13 & Peeking           & Classification    &     & \C & \C &     & \C         \\
        Bao et al.~\cite{bao2013popularity}                         & WWW '13    & Peeking           & Regression        &     & \C & \C &     &            \\
        Jenders et al.~\cite{jenders2013analyzing}                  & WWW '13    & \textit{Ex-ante}  & Classification    &     &    &    & \C  & \C         \\
        Kupavskii et al.~\cite{kupavskii2013predicting}             & ICWSM '13  & Both              & Regression        & \C  & \C & \C & \C  & \C         \\
        Romero et al.~\cite{romero2013interplay}                    & ICWSM '13  & Peeking           & Classification    &     & \C & \C &     &            \\
        Lakkaraju et al.~\cite{lakkaraju2013s}                      & ICWSM '13  & \textit{Ex-ante}  & Regression        &     &    &    &     & \C         \\
        Cui et al.~\cite{cui2013cascading}                          & KDD '13    & Peeking           & Classification    &     &    & \C & \C  &            \\
        Weng et al.~\cite{weng2013virality}                         & Sci. Rep. '13& Peeking         & Classification    &     & \C & \C &     &            \\
        Li et al.~\cite{li2013popularity}                           & CIKM '13   & Peeking           & Regression        & \C  & \C & \C &     &            \\
        Tatar et al.~\cite{tatar2014popularity}                     & SNAM '14   & Peeking           & Regression        & \C  &    &    &     &            \\
        Castillo et al.~\cite{castillo2014characterizing}           & CSCW '14   & Peeking           & Regression        & \C  &    &    & \C  & \C         \\
        Abisheva et al.~\cite{abisheva2014watches}                  & WSDM '14   & Peeking           & Both              &     &    & \C & \C  &            \\
        McParlane et al.~\cite{mcparlane2014nobody}                 & ICMR '14   & \textit{Ex-ante}  & Classification    &     &    &    & \C  & \C         \\
        Totti et al.~\cite{totti2014impact}                         & WebSci '14 & \textit{Ex-ante}  & Classification    &     &    &    & \C  & \C         \\
    \bottomrule
    \multicolumn{9}{r}{\textit{Continued in next page.}} \\
    \end{tabular}
\end{table}

\begin{table}\small
    \caption{Continued: Feature-based Models}
    \label{tab:feature-based-models-continued}
    \begin{tabular}{lrccccccc}
    \toprule
        \multirow{2}{*}[-3pt]{\textbf{Reference}}  & \multicolumn{1}{c}{\multirow{2}{*}[-3pt]{\textbf{Venue}}} & \multirow{2}{*}[-3pt]{\textbf{Strategy}}  & \multirow{2}{*}[-3pt]{\textbf{Formulation}}   & \multicolumn{5}{c}{\textbf{Feature}} \\ \cmidrule(lr){5-9}
        &       &                   &                   & \textbf{Te.} & \textbf{CS} & \textbf{GG} & \textbf{Us.} & \textbf{Co.}       \\ \midrule
        Gao et al.~\cite{gao2014effective}                          & WWW '14 & Peeking          & Both              & \C  & \C & \C &     &             \\ 
        Cheng et al.~\cite{cheng2014can}                            & WWW '14 & Peeking          & Classification    & \C  & \C & \C & \C  & \C          \\
        Khosla et al.~\cite{khosla2014makes}                        & WWW '14 & \textit{Ex-ante} & Regression        &     &    &    & \C  & \C          \\
        Weng et al.~\cite{weng2014predicting}                       & ICWSM '14 & Peeking & Classification    & \C  & \C & \C &     &             \\
        Bian et al.~\cite{bian2014predicting}                       & SIGIR '14 & Peeking & Classification    &     &    & \C & \C  & \C          \\
        Kong et al.~\cite{kong2014predictingbursts}                 & SIGIR '14 & Peeking & Both              & \C  & \C &    & \C  & \C          \\
        He et al.~\cite{he2014predicting}                           & SIGIR '14 & Peeking & Regression        & \C  &    & \C & \C  &             \\
        Gao et al.~\cite{gao2014popularity}                         & APWeb '14 & Peeking & Classification    & \C  & \C & \C &     &             \\
        Kong et al.~\cite{kong2014predicting}                       & JISIC '14 & Peeking & Classification    & \C  & \C &    &     &             \\
        Yu et al.~\cite{yu2014twitter}                              & ACM MM '14 & Peeking & Classification    & \C  &    & \C & \C  &            \\
        Wang et al.~\cite{wang2015burst}                            & AAAI '15 & Peeking & Classification    & \C  & \C & \C & \C  &              \\
        Liu et al.~\cite{liu2015effectively}                        & AAAI '15 & Peeking & Classification    & \C  &    &    &     &              \\
        Dong et al.~\cite{dong2015will}                             & WSDM '15 & \textit{Ex-ante} & Classification    &     &    & \C & \C  & \C  \\
        Kong et al.~\cite{kong2015towards}                          & JASIST '15 & Peeking & Both              & \C  & \C &    & \C  & \C         \\
        Yu et al.~\cite{yu2015lifecyle}                             & ICWSM '15 & Peeking & Regression        & \C  &    &    &     &             \\
        Alzahrani et al.~\cite{alzahrani2015network}                & SBP '15 & Peeking & Classification    &     & \C & \C &     &               \\
        Guo et al.~\cite{guo2015toward}                             & ASONAM '15 & Peeking & Classification    & \C  & \C & \C &     &            \\
        Bora et al.~\cite{bora2015role}                             & SNAM '15 & Peeking & Classification    &     & \C & \C & \C  &              \\
        Vallet et al.~\cite{vallet2015characterizing}               & CIKM '15 & Peeking & Classification    & \C  &    &    & \C  &              \\
        Gelli et al.~\cite{gelli2015image}                          & ACM MM '15 & \textit{Ex-ante} & Regression        &     &    &    & \C  & \C          \\
        Yi et al.~\cite{yi2016mining}                               & Physica A '16 & Peeking & Classification    & \C  & \C &    &     & \C          \\ 
        Dong et al.~\cite{dong2016can}                              & TBD '16 & Both              & Both              & \C  &    & \C & \C  & \C          \\
        Wu et al.~\cite{wu2016unfolding}                            & AAAI '16 & \textit{Ex-ante} & Regression        & \C  &    &    & \C  & \C          \\
        Martin et al.~\cite{martin2016exploring}                    & WWW '16 & \textit{Ex-ante} & Regression        &     &    &    & \C  & \C          \\
        Rizos et al.~\cite{rizos2016predicting}                     & WWW '16 & Peeking & Regression        & \C  & \C & \C &     &             \\ 
        Shulman et al.~\cite{shulman2016predictability}             & ICWSM '16 & Peeking & Classification    & \C  & \C & \C & \C  &             \\
        Krishnan et al.~\cite{krishnan2016seeing}                   & WebSci '16 & Peeking & Classification    & \C  & \C & \C & \C  &             \\
        Kong et al.~\cite{kong2016popularity}                       & CCIS '16 & Peeking & Classification    & \C  & \C & \C &     &             \\ 
        Guo et al.~\cite{guo2016comparison}                         & ASONAM '16 & Peeking & Both              & \C  & \C & \C &     &             \\
        Zhang et al.~\cite{zhang2016structure}                      & ECML PKDD '16 & Peeking & Classification    & \C  & \C & \C & \C  & \C          \\
        Chen et al.~\cite{chen2016micro}                            & ACM MM '16 & \textit{Ex-ante} & Regression        &     &    &    & \C  & \C          \\
        Mishra et al.~\cite{mishra2016feature}                      & CIKM '16 & Peeking & Both              & \C  &    &    & \C  &             \\ 
        Trzci{\'n}ski \& Rokita~\cite{trzcinski2017predicting}      & TMM '17 & Peeking & Regression        & \C  &    &    &     & \C          \\ 
        Hoang et al.~\cite{hoang2017gpop}                           & WWW '17 & Peeking & Regression        & \C  &    & \C & \C  &             \\ 
        Lu \& Syzmanski~\cite{lu2017predicting}                     & IPDPSW '17 & Peeking & Classification    &     &    &    & \C  & \C          \\ 
        Shafiq \& Liu~\cite{shafiq2017cascade}                      & Networking '17 & Peeking & Classification &   & \C&   &   &                 \\ 
        Lv et al.~\cite{lv2017multi}                                & ACM MM '17 & \textit{Ex-ante} & Regression        &     &    &    & \C  & \C          \\
        Xie et al.~\cite{xie2017whats}                              & BigData '17& Peeking & Classification & \C& \C& \C&   & \C                                    \\ 
        Luo \& Liu~\cite{luo2018real}                               & COLING '18 & \textit{Ex-ante}  & Classification    &     &    &    & \C  & \C          \\ 
        Jia et al.~\cite{jia2018predicting}                         & CN '18     & Both              & Classification    &     & \C & \C & \C  &             \\ 
        Wu et al.~\cite{wu2018beyond}                               & ICWSM '18  & \textit{Ex-ante}  & Regression        &     &    &    & \C  & \C          \\ 
        Zhao et al.~\cite{zhao2018comparative}                      & IJCAI '18  & \textit{Ex-ante}         & Classification    &     &    &    &     & \C          \\ 
        Kong et al.~\cite{kong2018exploring}                        & SMC '18    & Peeking                  & Classification    & \C  & \C &  \C  & \C  &           \\ 
        Tsugawa~\cite{tsugawa2019empirical}                         & ICWSM '19  & Peeking & Classification    &     &    & \C & \C  & \C          \\ 
        & & & & 51 & 32 & 39 & 53 & 49 \\
        \multicolumn{4}{l}{Total: 89} & \footnotesize{57\%} & \footnotesize{36\%}& \footnotesize{44\%}& \footnotesize{60\%}& \footnotesize{55\%} \\
    \bottomrule
    \end{tabular}
\end{table}

\begin{table}\small
  \caption{Temporal Features}
  \label{tab:temporal-features}
  \begin{tabular}{@{} lp{11cm} @{}}
  \toprule
      \textbf{Feature}                        & \textbf{Description} \\ \midrule
      \textit{change\_rate}                   & The change rate of early popularity before observation time $t_o$. \cite{cheng2014can,figueiredo2013prediction,figueiredo2014dynamics,krishnan2016seeing,liu2015effectively,rizos2016predicting,shafiq2017cascade,vallet2015characterizing,wang2015burst,xie2017whats} \\
      \textit{dormant\_period}                & Dormant period refers to the time period before the item getting adopted. \cite{gao2019taxonomy,kong2014predictingbursts,kong2015towards} \\
      \textit{local\_peaks}                   & Local peak means in this time interval the increasing speed of popularity is larger than its neighbor intervals' speed. \cite{wang2015burst} \\
      \textit{maximum\_interval}              & Maximum time interval, i.e., $\text{max}([t_j - t_{j-1}]_j)$. \cite{gao2014effective,gao2019taxonomy} \\ 
      \textit{mean\_$t$, stddev\_$t$}         & Mean, median, sum and/or standard deviation of time series. \cite{cheng2014can,gao2019taxonomy,kong2014predictingbursts,kong2015towards,krishnan2016seeing,lee2010approach,liu2015effectively,rizos2016predicting,shulman2016predictability,xie2017whats,zaman2014bayesian,zhang2016structure} \\ 
      \textit{$P_i(t_j)$}                     & Popularity of information item $I_i$ at time $t_j$ or incremental\ popularity at a time interval. \cite{bao2013popularity,bora2015role,borghol2012untold,castillo2014characterizing,cheng2013understanding,dong2016can,figueiredo2013prediction,figueiredo2014dynamics,gupta2012predicting,he2014predicting,kong2014predicting,kong2014predictingbursts,kong2016popularity,kupavskii2013predicting,lee2010approach,lu2014predicting,ma2012will,ma2013predicting,mishra2016feature,oghina2012predicting,rizos2016predicting,suh2010want,szabo2010predicting,tatar2011predicting,tsagkias2010news,tsugawa2019empirical,tsur2012s,vallet2015characterizing} \\
      \textit{peek\_fraction}                 & The maximum of incremental popularity in all time intervals divided by the total popularity $P(t_o)$ at observation time. \cite{figueiredo2013prediction,figueiredo2014dynamics} \\ 
      \textit{publication\_time}              & Publication time of an information item, e.g., year, month, week, day, hour, minute, etc. \cite{artzi2012predicting,borghol2012untold,figueiredo2013prediction,lakkaraju2013s,lv2017multi,martin2016exploring,petrovic2011rt,rowe2011predicting,totti2014impact,tsagkias2009predicting,vallet2015characterizing,wang2015burst,yan2016sth} \\
      \textit{stage}                          & Whether an item published at an early stage or late stage during observation time. \cite{yang2010predicting} \\ 
      \textit{$t_1$}                          & $t_1$ measures speed how soon the first participant $u_1$ will join in the cascade. \cite{bao2013cumulative,galuba2010outtweeting,lee2010approach} \\
      \textit{$t_k$}                          & $t_k$ measures the how soon a cascade reaches size $k$. \cite{bora2015role,cheng2014can,guo2015toward,krishnan2016seeing,rizos2016predicting,shulman2016predictability,weng2014predicting,xie2017whats,zhang2016structure} \\
      \textit{time\_series\_vector}           & $k$-dimensional time series vector in fixed $k$ number observation setting, or vector in fixed time observation setting. \cite{cheng2014can,gao2014effective,gao2014popularity,gao2019taxonomy,gursun2011describing,he2014predicting,hong2011predicting,trzcinski2017predicting} \\ 
      \textit{$[(t_j - t_{j-1})]_j$}          & Time series between the $j$-th participant $u_j$ and $(j-1)$-th participant $u_{j-1}, j\in [1, k]$, which is a $k$-dimensional time vector. This can also be extended to $[(P(t_j)-P(t_{j-1}))]$, where $[t_{j-1}, t_j]$ is a predefined time interval. \cite{bao2013cumulative,cheng2014can,figueiredo2013prediction,gao2014popularity,gao2019taxonomy,hong2011predicting,kong2014predicting,kong2018exploring,mishra2016feature,pinto2013using,wang2015burst,weng2014predicting,yi2016mining,yu2014twitter,zaman2014bayesian} \\ 
      \textit{time\_series\_cluster}          & Which trend cluster/shape a sequence of early time series belongs to. \cite{figueiredo2013prediction,figueiredo2014dynamics,gao2019taxonomy,gursun2011describing,kong2014predictingbursts,kong2015towards,kong2018exploring,yang2011patterns,yu2015lifecyle} \\
  \bottomrule
  \end{tabular}
\end{table}

\begin{table}\small
  \caption{Cascade Graph, $r$-reachable Graph, and Global Graph Features}
  \label{tab:structural-features}
  \begin{tabular}{lp{10.5cm}}
  \toprule
      \textbf{Feature}     & \textbf{Description} \\ \midrule
      \textit{authority/hub\_score}                       & Authority or hub scores of nodes in graph $\mathcal{G}$. \cite{gao2014effective,gao2014popularity,gao2019taxonomy,ma2012will,ma2013predicting,qiu2018deepinf,yan2011citation,yi2016mining,yu2014twitter} \\
      \textit{branching\_factor}                      & Branching factor is the number of children nodes at each node in tree cascade graphs $\mathcal{G}_c$. \cite{dow2013anatomy} \\
      \textit{centrality}                             & Centrality measured by a specific algorithm, e.g., centrality of eigenvector, closeness, and betweenness. \cite{alzahrani2015network,guo2016comparison,qiu2018deepinf,yi2016mining,zhang2016structure} \\
      \textit{clustering\_coefficient} & Clustering coefficient of graph $\mathcal{G}$. \cite{cheng2013understanding,galuba2010outtweeting,gao2014effective,gao2014popularity,gao2019taxonomy,hong2011predicting,jia2018predicting,qiu2018deepinf,shafiq2017cascade} \\ 
      \textit{community}                              & Communities detected by using specific algorithms. Example features including number of infected communities, intra-communities, Gini impurity, etc. \cite{cao2020popularity,guo2015toward,li2017deepcas,shafiq2017cascade,tsugawa2019empirical,weng2013virality,weng2014predicting,xie2017whats} \\
      \textit{connect\_component}                     & Connect component in the graph $\mathcal{G}$, measured by number, component size, maximum component size, etc. \cite{bora2015role,gao2014effective,gao2014popularity,gao2019taxonomy,ma2012will,ma2013predicting,qiu2018deepinf,romero2013interplay,shulman2016predictability,ugander2012structural} \\
      \textit{degree}                                 & Degrees (both in- and out-degrees in directed graph) of nodes in graph $\mathcal{G}$. \cite{abisheva2014watches,alzahrani2015network,bakshy2011everyone,bora2015role,cao2020popularity,cui2013cascading,dong2015will,dow2013anatomy,cheng2014can,galuba2010outtweeting,gao2019taxonomy,hong2011predicting,jamali2009digging,jia2018predicting,kong2014predicting,kong2014predictingbursts,kong2015towards,kong2016popularity,kong2018exploring,krishnan2016seeing,li2017deepcas,rizos2016predicting,shafiq2017cascade,shulman2016predictability,tsugawa2019empirical,vu2011dynamic,wang2015burst,xie2017whats,yang2012we,yi2016mining,yu2014twitter,zaman2014bayesian,zhang2016structure} \\
      \textit{density}                                & Density of graph, defined as the number of edges divided by all possible edges in $\mathcal{G}$, i.e., $|\mathcal{E}|/(|\mathcal{V}| \times (|\mathcal{V}|-1))$. \cite{bora2015role,gao2014effective,gao2014popularity,kong2014predictingbursts,kong2015towards,kong2016popularity,krishnan2016seeing,li2017deepcas,ma2012will,ma2013predicting,qiu2018deepinf,romero2013interplay,shulman2016predictability,wang2015burst,yang2012we,yi2016mining} \\
      \textit{depth}                                  & Depth of the path from node $u_0$ to node $u_j$ in graph $\mathcal{G}$. \cite{bao2013cumulative,bao2013popularity,cheng2014can,cheng2013understanding,galuba2010outtweeting,gao2014effective,gao2014popularity,guo2016comparison,jamali2009digging,kong2014predicting,kong2014predictingbursts,kong2016popularity,kong2018exploring,krishnan2016seeing,rizos2016predicting,shafiq2017cascade,shulman2016predictability,weng2014predicting,yi2016mining,zaman2014bayesian,zhang2016structure} \\
      \textit{direct\_connect}                        & Number of nodes in cascade graph $\mathcal{G}_c$ that are directly connected to the root node $u_0$, i.e., $|\{u_j|\text{dist}(u_0, u_j) = 1, 1 \leq j \leq N\}|$. \cite{cheng2014can,krishnan2016seeing,shulman2016predictability} \\
      \textit{edge\_density}                          & Edge density is defined by the ratio of number of edges to the number of all possible edges. \cite{bao2013popularity,cao2020popularity,gao2019taxonomy,kong2014predicting,kong2018exploring,li2017deepcas} \\ 
      \textit{edges\_$\mathcal{G}_c$}                 & Number of (weighted/unweighted) edges in cascade graph $\mathcal{G}_c$. \cite{bora2015role,cao2020popularity,cheng2014can,krishnan2016seeing,romero2013interplay} \\ 
      \textit{edges\_$\mathcal{G}_c^1$}               & Number of (weighted/unweighted) edges in $1$-reachable graph $\mathcal{G}_c^1$. \cite{cao2020popularity,cheng2014can,krishnan2016seeing,gao2019taxonomy,li2017deepcas,romero2013interplay,shulman2016predictability} \\
      \textit{indirect\_connect}                      & Number of nodes in cascade graph $\mathcal{G}_c$ that are indirectly connected to the root node $u_0$, i.e., $|\{u_j|\text{dist}(u_0, u_j) > 1, 1 \leq j \leq N\}|$. \cite{cheng2014can,krishnan2016seeing} \\
      \textit{leaf\_nodes\_$\mathcal{G}_c$}           & Number of leaf nodes in cascade graph $\mathcal{G}_c$. \cite{cao2017deephawkes,cao2020popularity,krishnan2016seeing,li2017deepcas} \\
      \textit{nodes\_$\mathcal{G}_c$}                 & Number of nodes in graph $\mathcal{G}_c$, i.e., $|\mathcal{V}_c|$. \cite{abisheva2014watches,bao2013cumulative,cao2020popularity,gao2014popularity,jia2018predicting,kong2014predictingbursts,kong2015towards,krishnan2016seeing,ma2012will,ma2013predicting,romero2013interplay,shafiq2017cascade,weng2013virality,weng2014predicting,yi2016mining} \\
      \textit{nodes\_$\mathcal{G}_c^1$}               & Number of nodes in $1$-reachable graph $\mathcal{G}_c^1$, i.e., $|\mathcal{V}_c^1|$. \cite{abisheva2014watches,cao2020popularity,cheng2014can,gao2014effective,gao2014popularity,guo2015toward,guo2016comparison,li2017deepcas,ma2012will,ma2013predicting,romero2013interplay,weng2013virality,weng2014predicting,xie2017whats} \\
      \textit{nodes\_$\mathcal{G}_c^2$}               & Number of nodes in $2$-reachable graph $\mathcal{G}_c^2$, i.e., $|\mathcal{V}_c^2|$ \cite{gao2019taxonomy,guo2016comparison,krishnan2016seeing,shulman2016predictability,weng2014predicting} \\
      \textit{pagerank}                               & PageRank. \cite{alzahrani2015network,dong2015will,dong2016can,he2014predicting,hong2011predicting,jia2018predicting,kupavskii2012prediction,kupavskii2013predicting,qiu2018deepinf,romero2011influence,wang2015burst,weng2014predicting,yang2012we,yi2016mining,yu2014twitter} \\
      \textit{reciprocity}                            & Reciprocity of directed graph $\mathcal{G}$ (cf.~\cite{garlaschelli2004patterns}, defined as the ratio of number of reciprocal links to the total number of links, i.e., $(\sum_{i, j}^{|\mathcal{E}|}\mathbbm{1}(A_{i, j} = A_{j, i}))/|\mathcal{E}|$. \cite{gao2014effective,gao2014popularity,gao2019taxonomy,hong2011predicting} \\
      \textit{similarity\_$\mathcal{G}_c$}            & Similarity of two cascade graphs measured by a specific algorithm, e.g., graph edit distance or vertex/edge overlap. \cite{wang2015burst} \\
      \textit{structural\_virality}                   & Structural virality is calculated by Wiener index (cf.~\cite{goel2015structural}). \cite{kong2018exploring,rizos2016predicting,wang2015burst} \\
      \textit{$|\mathcal{V}_c \cap \mathcal{N}_g(u_j)|$}& Number of nodes who in cascade graph $\mathcal{G}_c$ are also neighboring nodes to $u_j$ in global graph $\mathcal{G}_g$, e.g., how many followers of $u_j$ retweet $u_j$'s tweet in Twitter, sometimes refer to the retweet ratio $|\mathcal{V}_c| \cap \mathcal{N}_g(u_j)| / |\mathcal{N}_g(u_j)|$.  \cite{bakshy2011everyone,cheng2014can,krishnan2016seeing,kupavskii2013predicting,petrovic2011rt,ruan2012prediction,szabo2010predicting,zhang2016structure}. \\
      \textit{subcascade\_graph}                    & In tree cascade graph $\mathcal{G}_c$, for each node in $\mathcal{V}_c$, its subcascade graph is composed of all the descendants of this node. \cite{cao2020popularity,dow2013anatomy,shafiq2017cascade} \\
      \textit{triangles}                            & Number of triangles in graph $\mathcal{G}$. \cite{cao2020popularity,krishnan2016seeing,ma2012will,ma2013predicting,li2017deepcas,romero2011differences,vu2011dynamic} \\
  \bottomrule
  \end{tabular}
\end{table}

\begin{table}\small
  \caption{User/Item Features}
  \label{tab:user-features}
  \begin{tabular}{lp{10.5cm}}
  \toprule
      \textbf{Feature}                                & \textbf{Description} \\ \midrule
      \textit{activity\_$u_j$}                        & Activity measures the frequency of user $u_j$'s activities. \cite{cheng2014can,guille2012predictive,kong2014predictingbursts,krishnan2016seeing,lu2014predicting,rowe2011predicting,shulman2016predictability,wu2018beyond,yang2012we,yu2014twitter,yu2015micro} \\ 
      \textit{age\_$u_j$\_account}                    & Age of user $u_j$'s account from item publication date to the account creation date. \cite{bakshy2011everyone,borghol2012untold,cheng2014can,gao2019taxonomy,jia2018predicting,khosla2014makes,krishnan2016seeing,kupavskii2012prediction,kupavskii2013predicting,lu2014predicting,martin2016exploring,mishra2016feature,rowe2011predicting,shulman2016predictability,stieglitz2012political,suh2010want} \\
      \textit{age\_$u_j$}                             & Age of user $u_j$. \cite{cheng2014can,dow2013anatomy,yuan2016will} \\
      \textit{age\_$I_i$}                             & Age of item $I_i$. \cite{borghol2012untold,cheng2013understanding,dong2016can,yang2012we} \\
      \textit{attractiveness\_$I_i$}                  & Attractiveness of item $I_i$ is defined as the ratio of users who adopt this item after seeing this item. \cite{li2013popularity} \\
      \textit{favorites\_$u_j$}                       & Number of favorite items user $u_j$ likes. \cite{kupavskii2012prediction,kupavskii2013predicting,lu2014predicting,petrovic2011rt,suh2010want,vallet2015characterizing} \\
      \textit{followees\_$u_j$}                       & Number of users who follow user $u_j$, this is a special case of in-degree of user $u_j$ in global graph $\mathcal{G}_g$ (follower/followee graph). \cite{artzi2012predicting,bakshy2011everyone,castillo2014characterizing,chen2016micro,cheng2014can,gao2019taxonomy,huberman2008social,kong2018exploring,lu2014predicting,rowe2011predicting,suh2010want,totti2014impact,tsur2012s,vallet2015characterizing,wang2015burst,xie2015modelling,yu2015micro,zaman2010predicting,zhao2015seismic} \\
      \textit{followers\_$u_j$}                       & Number of followers to user $u_j$, this is a special case of out-degree of user $u_j$ in global graph $\mathcal{G}_g$ (follower/followee graph). \cite{abisheva2014watches,artzi2012predicting,bakshy2011everyone,bielski2018understanding,borghol2012untold,castillo2014characterizing,chen2016micro,cheng2014can,gupta2012predicting,he2014predicting,huberman2008social,jenders2013analyzing,kobayashi2016tideh,kong2014predictingbursts,kong2015towards,kong2018exploring,krishnan2016seeing,kupavskii2012prediction,kupavskii2013predicting,lu2014predicting,luo2018real,martin2016exploring,mishra2016feature,petrovic2011rt,rizoiu2018sir,romero2011influence,rowe2011predicting,ruan2012prediction,shulman2016predictability,stieglitz2012political,suh2010want,tsur2012s,vallet2015characterizing,wang2015burst,weng2014predicting,xie2015modelling,yu2015micro,zaman2010predicting,zaman2014bayesian,zhang2016structure,zhao2015seismic} \\ 
      \textit{friends\_$u_j$}                         & Number of friends of user $u_j$, where \textit{friend} is a reciprocal relationship between two users. \cite{artzi2012predicting,borghol2012untold,castillo2014characterizing,cheng2014can,gao2019taxonomy,gupta2012predicting,huberman2008social,khosla2014makes,kupavskii2012prediction,kupavskii2013predicting,luo2018real,martin2016exploring,mcparlane2014nobody,mishra2016feature,petrovic2011rt,suh2010want,tsur2012s,vallet2015characterizing,wu2016unfolding} \\
      \textit{gender\_$u_j$}                          & Gender of user $u_j$. \cite{abisheva2014watches,cheng2014can,dow2013anatomy,jia2018predicting,krishnan2016seeing,mcparlane2014nobody,totti2014impact,wang2015burst,yuan2016will,zhang2016structure} \\
      \textit{$h$-index}                              & Hirsch index of users' past publications. \cite{dong2015will,rizos2016predicting,romero2011influence,yan2011citation} \\
      \textit{historical\_items\_$u_j$}               & Number of historical items a user $u_j$ published. \cite{bakshy2011everyone,borghol2012untold,chen2016micro,dong2015will,hong2011predicting,jia2018predicting,khabiri2009analyzing,khosla2014makes,kong2014predictingbursts,kong2015towards,kupavskii2013predicting,lu2014predicting,martin2016exploring,mcparlane2014nobody,mishra2016feature,petrovic2011rt,rowe2011predicting,suh2010want,totti2014impact,vallet2015characterizing,wang2015burst,wu2016unfolding,yan2011citation,yang2010predicting,yang2010understanding,yang2012we,yu2014twitter,zhang2016structure} \\
      \textit{influence\_$u_j$}                       & User $u_j$'s influence measured by a specific criteria. \cite{krishnan2016seeing,romero2011influence,ruan2012prediction,wang2015learning,yang2010predicting,yang2012we} \\
      \textit{interests\_$u_j$}                       & User $u_j$'s interests/preferences in topics or items. \cite{abisheva2014watches,bian2014predicting,yang2010understanding,yang2012we} \\
      \textit{location\_$u_j$}                        & The location (country/region) of user $u_j$. \cite{abisheva2014watches,bora2015role,tsur2012s,wang2015burst,yuan2016will} \\
      \textit{name\_$u_j$}                            & Name of the user $u_j$. \cite{zaman2010predicting} \\
      \textit{passivity\_$u_j$}                       & Passivity/susceptibility measures how difficulty a user $u_j$ to be influenced. \cite{kong2015towards,lu2017predicting,romero2013interplay,wang2015learning} \\ 
      \textit{past\_success}                          & Past success or historical popularity of users/items. \cite{abisheva2014watches,artzi2012predicting,bakshy2011everyone,borghol2012untold,chen2016micro,dong2015will,hong2011predicting,jia2018predicting,khabiri2009analyzing,khosla2014makes,kong2014predictingbursts,kong2015towards,kupavskii2012prediction,kupavskii2013predicting,lakkaraju2011attention,luo2018real,lv2017multi,martin2016exploring,mishra2016feature,romero2011influence,tsugawa2019empirical,wu2016unfolding,wu2018beyond,yan2011citation,yu2014twitter} \\ 
      \textit{profile\_views\_$u_j$}                  & The number of profile views of user $u_j$. \cite{khabiri2009analyzing} \\
      \textit{rating\_$I_i$}                          & The rating of information item $I_i$. \cite{borghol2012untold,cheng2013understanding} \\
      \textit{referrer\_$I_i$}                        & Referrer means a external source or link that pointed to the information item $I_i$, e.g., a link refer to a video.  \cite{abisheva2014watches,borghol2012untold,castillo2014characterizing,figueiredo2013prediction,figueiredo2014dynamics} \\ 
      \textit{relevance\_$u_j$\_$I_i$}                & How relevance between user $u_j$ and item $I_i$. \cite{dong2015will,yang2012we} \\
      \textit{similarity\_$I_i$}                      & How many similar items or even clones/duplicates with regard to item $I_i$, or how uniqueness $I_i$ is. \cite{borghol2012untold,cheng2014can,coscia2014average,lakkaraju2013s,lu2017predicting,wang2018learning} \\ 
      \textit{type\_$I_i$}                            & Which type the information item $I_i$ belongs to. \cite{jia2018predicting,petrovic2011rt,totti2014impact,yang2010predicting} \\
      \textit{type\_$u_j$}                            & Which type the user $u_j$ belongs to, e.g., verified or premium account, personal or organization account. \cite{chen2016micro,cheng2014can,khosla2014makes,kupavskii2012prediction,kupavskii2013predicting,lu2014predicting,luo2018real,mcparlane2014nobody,petrovic2011rt,wang2015burst,wu2016unfolding,zhang2016structure} \\
  \bottomrule
  \end{tabular}
\end{table}

\end{document}